\documentclass[aps,prd,twocolumn,nofootinbib,superscriptaddress]{revtex4-2}
\usepackage[utf8]{inputenc}
\usepackage{color}
\usepackage{graphicx}
\usepackage{amsmath}
\usepackage{amssymb}
\usepackage{bm}
\usepackage{acronym}
\usepackage{ifthen}
\usepackage{blindtext}
\usepackage[normalem]{ulem}
\usepackage{comment}
\usepackage{todonotes}
\usepackage{hyperref}
\hypersetup{
colorlinks=true,
linkcolor=blue,
filecolor=magenta,
urlcolor=blue,
citecolor=blue
}
\usepackage{orcidlink}
\usepackage{etoolbox}
\usepackage{fancyhdr}
\usepackage{xspace}
\usepackage{textcomp}
\usepackage{multirow}
\usepackage{enumitem}
\usepackage[caption=false]{subfig}
\usepackage{tabularx}
\usepackage{orcidlink}
\usepackage{soul}
\usepackage{listings}
\usepackage[ruled,vlined]{algorithm2e}

\SetCommentSty{mycommfont}
\usepackage{svg}

\newacro{S/N}{signal-to-noise ratio}

\newacro{PN}{post-Newtonian}

\newacro{O3}{the third observing run}

\newacro{O2}{the second observing run}

\newacro{CW}{\emph{Continuous Wave}}

\newcommand{\bea}{\begin{eqnarray}}
\newcommand{\eea}{\end{eqnarray}}
\newcommand{\be}{\begin{equation}}
\newcommand{\ee}{\end{equation}}

\newcommand{\td}{\mathrm{d}}
\newcommand{\Ms}{M_{\odot}}
\newcommand{\Mc}{\mathcal{M}_c}
\newcommand{\Mcmin}{\mathcal{M}_c^{\min}}
\newcommand{\Mcmax}{\mathcal{M}_c^{\max}}
\newcommand{\phic}{\phi_{\mathrm{corr}}}
\newcommand{\Tcoh}{T_{\mathrm{coh}}}
\newcommand{\CR}{\mathrm{CR}}
\newcommand{\Tobs}{T_\mathrm{obs}}
\newcommand{\Tfft}{T_\text{coh}}
\newcommand{\tmerg}{t_\text{merg}}

\newcommand{\dGC}{d_{\mathrm{GC}}}
\newcommand{\fmin}{f_0^{\min}}
\newcommand{\fmax}{f_0^{\max}}
\newcommand{\mTfft}{T_\mathrm{coh}^{\max}}
\newcommand{\minTfft}{T_\mathrm{coh}^{\min}}
\newcommand{\ximin}{\xi^{\min}}
\newcommand{\ximax}{\xi^{\max}}

\newcommand{\Trun}{T_\mathrm{run}}
\newcommand{\tarTfft}{T_\mathrm{coh}^{\mathrm{tar}}}

\newtoggle{fullauthorlist}
\toggletrue{fullauthorlist}

\newtoggle{endauthorlist}
\toggletrue{endauthorlist}

\begin{document}
\author{M. Andr\'es-Carcasona\orcidlink{0000-0002-8738-1672}}
\email{mandres@ifae.es}
\affiliation{Institut de Física d'Altes Energies (IFAE), The Barcelona Institute of Science and Technology, Campus UAB, E-08193 Bellaterra (Barcelona), Spain}
\author{O.J. Piccinni\orcidlink{0000-0001-5478-3950}}
\email{ornellajuliana.piccinni@anu.edu.au}
\affiliation{OzGrav, Australian National University, Canberra, Australian Capital Territory 0200, Australia}
\author{M. Mart\'inez\orcidlink{0000-0002-3135-945X}}
\affiliation{Institut de Física d'Altes Energies (IFAE), The Barcelona Institute of Science and Technology, Campus UAB, E-08193 Bellaterra (Barcelona), Spain}
\affiliation{Catalan Institution for Research and Advanced Studies (ICREA), E-08010 Barcelona, Spain}
\author{Ll. M. Mir\orcidlink{0000-0002-4276-715X}}
\affiliation{Institut de Física d'Altes Energies (IFAE), The Barcelona Institute of Science and Technology, Campus UAB, E-08193 Bellaterra (Barcelona), Spain}

\title{New approach to search for long transient gravitational waves from inspiraling compact binary systems}

\date{\today}

\begin{abstract}

The search for gravitational waves generated by the inspiral phase of binaries of light compact objects holds significant promise in testing the existence of primordial black holes and/or other exotic objects. In this paper, we present a new method to detect such signals exploiting some techniques typically applied in searches for continuous quasi-monochromatic gravitational waves. We describe the signal model employed and present a new strategy to optimally construct the search grid over the parameter space investigated, significantly reducing the search computing cost. Additionally, we estimate the pipeline sensitivity corroborating the results with software injections in real data from the LIGO third observing run. The results show that the method is well suited to detect long-transient signals and standard continuous gravitational waves. 
According to the criteria used in the grid construction step, the method can be implemented to cover a wide parameter space with slightly reduced sensitivity and lower computational cost or to focus on a narrower parameter space with increased sensitivity at a higher computational expense.
The method shows an astrophysical reach up to the Galactic Center (8kpc) for some regions of the parameter space and given search configurations.

\end{abstract}

\maketitle

\section{Introduction}
The detection of gravitational waves (GWs) by the LIGO, Virgo and KAGRA collaborations (LVK) \cite{FirstGWDet} has increased the interest in searching for sub-solar mass compact objects, as their potential primordial origin offers insights into new physics. Primordial black holes (PBHs), hypothetical remnants of the early universe \cite{Carr:1974nx,Carr:1975qj}, are compelling because they are viable dark matter (DM) candidates and can shed light on the early universe condition and the formation of large-scale structures \cite{LISACosmologyWorkingGroup:2023njw}. Unlike black holes formed through stellar evolution, PBHs could span a wide range of masses, from microscopic to supermassive \cite{Carr:2017edp,Carr:2017jsz,Carr:2019kxo,Carr:2020xqk}.
In principle, the direct detection of PBHs could be achieved through their gravitational wave signatures. When two PBHs merge, they emit GWs detectable by Earth-based interferometers~\cite{LIGOScientific:2014pky,VIRGO:2014yos,KAGRA:2018plz}. PBHs in binary systems could be identified by their unique merger rates, low masses and waveform characteristics, distinct from those of stellar-mass black holes~\cite{Sasaki:2016jop, Bird:2016dcv}. Population analyses of the GW events from the LVK transient catalogues~\cite{Abbott2021GWTC-3:Run,Abbott2020GWTC-2:Run,Abbott2019GWTC-1:Runs} 
have not yet found any significant evidence of the presence of PBHs~\cite{Hutsi:2020sol, Hall:2020daa, Wong:2020yig, Franciolini:2021tla, DeLuca:2021wjr,Franciolini:2022tfm,Andres-Carcasona:2024wqk}. 
The detection of a merger with at least one sub-solar component could be a definitive evidence of their existence. However, no PBH signal has been detected until now~\cite{LIGOScientific:2018glc, LIGOScientific:2019kan, Nitz:2020bdb, Phukon:2021cus, Nitz:2021mzz, Nitz:2021vqh, Nitz:2022ltl, Miller:2020kmv, Miller:2021knj, Morras:2023jvb, Mukherjee:2021ags, Mukherjee:2021itf, Andres-Carcasona:2022prl, Miller:2024fpo,Miller:2024jpo_2}. 

Another direct detection approach involves microlensing surveys. PBHs passing in front of a background of stars can cause a temporary increase in the star's brightness due to gravitational lensing. Projects like OGLE, MACHO, and EROS have conducted extensive microlensing surveys to constrain the abundance of PBHs in various mass ranges~\cite{Niikura:2019kqi, EROS-2:2006ryy, MACHO:2000qbb}.

The most interesting regions to search for PBH signals are the ones where the DM abundance is expected to be high, such as globular clusters, galactic centers, dwarf spheroidal galaxies, DM halos or cosmic web filaments. The Milky Way center, harbouring the supermassive black hole Sagittarius A*~\cite{balick1974intense,ghez2003first,Genzel:2010:SgrA,EventHorizonTelescope:2019dse}, is an ideal target for this investigation. The dense environment around the galactic center increases the likelihood of encounters between PBHs and other dark compact objects, leading to a higher chance of forming binary systems and enhancing the prospects of GWs detection~\cite{Pujolas:2021yaw}.

If PBHs exist in binaries, they could produce long-transient GWs during the inspiral phase of the coalescence~\cite{MaggioreBible}, emitting in the frequency band of Earth-based detectors. Typical methods used to search for compact binary coalescences (CBC) signals~\cite{Abbott2019GWTC-1:Runs,Abbott2020GWTC-2:Run,Abbott2021GWTC-3:Run} are not optimal for signals that last longer than minutes and, therefore, techniques to search for more persistent signals might be preferred. Continuous GWs (CWs) lie in this category of long-lasting signals, typically having durations comparable with the run (from months to years) \cite{Riles_Rev}. Several methods for the search of these fainter signals, deeply buried in the noise, have been developed over the years (see Refs.~\cite{Riles_Rev,Wette20yrReview,Tenorio_Rev,Ornella_Review} for some recent reviews). These methods are routinely applied for the search of standard CW signals like those emitted by isolated non-axisymmetric spinning neutron stars~\cite{CW_GC_O2,LIGOScientific:2021mwx, LIGOScientific:2021yby, LIGOScientific:2020lkw, LIGOScientific:2020qhb, KAGRA:2021una, LIGOScientific:2021ozr, LIGOScientific:2021inr, KAGRA:2022dqk, LIGOScientific:2022enz,CW_allSky_O3,CW_GC_O3,CW_Pulsars_O3,Pitkin_DualHarmonic,CW_TwoHarmonics,CW_TwoHarmonics2,CW_LongTransients_O3}.

This paper outlines a new method to detect such long-transient signals originating from the inspiral of light compact objects. The main novelty, with respect to existing methods \cite{Horowitz:2019pru,Miller_2021,Miller:2021knj,Miller:2024fpo,Alestas:2024ubs,velcani2024thesis,Miller:2024khl}, 
lies in the exploitation of the degeneracies of the signal waveform. This allows for the reduction of the problem's dimensionality and, hence, the decrease of the computing cost of the search.

The paper is organized as follows. Section~\ref{sec:signal} introduces the signal that is being targeted and the model used in this work. Section~\ref{sec:method} provides an overview of the proposed method and the framework for the data analysis used. Section~\ref{sec:cand} describes how the candidates are selected and vetoed. Section~\ref{sec:sens} shows the sensitivity estimate of the method. Finally, Section~\ref{sec:inj} shows the results of an injection campaign over real O3 data.

\section{Signal}
\label{sec:signal}

During the inspiral phase of a binary system of chirp mass $\Mc$, the two orbiting objects approach each other while radiating GWs. During this phase, their orbital frequency, and hence their GW frequency $f(t)$, increases as the objects get closer. In the Post-Newtonian (PN) formalism \cite{Cutler:1994ys,Poisson:1995ef,MaggioreBible}, the 0PN approximation is the leading-order correction to the Newtonian gravitational dynamics. It can describe the relativistic effects in the case of low velocities and weak gravitational fields. Since the masses considered in this work are small ($<10^{-3}~\Ms$) and we are targeting objects far from the coalescence, this approximation is within its range of validity.
The signal amplitude is given by:
\begin{equation}\label{eq:hsignal}
    h(t)=\frac{4}{d}\left( \frac{G\Mc}{c^2} \right)^{5/3}\left( \frac{\pi f(t)}{c}\right)^{2/3}~,
\end{equation}
being $d$ the luminosity distance to the source.
The frequency evolution, for the case of objects in circular orbits, can be expressed as a power law ~\cite{MaggioreBible}:
\begin{equation}
\label{eq:powerlaw}
    \dot{f}=\frac{96}{5}\pi^{8/3}\left( \frac{G\Mc}{c^3} \right)^{5/3}f^n(t)=kf^n(t),
\end{equation}
with $n=11/3$.
Integrating Eq.~\eqref{eq:powerlaw}, the frequency of the emitted signal is
\begin{equation}\label{eq:f(t)}
    f(t) =\left[ f_0^{1/\alpha}+\frac{k}{\alpha}(t-t_0) \right]^{\alpha}~,
\end{equation}
where $\alpha = \frac{1}{1-n}=-3/8$ and $f_0$ is the frequency at a reference time, $t_0$. 
It follows that the time to the merger will be
\begin{equation}
    \tmerg=\frac{f_0^{1/\alpha}}{(n-1)k}~.
\end{equation}

\begin{figure}[htbp]
    \centering
    \includegraphics[width=\columnwidth]{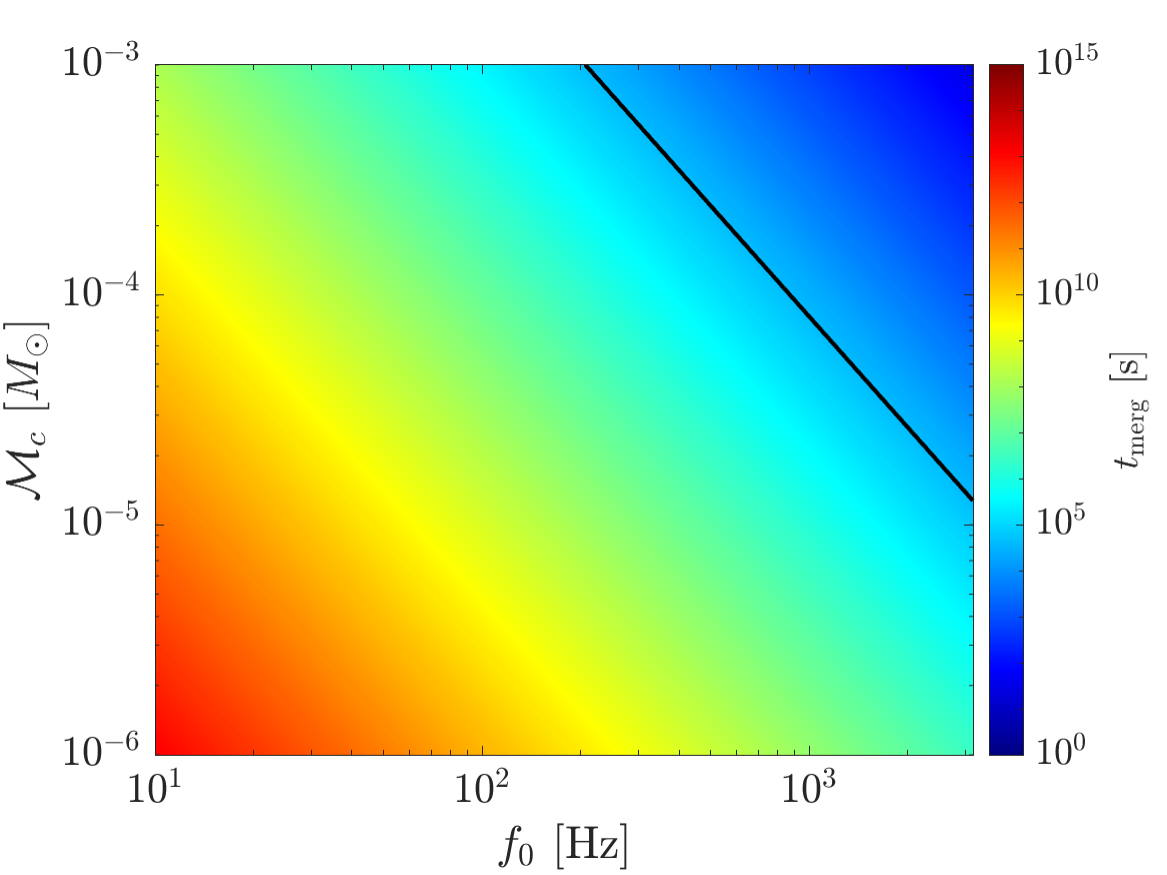}
    \caption{Merging time since $t_0$ in seconds. The black line indicates the curve corresponding to a $\tmerg=0.5$ days.}
    \label{fig:tmerg}
\end{figure}

Figure~\ref{fig:tmerg} shows the merging time for systems emitting in the frequency range of Earth-based detectors and for chirp masses between $[10^{-6}, 10^{-3}]~\Ms$. The typical duration of these signals is of the order $\mathcal{O}$(hours-months). As a reference, all those systems with a greater merging time than $0.5$ days lie below the black line. Large frequency variations are expected (e.g. $\sim 10^{-2}~\rm{Hz/s}$ for a system with a chirp mass of $10^{-3}~\Ms$ at around 300 Hz), well above the typical values covered in standard CW scenarios (the maximum frequency variation is - in absolute value - $10^{-8}~\rm{Hz/s}$). On the other hand, systems with small chirp masses have frequency variations comparable with more standard CWs.\footnote{It should be noted, however, that in standard CW searches for spinning neutron stars,  the signal is modelled with a spin-down rather than a spin-up, but from a technical point of view some search methods can be directly applied to spinning up sources with minimal adjustments.} For instance, variations of $\sim 10^{-9}~\rm{Hz/s}$ are expected from sources with $\Mc\sim 10^{-6}~\Ms$ at 100 Hz.

Due to the Earth's orbital and rotational motions, the signal that reaches the detector will be affected by the Doppler effect, resulting in a frequency modulation of:
\begin{equation}
    f_{d}(t)=f(t)\left( 1+\frac{\vec{v}\cdot \hat{n}}{c} \right)~,
\end{equation}
where $\vec{v}$ is the detector velocity and $\hat{n}$ is the unitary vector pointing to the source, in the Solar System barycenter reference frame. The Doppler contribution can be neglected for sufficiently short signals or particular choices of the search setup (see Sec.~\ref{sec:peakmap}).

\section{Method}
\label{sec:method}
As already mentioned, several CW detection methods exist, each prioritizing either the sensitivity, the computing cost or the pipeline's robustness.
Typical CW signals are characterized by their long duration $\mathcal{O}$(months or years), and a low signal-to-noise ratio if compared with transient signals emitted e.g. during the coalescence of two compact objects.  
The category of signals we are interested in this work lies in between transients and purely CWs with a duration of the order $\mathcal{O}$(hours-days) (See Sec.~\ref{sec:signal}). 
Optimal methods like matched filtering - typically applied for the search for transients CBC signals -  might be unfeasible for wide parameter space searches. This also applies to the long-transient regime considered in this work.
For this reason, we developed a new semi-coherent method to detect long-transient CW signals, exploiting the signal's parameter degeneracy to reduce the computing cost of the search. 
The method leverages on some standard CW search techniques, like the heterodyne demodulation procedure - adapted to the case of study - to increase the signal-to-noise ratio~\cite{BSD,Bayesian_CW}. 
In the following subsections, we will explain each step of the proposed method to search for this family of CW-transients signals.

\subsection{Data framework}
\label{sec:data}
For this work, the Band Sampled Data (BSD) framework and libraries are used \cite{BSD}. The BSD files are band-limited time series, down-sampled and divided into frequency sub-bands, partially cleaned from instrumental disturbances. 
The most recent version of the high-latency O3 calibrated gated data~\cite{Sun:2020wke,GWOSCO3,T2000384,PhysRevD.101.042003} was used for the injection analyses described in Sec.~\ref{sec:inj}. Additionally, only science segments - defined as time intervals during which the detector operated nominally with acceptable noise levels - were selected~\cite{T2300068}. 
The BSD files benefit from the removal of short-duration noise transients as in~\cite{Astone:2005fj}.
The data structures contain the reduced-analytic strain-calibrated time series of the LIGO and Virgo detectors. This data package allows for fast and flexible data management that reduces the computational cost of the analysis. 
In particular, the framework allows to construct input data either as long segments over small frequency bands or as short periods covering large frequency bands.  To date, several CW searches have been performed using the BSD library, mostly using the long-time/narrow-frequency band configuration~\cite{CW_GC_O2,CW_GC_O3,CW_SNR_O3a,CW_known_O1_O2_O3, CW_known_O2_O3,CW_ensemble_O3}, including those for DM candidates from superradiant boson clouds \cite{CW_BC_O3} or dark photon DM \cite{CW_DPDM_O3}.  

As shown in Fig.~\ref{fig:tmerg} we are interested in signals with durations considerably shorter than standard CW signals. Therefore, we will handle the data in the BSD short-time/wider-frequency band configuration. 
In practice, the data will be cut in time chunks of length $\Tobs$. The total number of periods to be analyzed will then be $N_T = \Trun/\Tobs$, where $\Trun$ is the total time of the observing run. This number is effectively reduced when each interferometer's duty cycle is considered. Typically, a detector is not taking data continuously or its quality is not good enough for science mode, therefore, in non-science periods there will be gaps in the data. A reasonable choice would be to discard those batches of time containing more than $40\%$ of non-science data, avoiding processing unnecessary chunks of time.

\subsection{Heterodyne correction}
As explained in Sec.~\ref{sec:signal}, long-transient signals may span a wider frequency band than standard CW sources.
To reduce its spread across the frequency band, we employ the heterodyne phase correction, which enhances the signal-to-noise ratio and improves detectability~\cite{BSD,Bayesian_CW} to demodulate the signal. This will allow us to use an excess power method to identify significant candidates in the frequency domain (see Sec.~\ref{sec:peakmap}). In the following, a short description of the heterodyne phase correction is provided.

For simplicity, the signal can be written as $s(t)=A(t)e^{i\phi(t)}$, where $A(t)$ is its amplitude and $\phi(t)$ its phase. The phase is related to the frequency evolution of the source as
\begin{equation}
    \frac{\td \phi}{\td t}=2\pi f(t)\, .
\end{equation}
Therefore, the phase evolution of a signal, $\phi_s(t)$, will be 
\begin{equation}
\label{eq:spin-up}
    \phi_s(t)=\frac{2\pi}{k(2-n)}f^{2-n}(t)\, .\\
\end{equation}
If we include the Doppler contribution the phase will be $\phi(t) = \phi_s(t) +\phi_d(t)$, and can be computed as
\begin{equation}
\label{eq:dopphet}
    \phi_d(t) = 2\pi \int_{t_0}^{t}f(t')\frac{\vec{v}\cdot \hat{n}}{c}~\td t'
\end{equation}
as discussed in Ref.~\cite{BSD}.

The measured and calibrated strain of an interferometer $h(t)$ will be a combination of the signal $s(t)$ and the noise $n(t)$.
In general, the noise can be non-stationary and non-Gaussian, but during short periods of time (of the order of some minutes at maximum), it can be assumed as stationary.

The heterodyne phase correction can now be applied to the strain data as
\begin{equation}
    h_{\mathrm{het}}(t) = h(t)e^{-i\phic(t)}~,
\end{equation}
where $\phic(t) = \phi(t)-2\pi f_0t$.
The effect of this correction is to modify the strain to 
\begin{equation}
    h_{\mathrm{het}}(t) = A(t)e^{i2\pi f_0 t}+n(t)e^{-i\phi(t)}~, 
\end{equation}
which implies that the signal has been converted to a purely monochromatic one (with a frequency of $f_0$) up to some residual modulations. These residual modulations could manifest in the case of incorrect modelling of the frequency evolution or if the assumed source sky position is incorrect (e.g. if the Doppler contribution is not negligible).

\subsection{$\xi-$space} 
As shown in Eq.~\eqref{eq:spin-up}, the signal phase, besides being a function of time, depends on the two source parameters $(f_0,\Mc)$ at the reference time $t_0$. At first glance, performing the heterodyne correction with different combinations of $(f_0,\Mc)$ might seem sufficient to complete the search.  However, this approach would not only be highly computationally demanding, but it also overlooks the fact that various parameter combinations can result in nearly identical frequency evolutions.
Indeed, multiple combinations of $(f_0,\Mc)$ might produce the same $\phi_s(t)$ and, hence, a single heterodyne correction would be enough to correct for more than one point in the $(f_0,\Mc)$ plane~\cite{Andres-Carcasona:2023PoS}.

Without loss of generality, we can define a new variable, $\xi$, that measures the relative increment of frequency during the observing time $\Tobs$:
\begin{equation} \label{eq:xi}
\xi(f_0,\Mc,\Tobs)=\left[ f_0^{1/\alpha}+\frac{k}{\alpha}\Tobs \right]^\alpha-f_0~.
\end{equation}

\begin{figure}[htbp]
    \centering
    \includegraphics[width=\columnwidth]{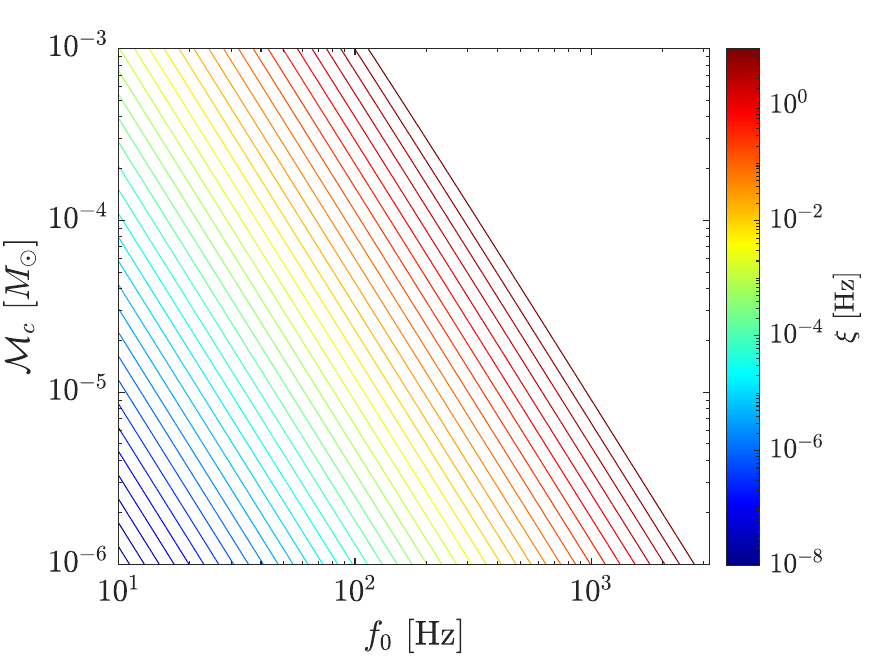}
    \caption{Iso-$\xi$ curves inside the region of the parameter space of interest with a fixed $\Tobs = 0.5$ days.}
    \label{fig:isoxi}
\end{figure}

This variable depends only on the reference frequency and chirp mass for a fixed observing time. Therefore, it maps the two-dimensional parameter space into the $\xi-$space. In Fig.~\ref{fig:isoxi} a plot containing the iso-$\xi$ curves is presented for a fixed $\Tobs$. Due to the 0PN model used for the frequency evolution, which is only valid for the inspiral part of the signal and not the merger, $\xi$ can diverge, as the frequency grows exponentially fast when it gets very close to the merger, or even take imaginary values if the argument inside the brackets becomes negative. We will therefore set a maximum frequency shift to avoid such divergences and imaginary values, these systems correspond to those with a merger time shorter than $\Tobs$. As $\Tobs$ increases, the magnitude of $\xi$ decreases and the slope of the curves $\Mc = \Mc(f_0,\xi,\Tobs)$ also decreases. Furthermore, higher frequencies and chirp masses in the upper right region of the parameter space become inaccessible, as $\xi$ takes imaginary values. Therefore, increasing $\Tobs$ allows more signals to merge within this time frame, which reduces the parameter space that can be effectively probed.

Equation~\eqref{eq:xi} can be inverted to find the values of $\Mc$ as a function of $\xi$ and the frequency $f_0$. This is
\begin{equation} \label{eq:Mc(xi)}
\Mc(\xi,f_0,\Tobs) = \frac{c^3}{G}\left[\frac{5\alpha}{96\pi^{8/3}}\frac{(\xi+f_0)^{1/\alpha}-f_0^{1/\alpha}}{\Tobs}\right]^{3/5}.
\end{equation}
While no analytical expression for $f_0=f_0(\xi,\Mc,\Tobs)$ exists, Eq.~\eqref{eq:Mc(xi)} can be numerically interpolated up to machine precision, to find the corresponding value of $f_0$.

\begin{figure}[htbp]
    \centering
    \includegraphics[width=1\columnwidth]{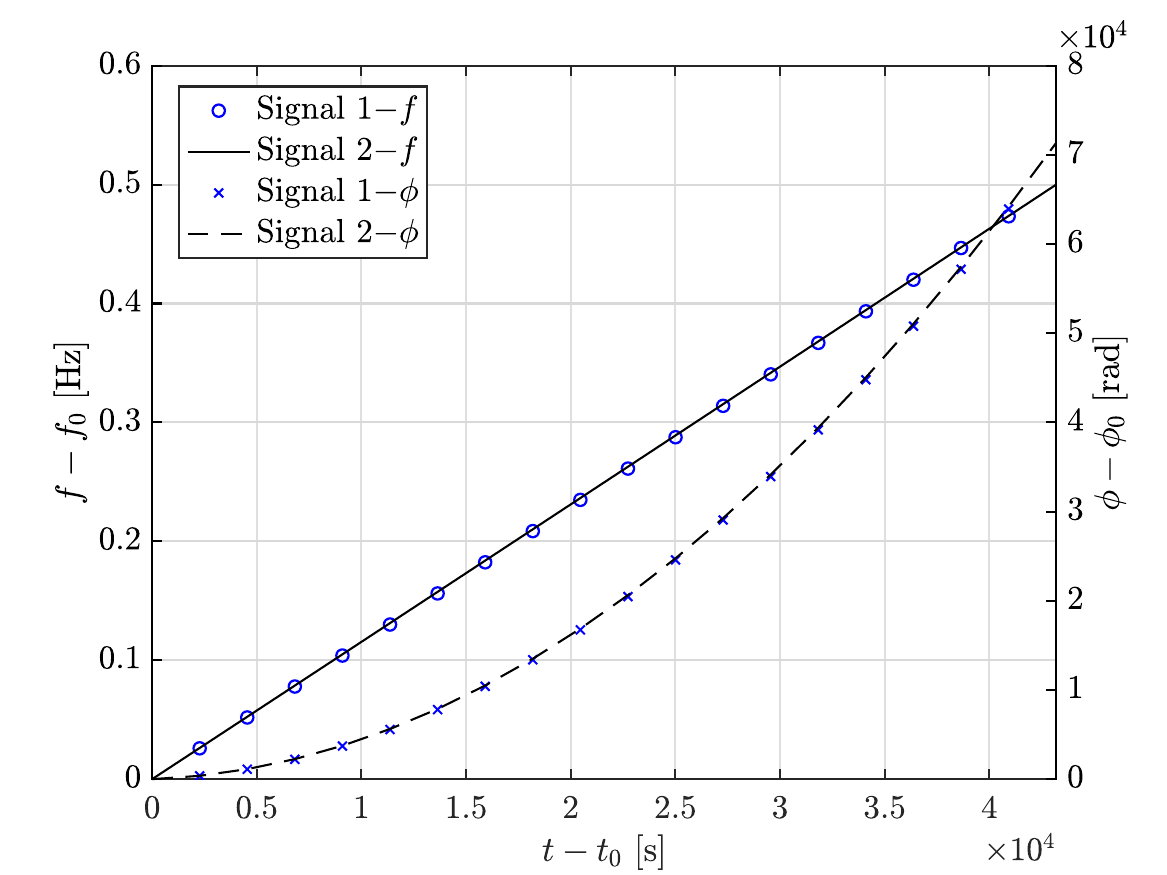}
    \caption{Comparison of the frequency and phase evolution of two signals with parameters $f_0 = 57.21~\rm Hz$, $\mathcal{M}_c=9.12\times 10^{-4}M_\odot$ (Signal 1) and $f_0 = 254.87~\rm Hz$, $\mathcal{M}_c=3.06\times 10^{-5}M_\odot$ (Signal 2). The signals lie on the same iso-$\xi$ curve $\xi=0.5~\rm Hz$. Here $\phi_0$ is the phase at $t_0$.}
    \label{fig:comparison_f_phi}
\end{figure}

In Fig.~\ref{fig:comparison_f_phi}  we show a comparison of the frequency and phase evolution for two signals belonging to the same iso-$\xi$ curve at $\xi= 0.5~\rm Hz$ with the following characteristics:
\begin{itemize}
    \item Signal 1: $f_0 = 57.21 ~\rm Hz$, $\mathcal{M}_c=8.12\times 10^{-4}M_\odot$.  
    \item Signal 2: $f_0 = 254.87~\rm Hz$, $\mathcal{M}_c=3.06\times 10^{-5}M_\odot$.
\end{itemize}

The frequencies and chirp masses of these signals are randomly chosen, but such that they keep the same value of $\xi$.

As it is observed, despite having very different parameters their relative frequency and phase evolve very similarly as the two signals belong to the same iso-$\xi$ curve. Given this phase evolution degeneracy, a unique heterodyne correction using a single set of $(f_0,\Mc)$ can be applied to the data to demodulate signals that belong to the same iso-$\xi$ curve.

\begin{figure*}[htbp]
    \centering
\includegraphics[width=1\textwidth]{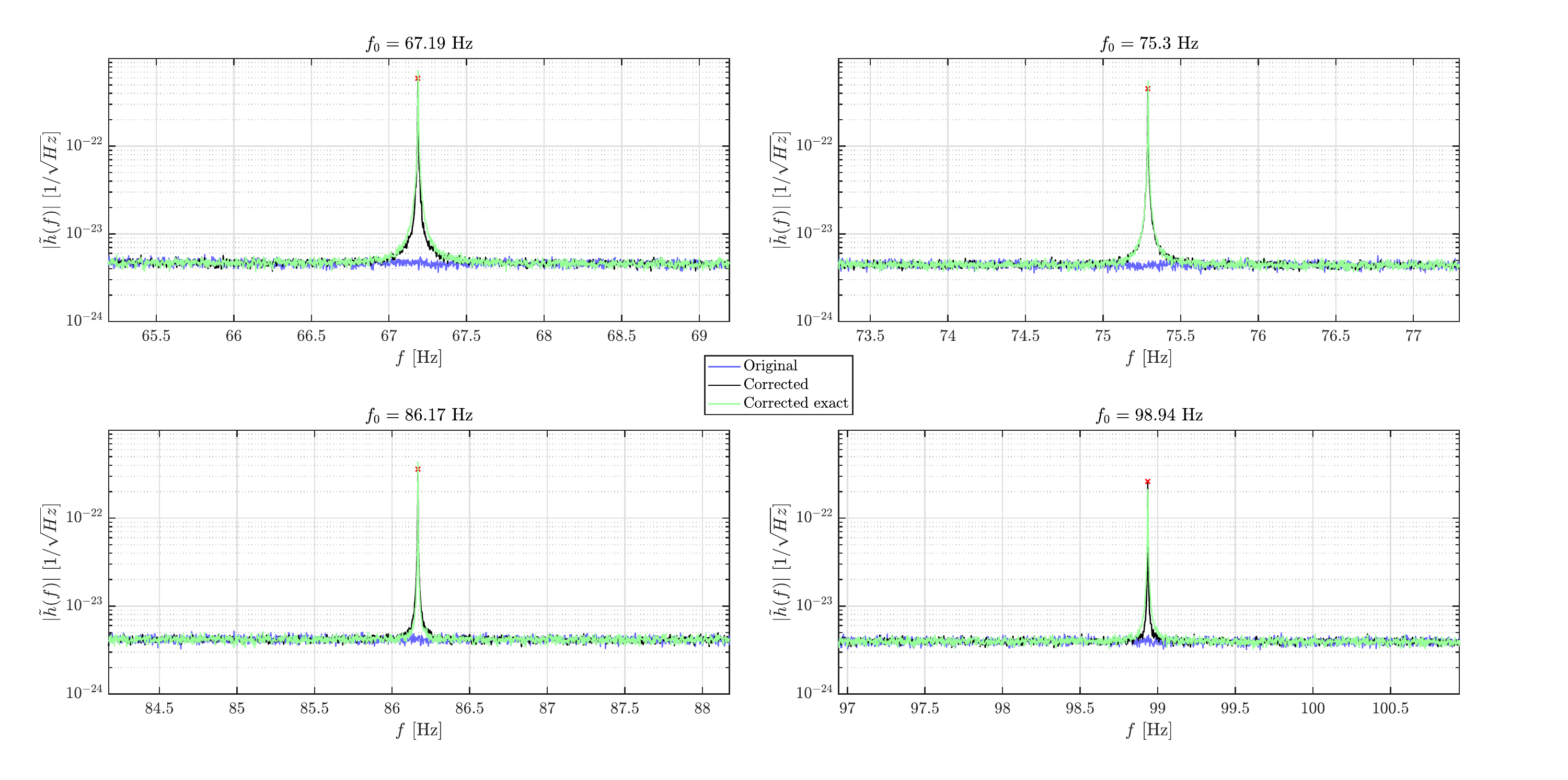}
    \caption{Power spectrum of the strain for injected signals with parameters in Tab.~\ref{tab:iso_xi}. For each signal, the red cross indicates the maximum of the power spectrum after the correction has been applied. The \textit{green} line shows the power spectrum after correcting the data with the exact parameters of the original injection. The \textit{black} line shows the power spectrum after correcting the data using $(\widehat{f_0},\widehat{\mathcal{M}}_c)$. The \textit{blue} curve is the original power spectrum when no demodulation is applied.}
    \label{fig:iso_xi_signals}
\end{figure*}

Multiple iso-$\xi$ injections with different $(f_0,\Mc)$ parameters are performed to show the method's validity. In Fig.~\ref{fig:iso_xi_signals}, we show the power spectra of four different injections belonging to $\xi=1~\rm Hz$ with a very high signal-to-noise ratio for illustrative purposes. In blue, we display the power spectra of the data containing a signal before applying the demodulation, while in black, we show the spectra after applying the same heterodyne correction to the four different datasets. Additionally, we include in green the power spectrum obtained correcting the signal with the exact parameters of the injection. The injection parameters are reported in Tab.~\ref{tab:iso_xi}. The reference time for the starting frequency $f_0$ is $t_0=1238198418$~s and the signals are injected in a chunk of data from the third observing run (O3) of the LIGO Livingston detector. We also report the degree of polarization $\eta= -2\cos\iota/(1+\cos^2 \iota)$, which is a function of the inclination angle of the binary's orbit, $\iota$.
For simplicity, we inject all the signals at the sky location of the Galactic Center ($\alpha=94.969^\circ$ and $\delta=17.3432^\circ$).

\begin{table}[htbp]
\begin{tabular}{c||c|c|c|c|c}
\textbf{ Signal} & $f_0$ [Hz]  &  $\Mc$ [$\Ms$]    & $d$ [kpc] & $\eta$ &$\psi$        \\ \hline \hline
 \textbf{1} & $67.19$ & $8.58\times 10^{-4}$ & \multirow{4}{*}{$0.1$}   & \multirow{4}{*}{$-0.85$} & \multirow{4}{*}{$8.9^\circ$}    \\ \cline{1-3}
 \textbf{2} & $75.30$ & $6.69\times 10^{-4}$ & & \\ \cline{1-3}
 \textbf{3} & $86.17$ & $4.98\times 10^{-4}$ & &  \\ \cline{1-3}
 \textbf{4} & $98.94$ &  $3.68\times 10^{-4}$ & & \\ \hline
  $(\widehat{f_0},\widehat{\mathcal{M}}_c)$ & $106.97$ & $3.10\times 10^{-4}$ & - & - & -\\ 
\hline \hline
\end{tabular}
\caption{Parameters of the 4 signals injected in the O3 Livingston data and shown in Fig.~\ref{fig:iso_xi_signals}. All the signals lie on the same $\xi$ curve of value $\xi = 1$ Hz. The last row reports the parameter used for the correction.}
\label{tab:iso_xi}
\end{table}

All of the power spectra in Figure~\ref{fig:iso_xi_signals}, show a clear peak at the frequency of the injected signal after applying the heterodyne demodulation using $(\widehat{f_0},\widehat{\mathcal{M}}_c)$.
Despite having very different parameters, the method is, in fact, capable of correctly demodulating signals along a single iso-$\xi$ line.
As a reference, we report the power spectra of the signal corrected using the exact parameters of the injection. We notice that no significant power loss is shown for these injections.

\subsection{Time-frequency map creation}
\label{sec:peakmap}
After the time series has been corrected using the heterodyne, it can be mapped into the time-frequency plane. This is done by creating the so-called \textit{peakmap}, which contains a collection of the most significant time-frequency peaks~\cite{Astone:2005fj}. These peaks are local maxima selected above a given threshold of the ratio between the modulus of the Fast Fourier Transform (FFT) and an estimation of the average spectrum (see Ref.~\cite{FreqHough}). The duration of the FFT, is referred to as the coherence time ($\Tcoh$), and the starting time corresponds to the time associated with the selected peak.
 
The threshold for selecting a peak significantly affects the probability of detecting a signal or missing it. In this method, the peakmap threshold is set to $2.5$, as in Ref.~\cite{FreqHough}. Once the peakmap is created, it is projected onto the frequency axis, and a histogram is generated showing the number of peaks $n$ in each frequency bin. The significance of the peaks in a frequency bin, which may indicate the presence of a signal, is determined by the critical ratio (CR), given by:
\begin{equation}\label{eq:CRstd}
    \CR = \frac{n-\mu}{\sigma}~,
\end{equation}
being $\mu$ the mean of $n$ and $\sigma$ its standard deviation. This is a standard statistic employed to select the first-level candidates (See e.g.~\cite{FreqHough}). All the candidates with $\CR$ values above a given threshold are selected and passed to the following post-processing phase discussed in section~\ref{sec:cand}.

\begin{figure*}[htbp]
    \centering
    \includegraphics[width=\textwidth]{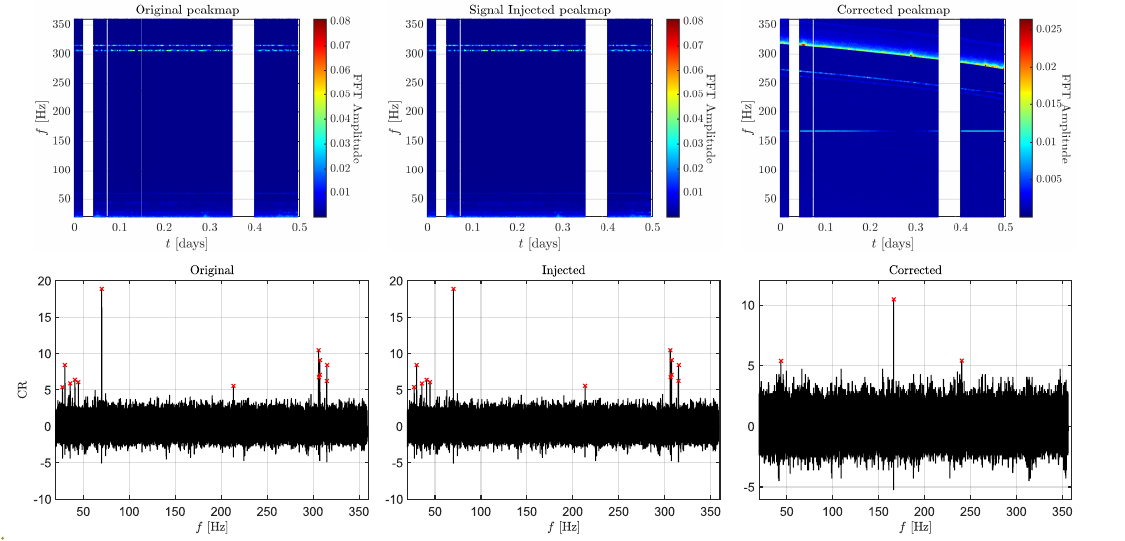}
    \caption{\textit{(Upper row}) Example of peakmaps built from real O3 data from LIGO Livingston. The left plot shows the original peakmap (noise-only), the middle plot contains the same data with an injected signal (uncorrected), and the right plot presents the results after applying the heterodyne correction. We use data between GPS times $t = 1238198418$~s and $t = 1238241618$~s, covering frequencies from $f = 20$~Hz to $f = 360$~Hz, with a coherence time $\Tcoh = 432$~s. The injection parameters are: $f_0 = 167.58$Hz, $\mathcal{M}_c = 5.67 \times 10^{-4}\Ms$, $d = 1$~kpc, $\alpha = 94.969^\circ$, and $\delta = 17.3432^\circ$.
    (\textit{Lower row}) The plots illustrate the CRs computed from each peakmap. Candidates with CR$ > 5$ are marked with red crosses in all three cases.} 
    \label{fig:PeakmapExample}
\end{figure*}

An example showing an injection of a signal in real O3 data of LIGO Livingston between GPS times $t=1238198418$~s and $t=1238241618$~s, frequencies $f=20$~Hz and $f=360$~Hz, and $\Tcoh=432~s$ is displayed in Fig.~\ref{fig:PeakmapExample}. The injection parameters are $f_0 =167.58$~Hz, $\mathcal{M}_c=5.67\times 10^{-4}~\Ms$, $d=1$~kpc. In this figure, the peakmaps and the CRs for the original data, the data with the signal injected and the data after the heterodyne correction has been applied are shown to demonstrate the effect of heterodyning. After the heterodyne (right-hand column), the signal becomes visible as a horizontal line around the reference frequency and it can also be identified as a high-significance candidate in the CR plot.

\subsubsection{Limits on the peakmap resolution} 
\label{sec:limits_Tcoh}
The natural frequency bin resolution of the peakmap has a width of $\delta f = 1/\Tfft$. In any semi-coherent search, the duration of the coherence time, $\Tfft$, impacts the sensitivity of the search, scaling as $1/\sqrt[4]{\Tfft}$, but also its computing cost, which scales at least as $T^2_{\rm coh}$. For this reason, given the computing limitations, using extremely long $\Tfft$ is generally not feasible and a maximum coherence time should be set. On the other hand, there are also restrictions on how short this length can be. If $\Tfft$ is too short, there will be fewer frequency bins which can lead to poor statistics when computing the CR. Furthermore, if the $\Tfft$ is too short, FFT-based methods are not optimal unless properly adapted, as they are more suitable for analyses of longer data chunks. Different arguments can be used to set the minimum coherence time. One possibility is to check the errors of the estimators of the mean and standard deviation used to compute the CR. As detailed in Appendix~\ref{appendix:tcohlim}, we find that for a search using $\Tobs= 0.5$~days, the minimum coherence time that can be used is $\sim 1.7$~s while the maximum is $\sim 432$~s. We note that, with these constraints, the Doppler modulation can be neglected.

\subsection{Grid on $\xi$: variable $\Tcoh$}\label{sec:gridxivarTcoh}

\begin{figure}[htbp]
    \centering
    \includegraphics[width = \columnwidth]{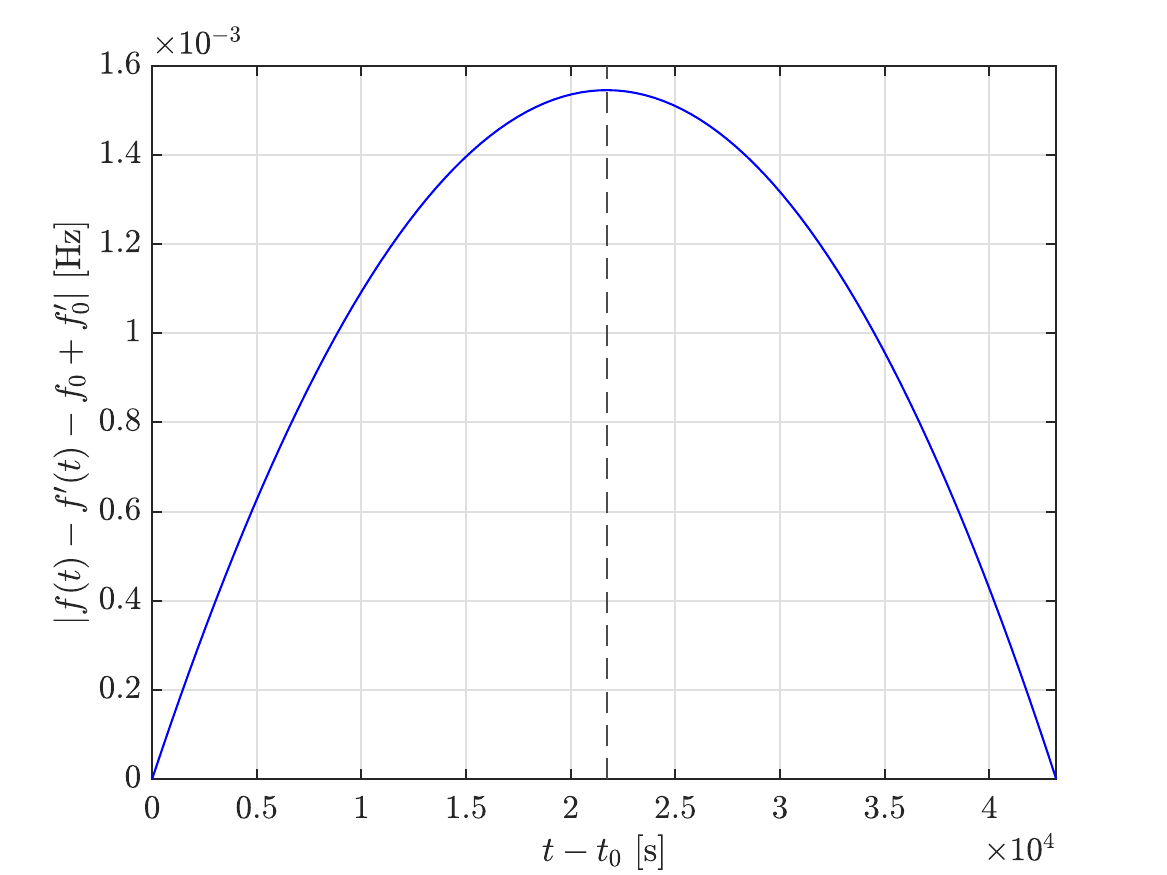}
    \caption{Comparison of the relative difference in the frequency evolution of two signals with parameters $f_0 = 57.21~\rm Hz$, $\mathcal{M}_c=9.12\times 10^{-4}M_\odot$ (Signal 1) and $f_0 = 254.87~\rm Hz$, $\mathcal{M}_c=3.06\times 10^{-5}M_\odot$ (Signal 2). The signals lie on the same iso-$\xi$ curve $\xi=0.5~\rm Hz$. The dashed vertical line is the time at $t=t_{\max}$ as in Eq.~\eqref{eq:tmax_Tcoh}.}
    \label{fig:Diff_freq_evolution}
\end{figure}

The optimal values for $\Tfft$ can be set assuming that,  after applying the partial correction using a given $(\widehat{f_0},\widehat{\mathcal{M}}_c)$, the signal will be fully confined in a single frequency bin, i.e.,  $\Tfft$  should satisfy the following condition, for two pairs $(f_0,\Mc)$ and $(f'_0,\Mc')$ of the same $\xi$ during $\Tobs$:

\begin{equation} \label{eq:maxt_Tfft}
    \mathcal{G}(\xi,f_0,f_0') = \max_{t\in[t_0,t_0+\Tobs]}\left| f(t)-f'(t)-f_0 +f_0' \right|\leq \frac{1}{\Tfft}~. 
\end{equation}
This can be interpreted as the maximum possible relative distance in frequency between two signals that have the same frequency increment in $\Tobs$ as depicted in Fig.~\ref{fig:Diff_freq_evolution}.

Equation~\eqref{eq:maxt_Tfft} has an analytical solution for $t$ at 
\begin{equation}\label{eq:tmax_Tcoh}
    t_{\max} = \frac{f_0^{1/\alpha}-\displaystyle\left( \frac{k'}{k}\right)^{\frac{1}{\alpha n}}f_0'^{1/\alpha}}{(1-n)\left[ \displaystyle\left( \frac{k'}{k}\right)^{\frac{1}{\alpha n}}k'-k\right]}~,
\end{equation}
where $k$ is the function of the chirp mass in Eq.~\eqref{eq:powerlaw}.

It should be noted that signals corresponding to more distant points on the iso-$\xi$ curve are less similar to each other (due to the dependence of $\dot{f}$ on $f_0$ and $\Mc$ (see Eq.~\eqref{eq:spin-up})). In particular, the coherence time must be adjusted to ensure effective signal demodulation for all points on the iso-$\xi$ curve. Specifically, a shorter coherence time is more optimal to cover the wider frequency variations. 

\begin{figure*}[ht]
    \centering
    \subfloat[Points of the grid.]{%
        \includegraphics[width=0.45\textwidth]{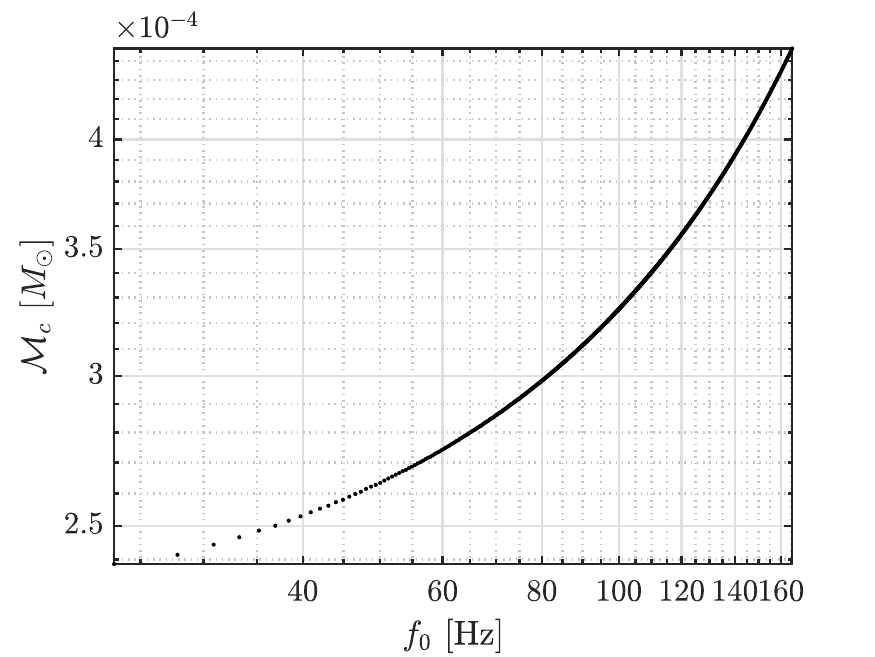}%
        \label{fig:GridVarTcoh}
    }
    \hfill
    \subfloat[Coherence time for the points of this grid.]{%
        \includegraphics[width=0.45\textwidth]{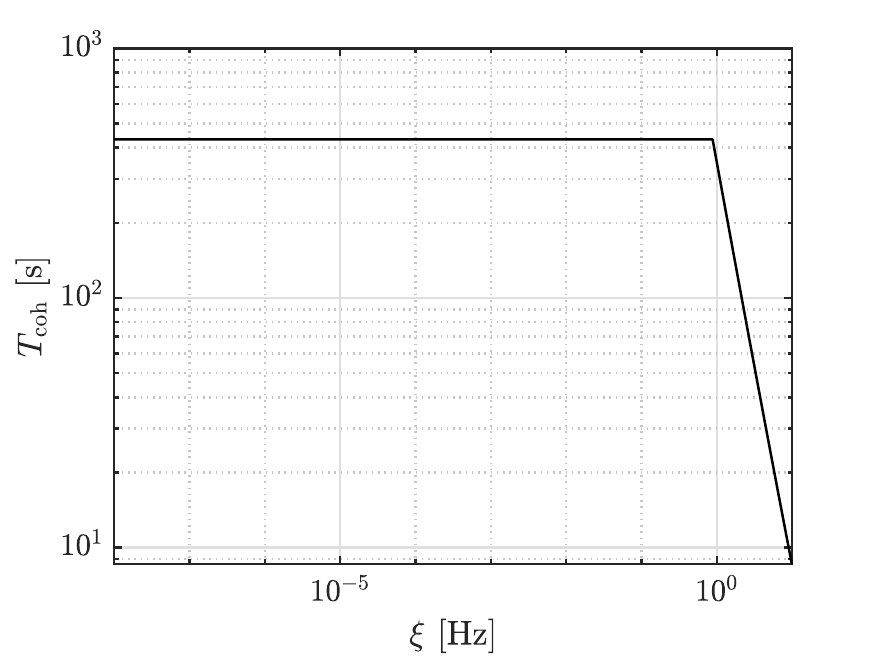}%
        \label{fig:TcohGridVarTcoh}
    }
    \caption{Grid with variable $\Tcoh$ and $\fmin = 10$ Hz, $\fmax = 300$ Hz, $\Mcmin = 10^{-6}~\Ms$, $\Mcmax = 10^{-3}~\Ms$, $\mTfft=432$ s, $\minTfft=1.7$ s, $\ximin=10^{-8}$ Hz, $\ximax = 10$ Hz.}
    \label{fig:AllGridVarTcoh}
\end{figure*}

Since we want to still use the maximum possible coherence time, a compromise is to choose the point where the variation of signal properties is minimal compared to the rest of the points on the same iso-$\xi$ (see Appendix~\ref{appendix:algo} for the details on the grid construction).
This approach is the one that creates the smallest possible grid that covers the full parameter space as it only places one point per iso-$\xi$ line while keeping the maximum possible $\Tfft$. A downside of this approach is that if $\Tfft \gtrsim 1/\Delta f_{d}^{\max}$ (Eq.~\eqref{eq:Doppler}) the Doppler modulation should be also included in the phase correction increasing the cost of the search. Furthermore, the limitations on the peakmap construction discussed in Sec.~\ref{sec:limits_Tcoh}, should also be taken into account. Additionally, for higher values of $\xi$, i.e. for signals with wide frequency variations, the placement of a single $(f_0,\Mc)$ point would allow for maximum values of the $\Tfft$ that are too small to meet the minimum coherence time requirements discussed in Sec.~\ref{sec:limits_Tcoh}. 
Hence, a maximum $\xi$ will be set for this type of grid, limiting the parameter space that can be investigated.
The grid obtained with $\fmin = 10$ Hz, $\fmax = 300$ Hz, $\Mcmin = 10^{-6}~\Ms$, $\Mcmax = 10^{-3}~\Ms$, $\mTfft=432$ s, $\minTfft=1.7$ s, $\ximin=10^{-8}$ Hz, $\ximax = 10$ Hz is displayed in Fig.~\ref{fig:GridVarTcoh}. It contains $851$ points in total. In Fig.~\ref{fig:TcohGridVarTcoh} we see that for smaller values of $\xi$, i.e. for signals having lower spin-up during $\Tobs$, the largest $\Tfft$ allowed by the grid is larger than the maximum $\Tfft$ needed for the peakmap construction (see Sec.~\ref{sec:limits_Tcoh}), hence a limiting $\Tfft$ is set at $432~$s. The cutoff at $\xi\gtrsim 10$ Hz corresponds to signals that cannot be handled with a single correction with the minimum $\Tcoh$ specified.

Given the limitations on the parameter space that can be covered by placing a single $(f_0,\Mc)$ point, an alternative grid can be constructed to densely cover the parameter space of interest, as discussed in the following section.

\subsection{Grid on $\xi$: fixed $\Tfft$}
\label{subsec:gridxi}

Let us now consider the case of a grid constructed using the maximum possible $\Tfft$ as discussed in Sec.~\ref{sec:limits_Tcoh}, dropping the requirement of having a single grid point per iso-$\xi$. In this setup, the only requirement is that $\Delta \xi = 1/\Tfft$.  For each iso-$\xi$ line, the grid is populated starting from the maximum frequency $f_i=\fmax$ of the parameter space of interest. The corresponding chirp mass value $\Mc^{i}$ is then obtained by evaluating $\Mc(\xi,f^{i}_0,\Tobs)$ (Eq.~\eqref{eq:Mc(xi)}). This will be the maximum chirp mass covered on that specific iso-$\xi$ line. 
As long as the condition in Eq.~\eqref{eq:maxt_Tfft} holds, no further pairs of $(f_0, \Mc)$ are placed on the same iso-$\xi$. If this condition is no longer valid, the next pair of $(f_0, \Mc)$, i.e. $(f^{i+1}_0, \Mc^{i+1})$ are placed by solving $\mathcal{G}(\xi,f^{i}_0,f^{i+1}_0) = 1/\Tfft$. This operation is iterated as long as the minimum frequency of the parameter space of interest is reached. The full operation is repeated for the next $\xi$ line. The complete algorithm is described in Appendix~\ref{alg:constantFFT}).  
This grid construction method, as it can keep the $\Tfft$ to any arbitrary level, can be used to reach a desired sensitivity at the expense of increasing the computational cost. 

\begin{figure}[htbp]
    \centering
    \includegraphics[width=\columnwidth]{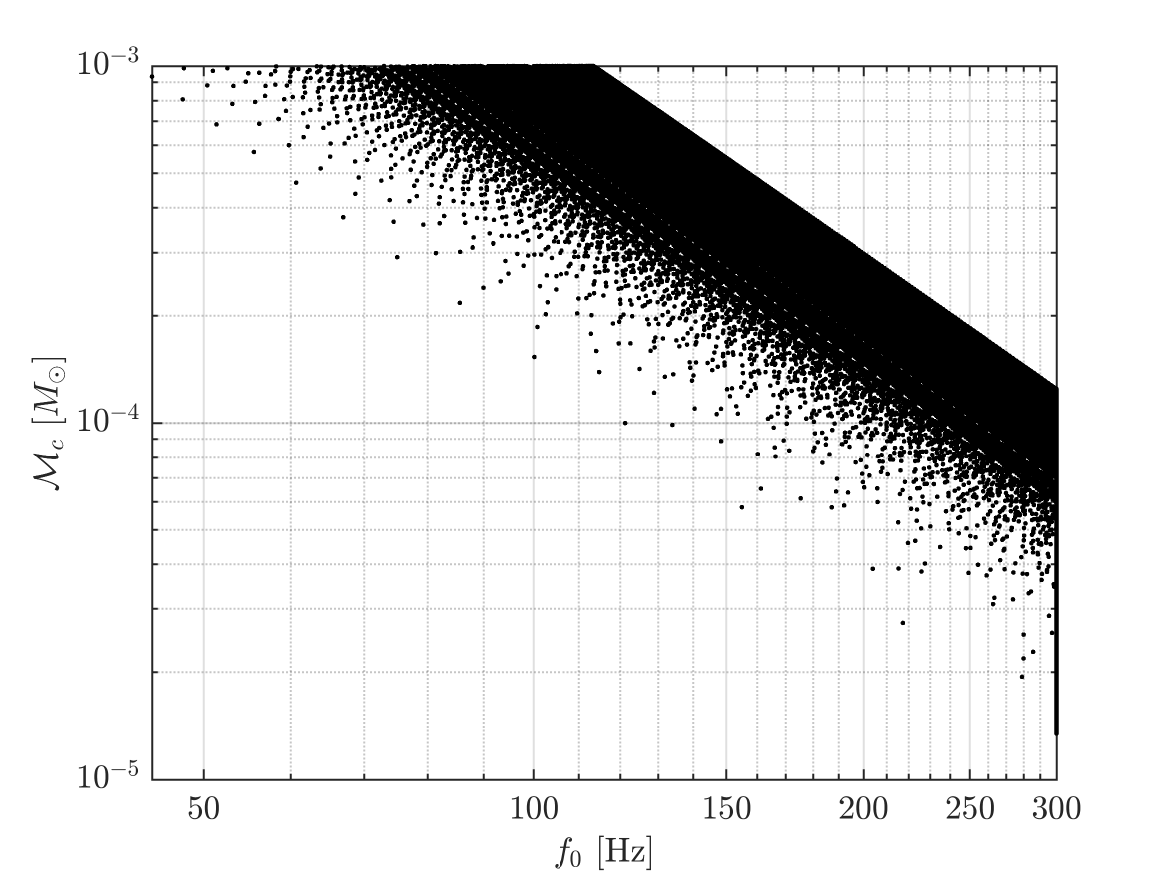}
    \caption{Grid with constant $\Tcoh$ and $\fmin = 10$ Hz, $\fmax = 300$ Hz, $\Mcmin = 10^{-6}~\Ms$, $\Mcmax = 10^{-3}~\Ms$, $\tarTfft=432$ s, $\ximin=10^{-8}$ Hz, $\ximax = 10$ Hz.}
    \label{fig:GridConstTcoh}
\end{figure}

In Fig.~\ref{fig:GridConstTcoh}, a grid with 
$\fmin = 10$ Hz, $\fmax = 300$ Hz, $\Mcmin = 10^{-6}~\Ms$, $\Mcmax = 10^{-3}~\Ms$, $\ximin=10^{-8}$ Hz and $\ximax = 10$ Hz is shown when keeping the target coherence time to $\tarTfft=432$ s. This grid contains a total of $1.75\times 10^5$ points.

To compare the number of points of this grid to other approaches, we can take as an example the work from Ref.~\cite{velcani2024thesis}, where a more standard grid has been constructed using a frequency spacing of $10^{-2}$ Hz and of chirp mass of $10^{-7}$~$\Ms$. The grid used in that study does not exploit the signal degeneracies and contains $2.9\times 10^8$ points. Therefore, our grid is already three orders of magnitude smaller than a typical grid size for a search of this kind.

\section{Candidates post-processing}
\label{sec:cand}
As detailed in Sec.~\ref{sec:peakmap}, once the search is performed on the dataset of duration $\Tobs$, a set of first-stage set of candidates is selected. Here, dataset refers to a data segment of duration  $\Tobs$ being analyzed. For a single detector there can be up to $N_{T}$ datasets (see Sec.~\ref{sec:data}). In case of two (or more) detectors, additional sets of candidates can be obtained. Below, we describe how to combine results from multiple datasets, whether from the same or different detectors.

\subsection{Coincidence between datasets}

For a given dataset, each candidate is associated with a pair of frequency and chirp mass, but, after heterodyning, the recovered frequency is that of the signal at the start of the chunk of data analyzed i.e., $t_i$, which may not coincide with the reference time $t_0$. Since most of the signals searched are longer than $\Tobs$, multiple triggers might arise for the same signal at different periods of $\Tobs$ analyzed. 

The different candidates from the various periods might belong to the same signal if they match the
same frequency evolution of Eq.~\eqref{eq:f(t)}. To check this, we update all the candidate parameters to the same reference time $t_0$ according to Eq.~\eqref{eq:f(t)}. 

Aligning all candidate parameters to the same reference time facilitates the tracking of data from different time segments and enables the combination of information from multiple interferometers. 
In case two candidates belong to the same signal, once updated to their value at $t_0$, they should be close in the $(f_0,\Mc)$ space. 

To quantify this, we can compute the distance between these candidates as the L$_2$-norm considering the spacing of the grid, $\td f_0$ and $\td \mathcal{M}_c$, as
\begin{equation}
    D = \sqrt{\left(\frac{f_0-f'_0}{\td f_0}\right)^2+\left( \frac{ \mathcal{M}_c-\mathcal{M}_c'}{\td \mathcal{M}_c}\right)^2}\,,
\end{equation}
where $\td f_0 = 1/\Tfft$ and taking into consideration that $\td \xi = \frac{\partial \xi}{\partial f_0} \td f_0+\frac{\partial \xi}{\partial \Mc} \td \Mc$ and that also equals $\td \xi = 1/\Tcoh$, the value of $\td \Mc$ can be obtained. From Eq.~\eqref{eq:xi} the partial derivatives can be evaluated and equal
\begin{align*}
\frac{\partial \xi}{\partial f_0} &= f_0^{\frac{1-\alpha}{\alpha}} \left( f_0^{\frac{1}{\alpha}} + \frac{k}{\alpha} T_{\text{obs}} \right)^{\alpha-1} - 1\, , \\
\frac{\partial \xi}{\partial \Mc} &= 32 \pi^{\frac{8}{3}} T_{\text{obs}} \left( \frac{G}{c^3} \right)^{\frac{5}{3}} \mathcal{M}_c^{\frac{2}{3}}  \left( f_0^{\frac{1}{\alpha}} + \frac{k}{\alpha} T_{\text{obs}} \right)^{\alpha-1}\, .
\end{align*}

The primed and unprimed quantities denote the recovered parameters for the two candidates (belonging to two different detectors or two data chunks of the same detector). 
A pair of candidates is found in coincidence when $D<4$, a typical value used in other searches \cite{FreqHough,CW_GC_O3}.

For real signals, we expect to have coincident candidates from two different detectors over the same $\Tobs$. For candidates from different $\Tobs$ and the same detector, we expect to have coincidences as long as their time to the merger is compatible with the total observing time (e.g. $\tmerg \geq 2\Tobs$ for 2 data chunks, etc.). Non-coincident candidates with $\tmerg \leq 2\Tobs$ are currently discarded. Requiring candidates to survive this stage before going to the next veto step will reduce the false positives of the search.

\subsection{Candidate vetoes and follow-up}
{\bf Vetoes:} Once the first level of coincident candidates is produced, these undergo a multi-stage veto procedure. 

One standard veto step is to check the consistency of the candidate's significance in each of the detectors~\cite{CW_GC_O3,CW_GC_O2,CW_allSky_O3}. This is achieved by discarding all coincident candidates where the weighted CR in the less sensitive detector is more than three times higher than in the more sensitive detector. Practically, assuming the noise spectral density in the first detector is worse, i.e., $S_{n_1}(f) > S_{n_2}(f)$, only the outliers satisfying ${\rm CR_{1}}/{\sqrt{S_{n_1}(f)}} < 3 {\rm CR_{2}}/{\sqrt{S_{n_2}(f)}}$ move on to the next post-processing step. This veto step applies only to candidates found in coincidence between datasets from different detectors, not to those arising from coincidences between two consecutive data chunks.

Instrumental artifacts, such as spectral lines and glitches, degrade the detector's data quality~\cite{Virgo:2022kwz, Virgo:2022ysc, DavisdetcharO3, CovaslinesO1O2, Sun:2020wke, Sun:2021qcg}. The impact of glitches is mitigated by applying various cleaning procedures to the data~\cite{BSD}. However, instrumental lines are not removed from the data before the analysis begins, hence some of the candidates that passed the first selection might have been produced by these lines. However, it is important to note that many of the signals we are targeting span a wide frequency band. Removing outliers whose frequency intersects with the frequency region contaminated by a known noise line~\cite{LIGO:2024kkz,LIGO:2021ppb,Virgo:2022ysc} is not advisable, as it would eliminate a significant portion of the selected candidates, potentially causing us to miss real signals.
A safer procedure is to verify that there are no known spectral lines in the same frequency bin of the candidate at $t_0$ before applying the heterodyne correction. We can study the peakmap projection of the original data and verify where the spectral lines are originally located. This is especially important for checking the presence of wandering or transient lines that might be present only in the data chunk $\Tobs$. This procedure should be repeated for each set of surviving candidates in every one of the $N_{T}$ analyzed chunks and for each detector.

Finally, another veto procedure involves checking if the candidate's significance, computed using the demodulated data with the parameters $(f_0,\Mc)$, increases compared to the significance computed on the same frequency bin using the uncorrected time series. This check applies to the coincidences between two detectors and between multiple chunks. 
Once the candidates survive these veto steps, they can pass to the follow-up stage.

{\bf Follow-up:} A potential follow-up stage involves creating a refined grid around the candidate’s parameters $(f_0,\Mc)$.  This grid can be uniform in $f_0$ and $\Mc$,  utilize longer  $\Tfft$ and $\Tobs$,  and possibly incorporate the Doppler modulation of the signal if required. The refined grid helps to accurately determine the true parameters of the signal. For candidates found in multiple subsequent chunks, we can estimate the total signal duration and apply more suitable matched-filter techniques. The details of the specific follow-up procedures to be used in a real search are beyond the scope of this paper and will be addressed in future work.

\section{Sensitivity estimate}
\label{sec:sens}
In this section, the theoretical sensitivity of the method is analytically estimated.
Following the approaches in \cite{FreqHough,CW_GC_O3}, for a peakmap-based search, the minimum detectable strain (at $\Gamma=0.95$ confidence level), is 
\begin{equation}\label{eq:hmin}
    h_{0,\min}(f)= \mathcal{B}\left(\frac{\Tfft}{\Tobs}\right)^{\frac{1}{4}}\sqrt{\frac{S_n(f)}{\Tfft}}\sqrt{\CR_{\mathrm{thr}}-\sqrt{2}\mathrm{erfc}^{-1}(2\Gamma)},
\end{equation}
where $S_n(f)$ is the power spectral density of the noise in the interferometer and $\mathcal{B}$ is a parameter that depends on the sky position, the threshold used for peak selection in the peakmap and the polarization of the signal, which for this case is averaged. For an all-sky search, this is equal to $\mathcal{B}=4.97$~\cite{FreqHough}.

Equating Eqs.~\eqref{eq:hmin} and \eqref{eq:hsignal}, the maximum reachable distance of the search is

\begin{multline}\label{eq:dmax}
    d_{\max}= \frac{4}{\mathcal{B}}\left( \frac{G\Mc}{c^2} \right)^{5/3}\left( \frac{\pi f_0}{c}\right)^{2/3}\left(\frac{\Tobs}{\Tfft }\right)^{1/4} \\ \times \sqrt{\frac{\Tfft}{S_n(f)}}\left[\CR_{\mathrm{thr}}-\sqrt{2}\mathrm{erfc}^{-1}(2\Gamma)\right]^{-1/2}~.
\end{multline}

\begin{figure*}[htbp]
    \centering
    \includegraphics[width=\textwidth]{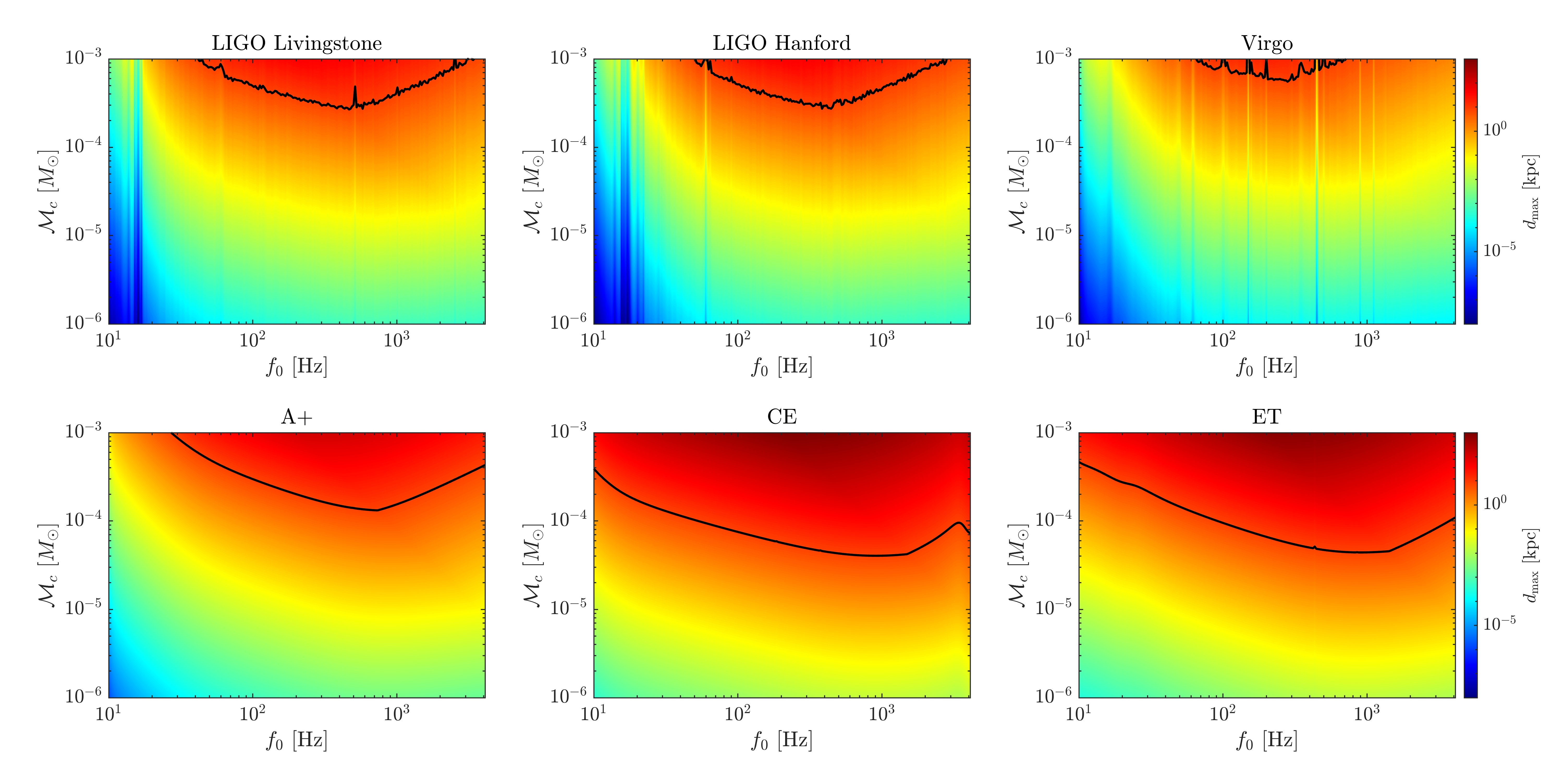}
    \caption{Maximum reachable distance computed using Eq.~\eqref{eq:dmax} for the LIGO and Virgo detectors using the O3b noise curves, design A+, design CE and design ET, $\Tfft=432$ s and $\Tobs=0.5$ days. The black line indicates where $d=8$ kpc, the distance to the GC.}
    \label{fig:dmaxplot}
\end{figure*}

In Fig.~\ref{fig:dmaxplot} the maximum reachable distance is shown as a function of the parameter space using a fixed $\Tcoh=432~$s and $\Tobs=0.5$ days and a $\CR_{\mathrm{thr}}=5$. The plots show the result using the noise curves of LIGO Livingston, Hanford and Virgo interferometers O3b-representative sensitivities, as in Ref.~\cite{Abbott2021GWTC-3:Run}, as well as those for the next-generation detectors like Einstein Telescope (ET)~\cite{ETcds,ETdesign} and Cosmic Explorer (CE)~\cite{CE1,CE2,CE3} and current detector's planned upgrades (Advanced LIGO plus - A+)~\cite{A+sensitivity}. 
These results indicate that the Galactic Center, situated at a distance of $\dGC=8~$kpc~\cite{CW_GC_O3, distGC1,distGC3,distGC4}, is already reachable in a fraction of the parameter space, with current generation detectors. However, because the Doppler correction is not applied due to the relatively short duration of $\Tcoh$, the search remains inherently omnidirectional.

\section{Tests with simulated signals}
\label{sec:inj}

The injection campaign is a fundamental aspect of addressing the pipeline's capability to detect our signals. The primary objective is to validate the method and its implementation, estimate its sensitivity comparing it to the theoretical one discussed in Sec.~\ref{sec:sens}, and measure the pipeline's efficiency. All the simulated signals will be injected in real detector data.

We randomly inject signals with starting frequency in the range $20-250$~Hz and a chirp mass range of $10^{-6}-10^{-3}~\Ms$. We fix the distance in this case to $1$~kpc, but the strain, $h_0$, will be different for each case as it strongly depends on the initial frequency and chirp mass. We use real data from the LIGO Livingston and Hanford O3 run, divided into two segments of 0.5 days each for the analysis. The reference time of the signals selected is at the beginning of the run, starting at a GPS time of $1238198418$~s.

A total of $10,000$ signals are injected sampled log-uniformly in initial frequency and chirp mass, uniformly in the polarization angle $\psi$, and the cosine of the inclination of the orbit, $\cos \iota$, relevant for the antenna patterns and amplitude of the signal that reaches the detector \cite{FreqHough}. The resulting strain will range between $7.6\times 10^{-27}$ and $4.6\times 10^{-23}$. We assume a fixed sky position corresponding to the GC.

We examine four different cases to assess the impact of using data from additional interferometers or extending the analysis duration to multiple chunks of duration $\Tobs$. These are:
\begin{itemize}
    \item Case 1: Only LIGO Livingston data, for a total analysis time of $0.5$ days.
    \item Case 2: Only LIGO Livingston data, for a total analysis time of $1$ days (2 contiguous chunks of duration $\Tobs$).
    \item Case 3: LIGO Livingston and LIGO Hanford data, for a total analysis time of $0.5$ days.
    \item Case 4: LIGO Livingston and LIGO Hanford data, for a total analysis time of $1$ days (2 contiguous chunks of duration $\Tobs$).
\end{itemize}

\begin{figure}[htbp]
    \centering
    \includegraphics[width=1\columnwidth]{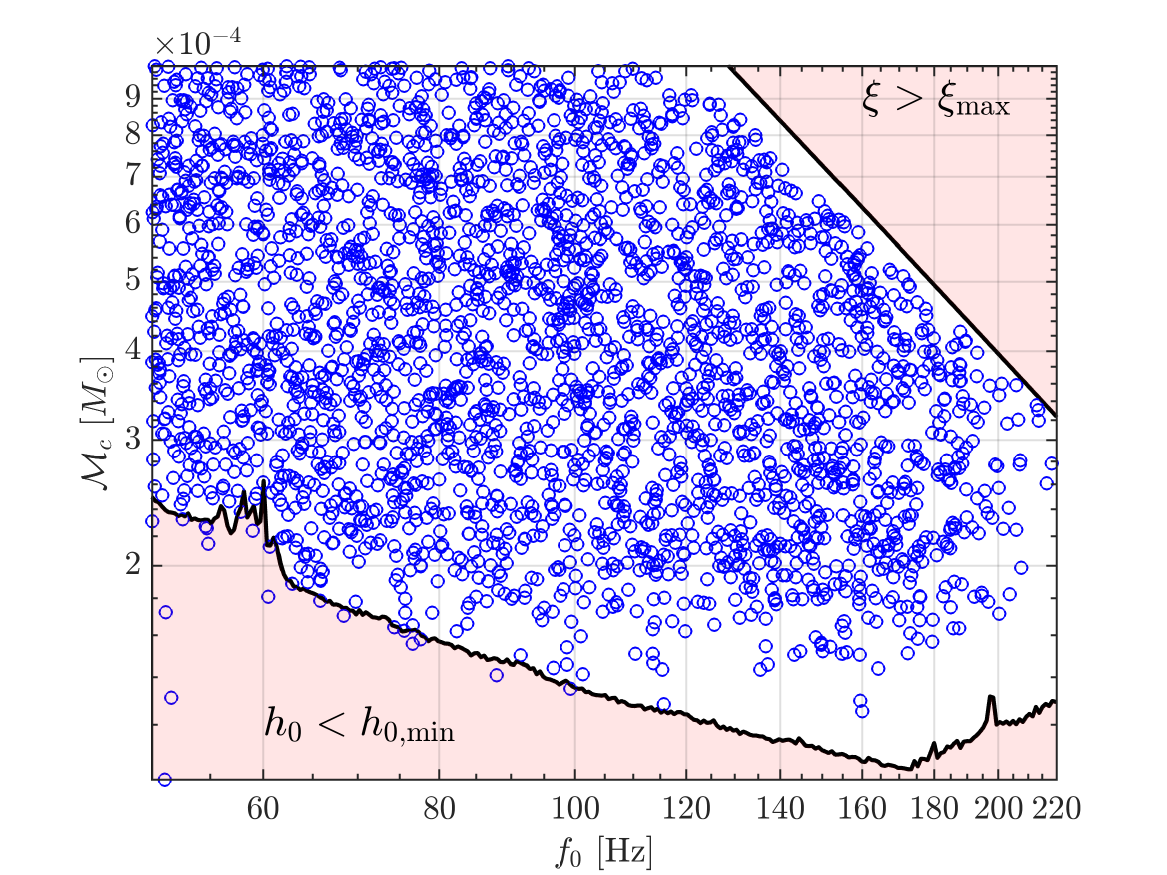}
    \caption{Recovered injections for (Case 1) along the theoretical limits based on the sensitivity of the method and the maximum $\xi$ searched.}
    \label{fig:recovered_injections}
\end{figure}

The search is performed with the grid with variable $\Tcoh$ described in Sec.~\ref{sec:gridxivarTcoh}, as it is the most computationally inexpensive. A plot showing the recovered injections is displayed in Fig.~\ref{fig:recovered_injections} alongside the theoretical limits of the search. The lower limit is governed by the 95\% confidence level minimum detectable strain, addressed in Sec.~\ref{sec:sens}, while the upper one is dictated by the maximum value of the variable $\xi$ used for the search.

Several factors can explain recovered candidates below the 95\% confidence level sensitivity curve. First, noise artifacts in the data can occasionally mimic true signals, passing through the detection pipeline and appearing below the sensitivity threshold. If this were the case, these candidates would likely be discarded in the postprocessing and follow-up steps, which have not been applied to the candidates in Fig.~\ref{fig:recovered_injections}. 
Second, the formula in Eq.~\eqref{eq:hmin}, is directly applicable for the case of standard monochromatic CW signals and, despite the method described here has a similar statistical foundation as the work in~\cite{FreqHough}, the difference in signal modeling and signal correction might impact the final result leading to the recovery of signals at lower amplitudes than expected. 

In any case, the analytical estimation and the one computed with the injections are in good agreement.
Some signals are missed towards the high end of the frequency spectrum. The main reason for missing such signals is that the frequency evolution is so high that it can get out of the data frequency band being analyzed.

To evaluate the impact of using data from additional detectors or extending the analysis duration, we will use different metrics to evaluate the performance of our recovery process. To do so we need to quantify the true positives (TP), false positives (FP), and false negatives (FN).

One of such metrics is the $F_1$ score, defined as 
\begin{equation}
    F_1 = \frac{2~\mathrm{TP}}{2~\mathrm{TP}+\mathrm{FP}+\mathrm{FN}}\, ,
\end{equation}
which equally weights the precision, i.e. the fraction of true positives over the recovered signals TP/(TP+FP), and the recall (or true positive rate (TPR)) which is given by the fraction of injected signals that were correctly recovered TP/(TP+FN). 

\begin{figure*}[ht]
    \centering
    \subfloat[F1 score of the various cases computed with the injections.]{%
        \includegraphics[width=0.45\textwidth]{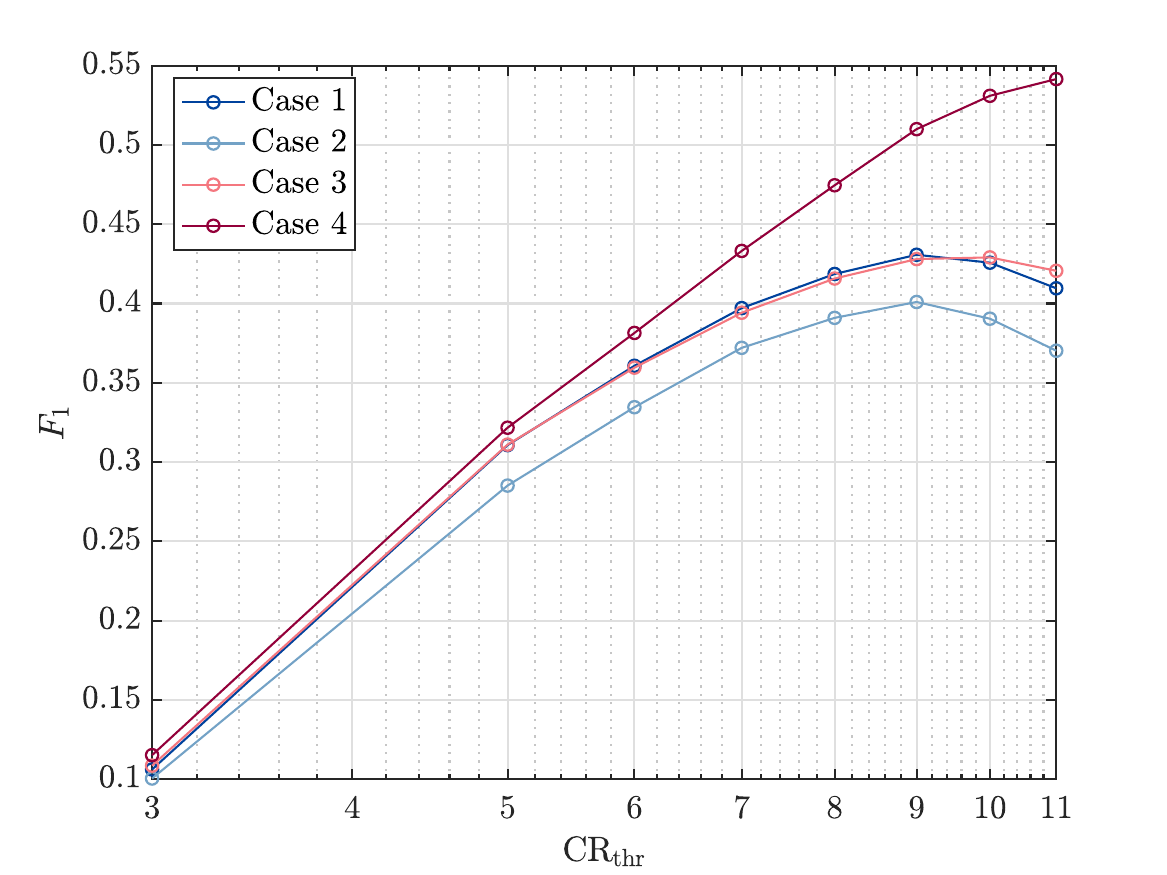}%
        \label{fig:F1score}
    }
    \hfill
    \subfloat[True positive rate of the various cases computed with the injections.]{%
        \includegraphics[width=0.45\textwidth]{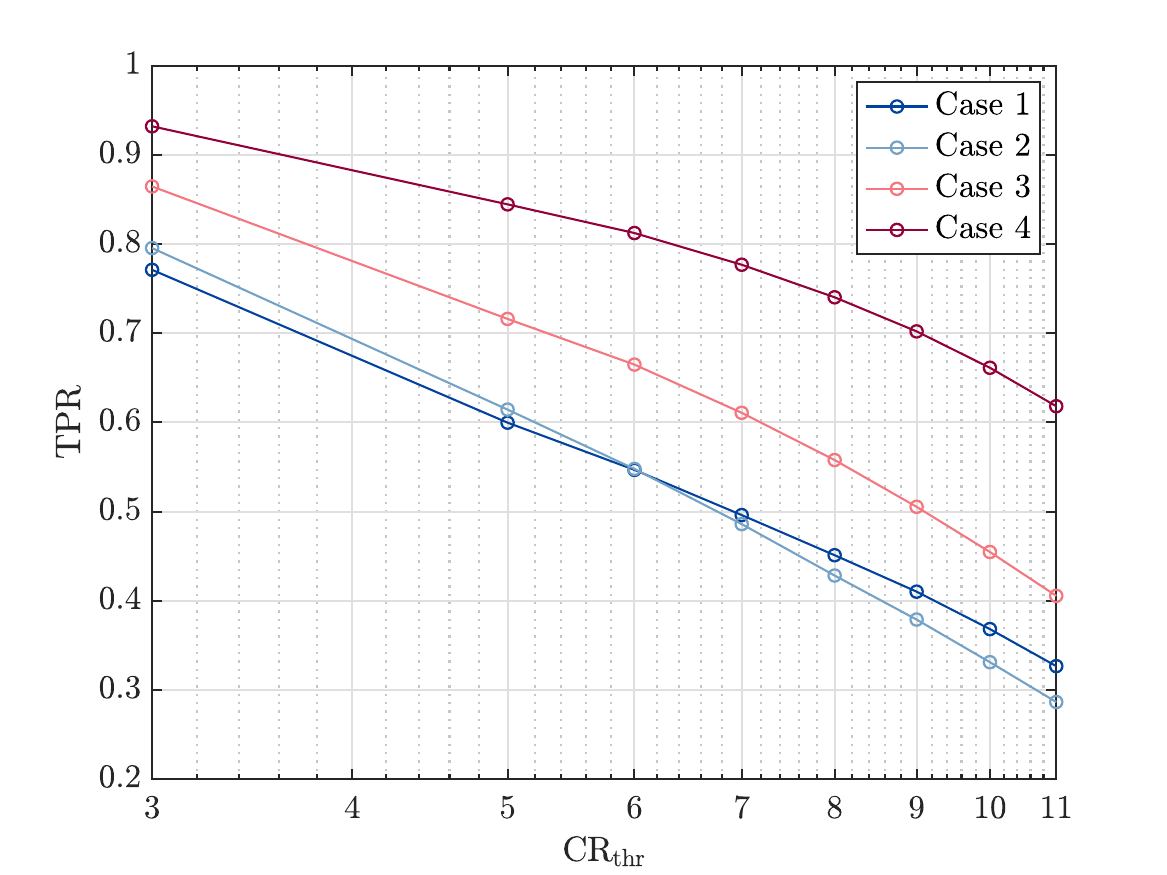}%
        \label{fig:TPR}
    }
    \caption{F1 score and true positive rate computed for the various cases with the injection set as described in the main text.}
    \label{fig:F1_and_TPR}
\end{figure*}

A process with a high $F_1$ value is generally considered better-performing. In Fig.~\ref{fig:F1score}, various values of the $F_1$ score are presented as a function of the CR threshold, $\mathrm{CR}_{\mathrm{thr}}$. As expected, for high CR thresholds, the number of recovered injections reduces. But this reduction is compensated by the highest confidence in the detections. Case 4 is the one displaying a better $F_1$, indicating that is the best-performing at recovering the true positives and reducing the false positives. It is also observed that there is a maximum value of the $F_1$ score for a given $\mathrm{CR}_{\mathrm{thr}}$. This occurs because, while increasing this threshold statistic reduces the number of false positives, it also decreases the number of true positives recovered. At a certain point, the reduction in false positives no longer compensates for the loss of true positives, leading to a decline in the overall $F_1$ score as too many signals are missed.

Similarly, a high TPR indicates that the process 
is better at capturing TP and minimizing FN. We display in Fig.~\ref{fig:TPR} the values of the TPR for different values of the critical ratio threshold. As seen with the F1 score, the more information is added (from another interferometer and/or from another piece of time) the more efficient is the selection process. At the same time, when the $\mathrm{CR}_{\mathrm{thr}}$ is increased, the TPR decreases. Being more restrictive with the selection criterion has as a consequence missing more injections while reducing the number of false positives.

\begin{figure}[htbp]
    \centering
    \includegraphics[width=1\columnwidth]{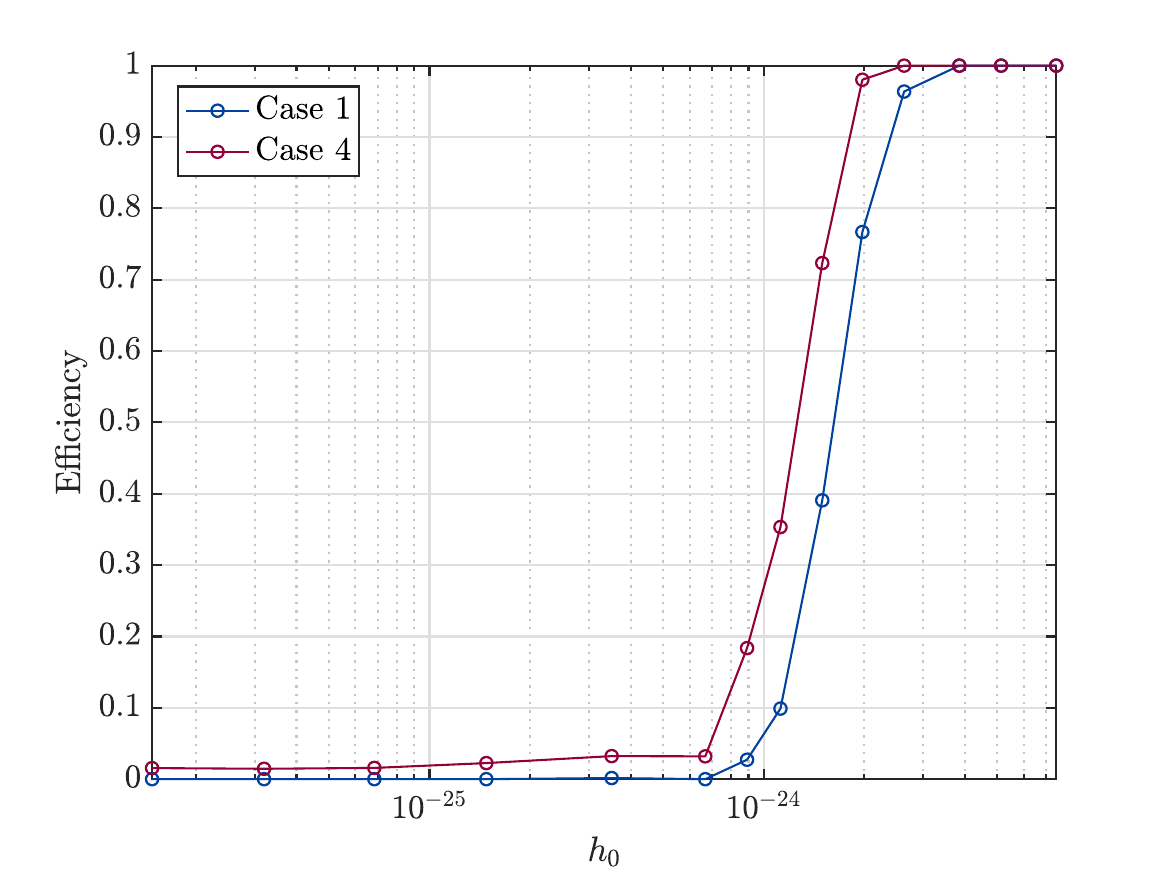}
    \caption{Efficiency as a function of the strain of the signal for two of the cases of the injections.}
    \label{fig:h0_sens}
\end{figure}

Lastly, we can compute the detection efficiency by estimating the fraction of recovered signal, as a function of the the signal strain. This is displayed in Fig.~\ref{fig:h0_sens} for the two limiting cases (Case 1) and (Case 4). As expected, adding the information across various detectors as well as more data chunks, increases the detection efficiency.

\section{Conclusions}
\label{sec:conc}
In this work, we have presented a novel semi-coherent method for detecting gravitational waves emitted by inspiraling binary systems of light compact objects, such as primordial black holes. The method is based on the heterodyne correction and includes a customized grid construction in the parameter space, which significantly reduces the computational cost while preserving high sensitivity. This is achieved by exploiting parameter degeneracies that characterize the gravitational wave signal. Our injection campaign, conducted using real LIGO O3 data, has demonstrated the effectiveness of the method, successfully recovering simulated signals across a broad parameter space. Furthermore, the results align well with theoretical sensitivity predictions, underscoring the method's robustness and practical utility for future gravitational wave searches.

\section*{Acknowledgements}

We would like to thank Cristiano Palomba for the valuable discussions at the early stages of the project. This research has made use of data or software obtained from the Gravitational Wave Open Science Center (gwosc.org), a service of the LIGO Scientific Collaboration, the Virgo Collaboration, and KAGRA. This material is based upon work supported by NSF's LIGO Laboratory which is a major facility fully funded by the National Science Foundation, as well as the Science and Technology Facilities Council (STFC) of the United Kingdom, the Max-Planck-Society (MPS), and the State of Niedersachsen/Germany for support of the construction of Advanced LIGO and construction and operation of the GEO600 detector. Additional support for Advanced LIGO was provided by the Australian Research Council. Virgo is funded, through the European Gravitational Observatory (EGO), by the French Centre National de Recherche Scientifique (CNRS), the Italian Istituto Nazionale di Fisica Nucleare (INFN) and the Dutch Nikhef, with contributions by institutions from Belgium, Germany, Greece, Hungary, Ireland, Japan, Monaco, Poland, Portugal, Spain. KAGRA is supported by Ministry of Education, Culture, Sports, Science and Technology (MEXT), Japan Society for the Promotion of Science (JSPS) in Japan; National Research Foundation (NRF) and Ministry of Science and ICT (MSIT) in Korea; Academia Sinica (AS) and National Science and Technology Council (NSTC) in Taiwan. 
This project has received funding from the European Union’s Horizon 2020 research and innovation programme under the Marie Skłodowska-Curie Grant Agreement No. 754510.
This work is partially supported by the Spanish MCIN/AEI/10.13039/501100011033 under the Grants
No. SEV-2016-0588, No. PGC2018-101858-B-I00, and No. PID2020-113701GB-I00, some of which include ERDF funds from the European Union, and by the MICINN with funding from the European Union NextGenerationEU (PRTR-C17.I1) and by the Generalitat de Catalunya. IFAE is partially funded by the CERCA program of the Generalitat de Catalunya.
MAC is supported by the 2022 FI-00335 grant. This research is supported by the Australian Research Council Centre of Excellence for Gravitational Wave Discovery (OzGrav), Project Number CE230100016.

\appendix
\section{Limits on the coherence time}\label{appendix:tcohlim}

In this Appendix, the various limits, either upper or lower, that can be considered for $\Tcoh$ are discussed. 
\subsection{Maximum coherence time}

One potential limit on the coherence time is settled by the statistics used and the resolution of the time-frequency map we want to achieve. The statistic is originally binomial but has a Gaussian approximation under the central limit theory validity. The time resolution of the peakmap is set by the coherence time, which also sets the frequency resolution of the map as $1/\Tfft$. To preserve the Gaussian approximation for the peak number counts when dealing with a binomial distribution, the following condition must be met given the number $N_t$ of FFTs: $N_t\min(p_0,1-p_0)>z^2$, where $z$ is the desired z-score and $p_0$ the probability of success of the binomial distribution (i.e. the probability of finding a peak which is $p_0= 0.0755$ as in \cite{FreqHough}). This ensures that the data is not too skewed and there are enough points for the approximation to be valid. Typically, the approximation is considered correct for $z^2>5$, corresponding to a probability coverage of $P(-z<X<z) = 0.97465$ for $z^2=5$  or $0.99842$ for $z^2=10$. This leads to a limit on the minimum number of FFTs to use as $N_t=60$ for $z^2=5$  and $N_t=130$  for $z^2=10$.  
In our case, since we assume $\Tobs=0.5$ days, it seems reasonable to use a value in between, for example, $N_t=100$ FFTs for the time resolution of the peakmap. 
This automatically translates into a maximum limit on the coherence time as $\Tfft=\Tobs/N_{t}$ which brings $\Tfft=432$~s.

Other restrictions for the maximum value of the FFT must be considered if we use longer $\Tobs$. One natural restriction to the $\Tfft$ is given by the duration of the sidereal day $T_{\rm sid}\simeq 86164~$s. The sidereal modulation, which can be appreciated for signals integrated over coherence times approaching $T_{\rm sid}$,  produces a splitting of the intrinsic signal frequency $f_0$ into five components: $f_0\pm l T^{-1}_{\rm sid}$ with $l=0,1,2$. If this modulation is not properly considered, the signal might be undetected \cite{Astone:2010zz,DOnofrio:2022zvw,DAntonio:2023jxm}.  
Another limitation to consider is the Doppler modulation. The grid construction for $\xi$ only considered the frequency modulation due to the spin-up without the Doppler contribution. Therefore, we need to ensure that, after the spin-up modulation is removed, a signal stays in a single frequency bin during the time $\Tcoh$ as discussed in~\cite{Frasca2005}:
\begin{equation}
\label{eq:Doppler}
    \Tcoh \leq T_\oplus\sqrt{\frac{c}{4\pi^2R_\oplus f_0}} \approx 10^{4}~\mathrm{s}~\sqrt{\frac{100~\mathrm{Hz}}{f_0}}~,
\end{equation}
where $T_\oplus$ is Earth's period of rotation and $R_\oplus$ is the rotational radius at the antenna's latitude. For frequencies around 300 Hz, the maximum allowed coherence time is approximately 5000~s. We have also assessed the impact of the Doppler effect over the entire observation period $\Tobs$ for signals at our maximum targeted frequencies ($300$ Hz). The maximum Doppler shift is always smaller than the minimum frequency bin used in our search (i.e. $1/432$ Hz) so we conclude that the Doppler modulation does not impact this specific search setup but is important to consider for future searches.

\subsection{Minimum coherence time}

The error of the estimators $\hat{\mu}$ and $\hat{\sigma}^2$ used to compute the CR can also be used to establish a minimum value of the coherence time. The standard error (SE) of the mean estimator, normalized by its true value, can be computed as
\begin{equation}
\mathrm{SE}_{\hat{\mu}} = \frac{\sigma}{\mu\sqrt{N_f}}\, .
\end{equation}
    
In this case, the values of $\mu$ and $\sigma$ are known from the binomial distribution that this random variable follows and equal $\mu = N_tp_0$ and $\sigma = \sqrt{N_tp_0(1-p_0)}$ \cite{FreqHough}. This states that
\begin{equation}
    \mathrm{SE}_{\hat{\mu}} = \sqrt{\frac{1-p_0}{N_{\rm tot}p_0}}\, ,
\end{equation}
which does not depend on $\Tcoh$. This is because $N_{\rm tot} = N_fN_t = \Delta f \Tcoh \times \Tobs/\Tcoh = \Delta f \Tobs$. The term $\Delta f=f_{\max}-f_{\min}$ denotes the bandwidth of the analysis.
    
\begin{figure}[htbp]
     \centering
     \includegraphics[width=\columnwidth]{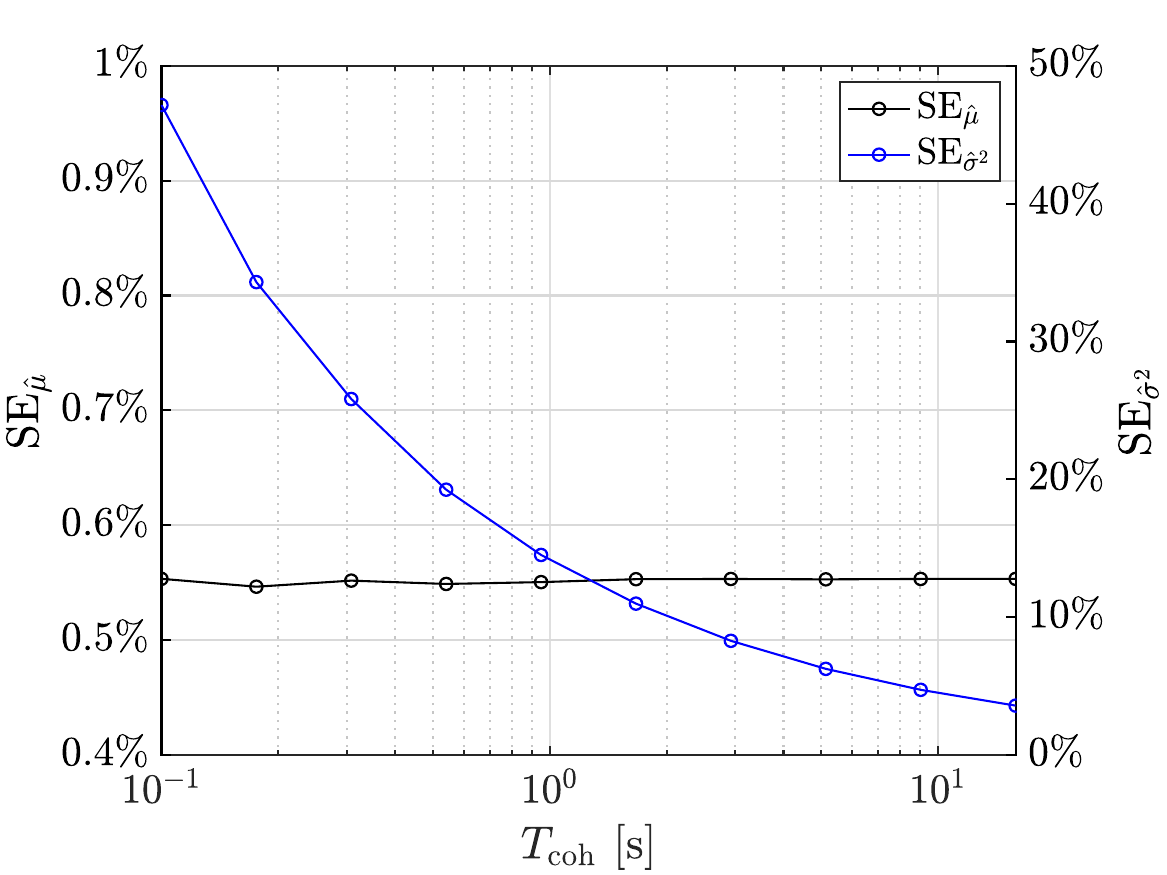}
     \caption{Evolution of the standard error for the estimators of the mean and variance for $\Tcoh$ for $\Tobs = 0.5$~days and $\Delta f = 100$ Hz.}
     \label{fig:SE_Tcoh}
 \end{figure}

The other error that needs to be taken into account is that of estimating the variance. The variance of the sample variance estimator is 
\begin{equation}
        \mathbb{V}\mathrm{ar}[S^2] = \frac{1}{N_f}\left(\mu_4 -\frac{N_f-3}{N_f-1}\mu_2^2\right)\, ,
\end{equation}
where $\mu_i$ indicates the $i$th central moment. For the binomial distribution here considered, $\mu_2 = \sigma^2$ and $\mu_4 = N_tp_0(1-p_0)[1+(3N_t-6)p_0(1-p_0)]$. This leads to a SE (which we define as the standard deviation of the error of the variance divided by the true variance) of
\begin{equation}
\mathrm{SE}_{\hat{\sigma}^2} = \frac{\sqrt{\mathbb{V}\mathrm{ar}[S^2]}}{\sigma^2}\, ,
\end{equation}
where we have used the fact that $S^2$ is an unbiased estimator of $\sigma^2$. This error depends on the coherence time so another choice to set the minimum coherence time is to ensure that both errors are at an acceptable level. The one of the mean is constant and does not set any constraint, but the one of the variance does increase the smaller the coherence time gets and can be used to set the lower limit. In Fig.~\ref{fig:SE_Tcoh} the evolution of both errors as a function of $\Tcoh$ for $\Tobs = 0.5$~days and $\Delta f = 100$ Hz is shown. For this choice of parameters, setting a maximum error of $10\%$ would imply a minimum coherence time of $\sim 1.7$ s.

\section{Pseudo-codes for the grid construction}
\label{appendix:algo}
In this Appendix, the pseudo-codes showing the detailed steps of the different algorithms explained in Secs.~\ref{sec:gridxivarTcoh} and \ref{subsec:gridxi} are presented. The one showing how the grid in $\xi$ is constructed with a variable coherence time is displayed in Alg.~\ref{alg:variableFFT} while the one for a fixed coherence time in Alg.~\ref{alg:constantFFT}. 

For both cases, the output is a collection of pairs of $(f_0,\Mc)$ that comprise the grid, and the vector of coherence times, $\Tcoh$, for each value of $\xi$. The variable coherence time algorithm also returns the vector of $\xi$ values that are assigned during the construction of the grid.

\subsection{Variable $\Tfft$}

The idea behind this algorithm is the following:
\begin{itemize}[label={}]
\item 1. First, define the limits on the new variable, $\xi_{\min}, \xi_{\max}$ and on the frequency $f_{0}^{\min}$, $f_{0}^{\max}$, which will set the overall parameter space to be searched. Set $\xi_0 = \xi_{\min}$ as the first element to start the iteration.
\item 2. For the iso-$\xi$ line, find the point that allows to have a longer $\Tfft$. This is equivalent to finding the signal that has less frequency residual with all the rest of the signals that lie on the same iso-$\xi$ curve. To do so, we recast it into the following optimization problem for a given $\xi_i$:
\begin{align*}
    \min_{f_0}& ~~\max(\mathcal{G}(\xi,f_0,f_0^{\min}),\mathcal{G}(\xi,f_0,f_0^{\max})) \\
    \mathrm{s.t.}& ~~f_0\in[f_0^{\min},f_0^{\max}]
\end{align*}
This optimization problem tries to minimize the maximum frequency deviation. Since the difference between two signals increases the further away they are along the same iso-$\xi$ curve, it is enough to compare each point to the signals that have the minimum and maximum frequencies, as they are the most limiting cases.
\item 3. Once the optimal point has been found, the cost function value, $\tilde{\mathcal{G}}$, represents the maximum variation of frequency between any two signals. Therefore, we set  the coherence time, as $T_{\mathrm{coh}}=1/{\tilde{\mathcal{G}}}$, which ensures that all frequency modulation is going to be contained within a single frequency bin.
\item 4. Prepare the next iteration by setting $i=i+1$ and the next value of the $\xi$ variable as $\xi_{i}=\xi_{i-1}+1/\Tfft$.
\item 5. Check if $\xi_{i}<\xi_{\max}$, for which one must return to step 2, or finish the code if either $1/\tilde{\mathcal{G}}<\minTfft$ or  $\xi_{i}>\xi_{\max}$ are true.
\end{itemize} 

Using this approach and setting $\fmin = 10$ Hz, $\fmax = 4096$ Hz, $\Mcmin = 10^{-6}~\Ms$, $\Mcmax = 10^{-3}~\Ms$, $\mTfft=432$ s, $\minTfft=1.7$ s, $\ximin=10^{-8}$ Hz, $\ximax = 10$ Hz, the grid displayed in Fig.~\ref{fig:AllGridVarTcoh} is obtained. It contains $851$ points in total.

\begin{algorithm}[htbp]
\SetAlgoLined

\SetKwInOut{Input}{Input}
\SetKwInOut{Output}{Output}
\Input{$\fmin$ ; $\fmax$ ; $\xi_{\min}$ ; $\xi_{\max}$ ; $\mTfft$ ; $\minTfft$} 
\Output{$\Tfft\in \mathbb{R}^{n_p}$; $f_0\in \mathbb{R}^{n_p}$ ; $\Mc\in \mathbb{R}^{n_p}$ ; $\xi\in\mathbb{R}^{n_p}$}
$\xi[1] \gets \xi_{\min}$

$i \gets 1$

\While{$\xi[i]<\xi_{\max}$}{
    Solve the following optimization problem:
    $$\begin{aligned}
    \min_{f_0} \quad & \max(\mathcal{G}(\xi[i],f_0,\fmin),\mathcal{G}(\xi[i],f_0,\fmax))\\
    \textrm{s.t.} \quad & f_0\in[f_0^{\min},f_0^{\max}]
    \end{aligned}$$

    The solution is $\tilde{\mathcal{G}}=\mathcal{G}(\xi[i],\tilde{f_0},\fmin)$ found at $\tilde{f_0}$
    
    $f_0[i]  \gets  \tilde{f_0}$
    
    $\Mc[i] \gets \Mc(\xi[i],\tilde{f_0})$
    
    \uIf{$1/\tilde{\mathcal{G}}\geq \mTfft$}
    {$\Tfft[i] \gets \mTfft$}
    \uElseIf{$1/\tilde{\mathcal{G}}< \minTfft$}
    {\textbf{break}}
    \Else{$\Tfft[i] \gets 1/\tilde{\mathcal{G}}$}

    $i\gets i+1$
    
    $\xi[i]\gets \xi[i-1]+1/\Tfft[i]$
}

$n_p \gets i$

 \caption{Pseudo-code for the construction of the grid with a variable FFT length}
 \label{alg:variableFFT}
\end{algorithm}

\subsection{Fixed $\Tfft$}

The idea behind this algorithm is the following:
\begin{itemize}[label={}]
\item 1. First, define the limits on the new variable, $\xi_{\min}, \xi_{\max}$, on the initial frequency $f_{0}^{\min}$, $f_{0}^{\max}$, and on the chirp mass $\Mcmin$, $\Mcmax$, which will set the overall parameter space to be searched; as well as the target coherence time, $\tarTfft$. Set vector of $\xi$ values can be already computed as it is evenly spaced with an increment of $1/\tarTfft$.
\item 2. For the iso-$\xi$ line, find whether with $\tarTfft$, a single point can absorb all the frequency evolution of the signals contained in the iso-$\xi$ line. We do this by checking whether $\mathcal{G}(\xi[i],f_0[k],\fmax)\leq 1/\tarTfft $. If this relation is hold, we then place a point at $\fmax$ and the corresponding chirp mass for that given initial frequency and $\xi$. 
\item 3. If that relation is not maintained, we need to place more than one point along the iso-$\xi$ line to contain all possible frequency evolution residuals inside a single frequency bin. We do so by solving the following optimization problem:
$$\begin{aligned}
\max_{f_0} \quad & f_0\\
\textrm{s.t.} \quad & f_0\in[f_0[k],f_0^{\max}]\\
  &\mathcal{G}(\xi[i],f_0[k],\fmax)\leq 1/\tarTfft
\end{aligned}$$
This optimization problem tries to place the next point as far away from the last point in the same iso-$\xi$ (thus maximizes $f_0$) while making sure that the residual between this new point and the last one placed is below $1/\tarTfft$. This optimization problem has its solution on the boundary of the feasible region, as the target function is monotonically increasing. This means that the computational cost can be slightly reduced by finding the point that satisfies $\mathcal{G}(\xi[i],f_0[k],\fmax) = 1/\tarTfft $ to avoid the optimization problem as, typically, the solution of the nonlinear equation will be faster.
    
\item 4. Go to point 2 until all iso-$\xi$ curves have been looped for.
\end{itemize}

\begin{algorithm}
\SetAlgoLined

\SetKwInOut{Input}{Input}
\SetKwInOut{Output}{Output}
\Input{$\fmin$ ; $\fmax$ ; $\xi_{\min}$ ; $\xi_{\max}$ ; $\Mcmin$ ; $\Mcmax$ ; $\tarTfft$}
\Output{$\Tfft\in \mathbb{R}^{n_p}$; $f_0\in \mathbb{R}^{n_p}$ ; $\Mc\in \mathbb{R}^{n_p}$}
$\xi \gets \mathrm{linspace}(\xi_{\min},\xi_{\max},\Delta \xi = 1/\tarTfft)$

$k \gets 1$

\For{$i \in[1,n_p]$}{
    \If{$1/\mathcal{G}(\xi[i],\fmin,\fmax)\geq \tarTfft$}{
    
    $f_0[k] \gets \fmax$

    $\Mc[k] \gets \Mc(\xi[i],f_0[k])$

    $k\gets k+1$
    
    \textbf{continue}
    }
    \While{$f_0[k]<\fmax$}{Solve the following optimization problem:

    $$\begin{aligned}
    \max_{f_0} \quad & f_0\\
    \textrm{s.t.} \quad & f_0\in[f_0[k],f_0^{\max}]\\
      &\mathcal{G}(\xi[i],f_0[k],\fmax)\leq 1/\tarTfft 
    \end{aligned}$$

    The solution is $\tilde{f_0}$
    
    $f_0[k]  \gets  \tilde{f_0}$
    
    $\Mc[k] \gets \Mc(\xi[i],\tilde{f_0})$
    
    $k\gets k +1 $
    }

}
$n_p\gets k$

\caption{Pseudo-code for the construction of the grid with a constant FFT length}
\label{alg:constantFFT}
\end{algorithm}

\newpage
\bibliography{ref.bib}{}

\begin{thebibliography}{111}%
\makeatletter
\providecommand \@ifxundefined [1]{%
 \@ifx{#1\undefined}
}%
\providecommand \@ifnum [1]{%
 \ifnum #1\expandafter \@firstoftwo
 \else \expandafter \@secondoftwo
 \fi
}%
\providecommand \@ifx [1]{%
 \ifx #1\expandafter \@firstoftwo
 \else \expandafter \@secondoftwo
 \fi
}%
\providecommand \natexlab [1]{#1}%
\providecommand \enquote  [1]{``#1''}%
\providecommand \bibnamefont  [1]{#1}%
\providecommand \bibfnamefont [1]{#1}%
\providecommand \citenamefont [1]{#1}%
\providecommand \href@noop [0]{\@secondoftwo}%
\providecommand \href [0]{\begingroup \@sanitize@url \@href}%
\providecommand \@href[1]{\@@startlink{#1}\@@href}%
\providecommand \@@href[1]{\endgroup#1\@@endlink}%
\providecommand \@sanitize@url [0]{\catcode `\\12\catcode `\$12\catcode `\&12\catcode `\#12\catcode `\^12\catcode `\_12\catcode `\%12\relax}%
\providecommand \@@startlink[1]{}%
\providecommand \@@endlink[0]{}%
\providecommand \url  [0]{\begingroup\@sanitize@url \@url }%
\providecommand \@url [1]{\endgroup\@href {#1}{\urlprefix }}%
\providecommand \urlprefix  [0]{URL }%
\providecommand \Eprint [0]{\href }%
\providecommand \doibase [0]{https://doi.org/}%
\providecommand \selectlanguage [0]{\@gobble}%
\providecommand \bibinfo  [0]{\@secondoftwo}%
\providecommand \bibfield  [0]{\@secondoftwo}%
\providecommand \translation [1]{[#1]}%
\providecommand \BibitemOpen [0]{}%
\providecommand \bibitemStop [0]{}%
\providecommand \bibitemNoStop [0]{.\EOS\space}%
\providecommand \EOS [0]{\spacefactor3000\relax}%
\providecommand \BibitemShut  [1]{\csname bibitem#1\endcsname}%
\let\auto@bib@innerbib\@empty
\bibitem [{\citenamefont {Abbott}\ \emph {et~al.}(2016)\citenamefont {Abbott} \emph {et~al.}}]{FirstGWDet}%
  \BibitemOpen
  \bibfield  {author} {\bibinfo {author} {\bibfnamefont {B.}~\bibnamefont {Abbott}} \emph {et~al.} (\bibinfo {collaboration} {LIGO Scientific Collaboration, Virgo Collaboration}),\ }\bibfield  {title} {\bibinfo {title} {Observation of gravitational waves from a binary black hole merger},\ }\href {https://doi.org/https://doi.org/10.1103/PhysRevLett.116.061102} {\bibfield  {journal} {\bibinfo  {journal} {Phys. Rev. Lett.}\ }\textbf {\bibinfo {volume} {116}},\ \bibinfo {pages} {061102} (\bibinfo {year} {2016})}\BibitemShut {NoStop}%
\bibitem [{\citenamefont {Carr}\ and\ \citenamefont {Hawking}(1974)}]{Carr:1974nx}%
  \BibitemOpen
  \bibfield  {author} {\bibinfo {author} {\bibfnamefont {B.~J.}\ \bibnamefont {Carr}}\ and\ \bibinfo {author} {\bibfnamefont {S.~W.}\ \bibnamefont {Hawking}},\ }\bibfield  {title} {\bibinfo {title} {{Black holes in the early Universe}},\ }\href {https://doi.org/10.1093/mnras/168.2.399} {\bibfield  {journal} {\bibinfo  {journal} {Mon. Not. Roy. Astron. Soc.}\ }\textbf {\bibinfo {volume} {168}},\ \bibinfo {pages} {399} (\bibinfo {year} {1974})}\BibitemShut {NoStop}%
\bibitem [{\citenamefont {{Carr}}(1975)}]{Carr:1975qj}%
  \BibitemOpen
  \bibfield  {author} {\bibinfo {author} {\bibfnamefont {B.~J.}\ \bibnamefont {{Carr}}},\ }\bibfield  {title} {\bibinfo {title} {{The primordial black hole mass spectrum}},\ }\href {https://doi.org/10.1086/153853} {\bibfield  {journal} {\bibinfo  {journal} {\apj}\ }\textbf {\bibinfo {volume} {201}},\ \bibinfo {pages} {1} (\bibinfo {year} {1975})}\BibitemShut {NoStop}%
\bibitem [{\citenamefont {Bagui}\ \emph {et~al.}(2023)\citenamefont {Bagui} \emph {et~al.}}]{LISACosmologyWorkingGroup:2023njw}%
  \BibitemOpen
  \bibfield  {author} {\bibinfo {author} {\bibfnamefont {E.}~\bibnamefont {Bagui}} \emph {et~al.} (\bibinfo {collaboration} {LISA Cosmology Working Group}),\ }\bibfield  {title} {\bibinfo {title} {{Primordial black holes and their gravitational-wave signatures}},\ }\href@noop {} {\  (\bibinfo {year} {2023})},\ \Eprint {https://arxiv.org/abs/2310.19857} {arXiv:2310.19857 [astro-ph.CO]} \BibitemShut {NoStop}%
\bibitem [{\citenamefont {Carr}\ \emph {et~al.}(2017{\natexlab{a}})\citenamefont {Carr}, \citenamefont {Tenkanen},\ and\ \citenamefont {Vaskonen}}]{Carr:2017edp}%
  \BibitemOpen
  \bibfield  {author} {\bibinfo {author} {\bibfnamefont {B.}~\bibnamefont {Carr}}, \bibinfo {author} {\bibfnamefont {T.}~\bibnamefont {Tenkanen}},\ and\ \bibinfo {author} {\bibfnamefont {V.}~\bibnamefont {Vaskonen}},\ }\bibfield  {title} {\bibinfo {title} {{Primordial black holes from inflaton and spectator field perturbations in a matter-dominated era}},\ }\href {https://doi.org/10.1103/PhysRevD.96.063507} {\bibfield  {journal} {\bibinfo  {journal} {Phys. Rev. D}\ }\textbf {\bibinfo {volume} {96}},\ \bibinfo {pages} {063507} (\bibinfo {year} {2017}{\natexlab{a}})}\BibitemShut {NoStop}%
\bibitem [{\citenamefont {Carr}\ \emph {et~al.}(2017{\natexlab{b}})\citenamefont {Carr}, \citenamefont {Raidal}, \citenamefont {Tenkanen}, \citenamefont {Vaskonen},\ and\ \citenamefont {Veerm\"ae}}]{Carr:2017jsz}%
  \BibitemOpen
  \bibfield  {author} {\bibinfo {author} {\bibfnamefont {B.}~\bibnamefont {Carr}}, \bibinfo {author} {\bibfnamefont {M.}~\bibnamefont {Raidal}}, \bibinfo {author} {\bibfnamefont {T.}~\bibnamefont {Tenkanen}}, \bibinfo {author} {\bibfnamefont {V.}~\bibnamefont {Vaskonen}},\ and\ \bibinfo {author} {\bibfnamefont {H.}~\bibnamefont {Veerm\"ae}},\ }\bibfield  {title} {\bibinfo {title} {{Primordial black hole constraints for extended mass functions}},\ }\href {https://doi.org/10.1103/PhysRevD.96.023514} {\bibfield  {journal} {\bibinfo  {journal} {Phys. Rev. D}\ }\textbf {\bibinfo {volume} {96}},\ \bibinfo {pages} {023514} (\bibinfo {year} {2017}{\natexlab{b}})}\BibitemShut {NoStop}%
\bibitem [{\citenamefont {Carr}\ \emph {et~al.}(2021)\citenamefont {Carr}, \citenamefont {Clesse}, \citenamefont {Garc{\'\i}a-Bellido},\ and\ \citenamefont {K{\"u}hnel}}]{Carr:2019kxo}%
  \BibitemOpen
  \bibfield  {author} {\bibinfo {author} {\bibfnamefont {B.}~\bibnamefont {Carr}}, \bibinfo {author} {\bibfnamefont {S.}~\bibnamefont {Clesse}}, \bibinfo {author} {\bibfnamefont {J.}~\bibnamefont {Garc{\'\i}a-Bellido}},\ and\ \bibinfo {author} {\bibfnamefont {F.}~\bibnamefont {K{\"u}hnel}},\ }\bibfield  {title} {\bibinfo {title} {Cosmic conundra explained by thermal history and primordial black holes},\ }\href {https://doi.org/10.1016/j.dark.2020.100755} {\bibfield  {journal} {\bibinfo  {journal} {Phys. Dark Universe}\ }\textbf {\bibinfo {volume} {31}},\ \bibinfo {pages} {100755} (\bibinfo {year} {2021})}\BibitemShut {NoStop}%
\bibitem [{\citenamefont {Carr}\ and\ \citenamefont {K{\"u}hnel}(2020)}]{Carr:2020xqk}%
  \BibitemOpen
  \bibfield  {author} {\bibinfo {author} {\bibfnamefont {B.}~\bibnamefont {Carr}}\ and\ \bibinfo {author} {\bibfnamefont {F.}~\bibnamefont {K{\"u}hnel}},\ }\bibfield  {title} {\bibinfo {title} {Primordial black holes as dark matter: recent developments},\ }\href {https://doi.org/10.1146/annurev-nucl-050520-125911} {\bibfield  {journal} {\bibinfo  {journal} {Annu. Rev. Nucl. Part. Sci.}\ }\textbf {\bibinfo {volume} {70}},\ \bibinfo {pages} {355} (\bibinfo {year} {2020})}\BibitemShut {NoStop}%
\bibitem [{\citenamefont {Aasi}\ \emph {et~al.}(2015)\citenamefont {Aasi} \emph {et~al.}}]{LIGOScientific:2014pky}%
  \BibitemOpen
  \bibfield  {author} {\bibinfo {author} {\bibfnamefont {J.}~\bibnamefont {Aasi}} \emph {et~al.} (\bibinfo {collaboration} {LIGO Scientific}),\ }\bibfield  {title} {\bibinfo {title} {{Advanced LIGO}},\ }\href {https://doi.org/10.1088/0264-9381/32/7/074001} {\bibfield  {journal} {\bibinfo  {journal} {Class. Quant. Grav.}\ }\textbf {\bibinfo {volume} {32}},\ \bibinfo {pages} {074001} (\bibinfo {year} {2015})},\ \Eprint {https://arxiv.org/abs/1411.4547} {arXiv:1411.4547 [gr-qc]} \BibitemShut {NoStop}%
\bibitem [{\citenamefont {Acernese}\ \emph {et~al.}(2015)\citenamefont {Acernese} \emph {et~al.}}]{VIRGO:2014yos}%
  \BibitemOpen
  \bibfield  {author} {\bibinfo {author} {\bibfnamefont {F.}~\bibnamefont {Acernese}} \emph {et~al.} (\bibinfo {collaboration} {Virgo}),\ }\bibfield  {title} {\bibinfo {title} {{Advanced Virgo: a second-generation interferometric gravitational wave detector}},\ }\href {https://doi.org/10.1088/0264-9381/32/2/024001} {\bibfield  {journal} {\bibinfo  {journal} {Class. Quant. Grav.}\ }\textbf {\bibinfo {volume} {32}},\ \bibinfo {pages} {024001} (\bibinfo {year} {2015})},\ \Eprint {https://arxiv.org/abs/1408.3978} {arXiv:1408.3978 [gr-qc]} \BibitemShut {NoStop}%
\bibitem [{\citenamefont {Akutsu}\ \emph {et~al.}(2019)\citenamefont {Akutsu} \emph {et~al.}}]{KAGRA:2018plz}%
  \BibitemOpen
  \bibfield  {author} {\bibinfo {author} {\bibfnamefont {T.}~\bibnamefont {Akutsu}} \emph {et~al.} (\bibinfo {collaboration} {KAGRA}),\ }\bibfield  {title} {\bibinfo {title} {{KAGRA: 2.5 Generation Interferometric Gravitational Wave Detector}},\ }\href {https://doi.org/10.1038/s41550-018-0658-y} {\bibfield  {journal} {\bibinfo  {journal} {Nature Astron.}\ }\textbf {\bibinfo {volume} {3}},\ \bibinfo {pages} {35} (\bibinfo {year} {2019})},\ \Eprint {https://arxiv.org/abs/1811.08079} {arXiv:1811.08079 [gr-qc]} \BibitemShut {NoStop}%
\bibitem [{\citenamefont {Sasaki}\ \emph {et~al.}(2016)\citenamefont {Sasaki}, \citenamefont {Suyama}, \citenamefont {Tanaka},\ and\ \citenamefont {Yokoyama}}]{Sasaki:2016jop}%
  \BibitemOpen
  \bibfield  {author} {\bibinfo {author} {\bibfnamefont {M.}~\bibnamefont {Sasaki}}, \bibinfo {author} {\bibfnamefont {T.}~\bibnamefont {Suyama}}, \bibinfo {author} {\bibfnamefont {T.}~\bibnamefont {Tanaka}},\ and\ \bibinfo {author} {\bibfnamefont {S.}~\bibnamefont {Yokoyama}},\ }\bibfield  {title} {\bibinfo {title} {{Primordial Black Hole Scenario for the Gravitational-Wave Event GW150914}},\ }\href {https://doi.org/10.1103/PhysRevLett.117.061101} {\bibfield  {journal} {\bibinfo  {journal} {Phys. Rev. Lett.}\ }\textbf {\bibinfo {volume} {117}},\ \bibinfo {pages} {061101} (\bibinfo {year} {2016})},\ \bibinfo {note} {[Erratum: Phys.Rev.Lett. 121, 059901 (2018)]},\ \Eprint {https://arxiv.org/abs/1603.08338} {arXiv:1603.08338 [astro-ph.CO]} \BibitemShut {NoStop}%
\bibitem [{\citenamefont {Bird}\ \emph {et~al.}(2016)\citenamefont {Bird}, \citenamefont {Cholis}, \citenamefont {Mu\~noz}, \citenamefont {Ali-Ha\"\i{}moud}, \citenamefont {Kamionkowski}, \citenamefont {Kovetz}, \citenamefont {Raccanelli},\ and\ \citenamefont {Riess}}]{Bird:2016dcv}%
  \BibitemOpen
  \bibfield  {author} {\bibinfo {author} {\bibfnamefont {S.}~\bibnamefont {Bird}}, \bibinfo {author} {\bibfnamefont {I.}~\bibnamefont {Cholis}}, \bibinfo {author} {\bibfnamefont {J.~B.}\ \bibnamefont {Mu\~noz}}, \bibinfo {author} {\bibfnamefont {Y.}~\bibnamefont {Ali-Ha\"\i{}moud}}, \bibinfo {author} {\bibfnamefont {M.}~\bibnamefont {Kamionkowski}}, \bibinfo {author} {\bibfnamefont {E.~D.}\ \bibnamefont {Kovetz}}, \bibinfo {author} {\bibfnamefont {A.}~\bibnamefont {Raccanelli}},\ and\ \bibinfo {author} {\bibfnamefont {A.~G.}\ \bibnamefont {Riess}},\ }\bibfield  {title} {\bibinfo {title} {{Did LIGO detect dark matter?}},\ }\href {https://doi.org/10.1103/PhysRevLett.116.201301} {\bibfield  {journal} {\bibinfo  {journal} {Phys. Rev. Lett.}\ }\textbf {\bibinfo {volume} {116}},\ \bibinfo {pages} {201301} (\bibinfo {year} {2016})},\ \Eprint {https://arxiv.org/abs/1603.00464} {arXiv:1603.00464 [astro-ph.CO]} \BibitemShut {NoStop}%
\bibitem [{\citenamefont {Abbott}\ \emph {et~al.}(2021{\natexlab{a}})\citenamefont {Abbott} \emph {et~al.}}]{Abbott2021GWTC-3:Run}%
  \BibitemOpen
  \bibfield  {author} {\bibinfo {author} {\bibfnamefont {R.}~\bibnamefont {Abbott}} \emph {et~al.} (\bibinfo {collaboration} {LIGO Scientific Collaboration, Virgo Collaboration, KAGRA Collaboration}),\ }\bibfield  {title} {\bibinfo {title} {{GWTC-3: Compact Binary Coalescences Observed by LIGO and Virgo During the Second Part of the Third Observing Run}},\ }\href {https://doi.org/10.48550/arXiv.2111.03606} {\bibfield  {journal} {\bibinfo  {journal} {arXiv:2111.03606 [gr-qc]}\ } (\bibinfo {year} {2021}{\natexlab{a}})}\BibitemShut {NoStop}%
\bibitem [{\citenamefont {Abbott}\ \emph {et~al.}(2020{\natexlab{a}})\citenamefont {Abbott} \emph {et~al.}}]{Abbott2020GWTC-2:Run}%
  \BibitemOpen
  \bibfield  {author} {\bibinfo {author} {\bibfnamefont {R.}~\bibnamefont {Abbott}} \emph {et~al.} (\bibinfo {collaboration} {LIGO Scientific Collaboration, Virgo Collaboration}),\ }\bibfield  {title} {\bibinfo {title} {{GWTC-2: Compact Binary Coalescences Observed by LIGO and Virgo During the First Half of the Third Observing Run}},\ }\href {https://doi.org/https://doi.org/10.1103/PhysRevX.11.021053} {\bibfield  {journal} {\bibinfo  {journal} {Phys. Rev. X}\ }\textbf {\bibinfo {volume} {11}},\ \bibinfo {pages} {021053} (\bibinfo {year} {2020}{\natexlab{a}})}\BibitemShut {NoStop}%
\bibitem [{\citenamefont {Abbott}\ \emph {et~al.}(2019{\natexlab{a}})\citenamefont {Abbott} \emph {et~al.}}]{Abbott2019GWTC-1:Runs}%
  \BibitemOpen
  \bibfield  {author} {\bibinfo {author} {\bibfnamefont {R.}~\bibnamefont {Abbott}} \emph {et~al.} (\bibinfo {collaboration} {LIGO Scientific Collaboration, Virgo Collaboration}),\ }\bibfield  {title} {\bibinfo {title} {{GWTC-1: A Gravitational-Wave Transient Catalog of Compact Binary Mergers Observed by LIGO and Virgo during the First and Second Observing Runs}},\ }\href {https://doi.org/https://doi.org/10.1103/PhysRevX.9.031040} {\bibfield  {journal} {\bibinfo  {journal} {Phys. Rev. X}\ }\textbf {\bibinfo {volume} {9}},\ \bibinfo {pages} {031040} (\bibinfo {year} {2019}{\natexlab{a}})}\BibitemShut {NoStop}%
\bibitem [{\citenamefont {H\"utsi}\ \emph {et~al.}(2021)\citenamefont {H\"utsi}, \citenamefont {Raidal}, \citenamefont {Vaskonen},\ and\ \citenamefont {Veerm\"ae}}]{Hutsi:2020sol}%
  \BibitemOpen
  \bibfield  {author} {\bibinfo {author} {\bibfnamefont {G.}~\bibnamefont {H\"utsi}}, \bibinfo {author} {\bibfnamefont {M.}~\bibnamefont {Raidal}}, \bibinfo {author} {\bibfnamefont {V.}~\bibnamefont {Vaskonen}},\ and\ \bibinfo {author} {\bibfnamefont {H.}~\bibnamefont {Veerm\"ae}},\ }\bibfield  {title} {\bibinfo {title} {{Two populations of LIGO-Virgo black holes}},\ }\href {https://doi.org/10.1088/1475-7516/2021/03/068} {\bibfield  {journal} {\bibinfo  {journal} {JCAP}\ }\textbf {\bibinfo {volume} {03}},\ \bibinfo {pages} {068}},\ \Eprint {https://arxiv.org/abs/2012.02786} {arXiv:2012.02786 [astro-ph.CO]} \BibitemShut {NoStop}%
\bibitem [{\citenamefont {Hall}\ \emph {et~al.}(2020)\citenamefont {Hall}, \citenamefont {Gow},\ and\ \citenamefont {Byrnes}}]{Hall:2020daa}%
  \BibitemOpen
  \bibfield  {author} {\bibinfo {author} {\bibfnamefont {A.}~\bibnamefont {Hall}}, \bibinfo {author} {\bibfnamefont {A.~D.}\ \bibnamefont {Gow}},\ and\ \bibinfo {author} {\bibfnamefont {C.~T.}\ \bibnamefont {Byrnes}},\ }\bibfield  {title} {\bibinfo {title} {{Bayesian analysis of LIGO-Virgo mergers: Primordial vs. astrophysical black hole populations}},\ }\href {https://doi.org/10.1103/PhysRevD.102.123524} {\bibfield  {journal} {\bibinfo  {journal} {Phys. Rev. D}\ }\textbf {\bibinfo {volume} {102}},\ \bibinfo {pages} {123524} (\bibinfo {year} {2020})},\ \Eprint {https://arxiv.org/abs/2008.13704} {arXiv:2008.13704 [astro-ph.CO]} \BibitemShut {NoStop}%
\bibitem [{\citenamefont {Wong}\ \emph {et~al.}(2021)\citenamefont {Wong}, \citenamefont {Franciolini}, \citenamefont {De~Luca}, \citenamefont {Baibhav}, \citenamefont {Berti}, \citenamefont {Pani},\ and\ \citenamefont {Riotto}}]{Wong:2020yig}%
  \BibitemOpen
  \bibfield  {author} {\bibinfo {author} {\bibfnamefont {K.~W.~K.}\ \bibnamefont {Wong}}, \bibinfo {author} {\bibfnamefont {G.}~\bibnamefont {Franciolini}}, \bibinfo {author} {\bibfnamefont {V.}~\bibnamefont {De~Luca}}, \bibinfo {author} {\bibfnamefont {V.}~\bibnamefont {Baibhav}}, \bibinfo {author} {\bibfnamefont {E.}~\bibnamefont {Berti}}, \bibinfo {author} {\bibfnamefont {P.}~\bibnamefont {Pani}},\ and\ \bibinfo {author} {\bibfnamefont {A.}~\bibnamefont {Riotto}},\ }\bibfield  {title} {\bibinfo {title} {{Constraining the primordial black hole scenario with Bayesian inference and machine learning: the GWTC-2 gravitational wave catalog}},\ }\href {https://doi.org/10.1103/PhysRevD.103.023026} {\bibfield  {journal} {\bibinfo  {journal} {Phys. Rev. D}\ }\textbf {\bibinfo {volume} {103}},\ \bibinfo {pages} {023026} (\bibinfo {year} {2021})},\ \Eprint {https://arxiv.org/abs/2011.01865} {arXiv:2011.01865 [gr-qc]} \BibitemShut {NoStop}%
\bibitem [{\citenamefont {Franciolini}\ \emph {et~al.}(2022{\natexlab{a}})\citenamefont {Franciolini}, \citenamefont {Baibhav}, \citenamefont {De~Luca}, \citenamefont {Ng}, \citenamefont {Wong}, \citenamefont {Berti}, \citenamefont {Pani}, \citenamefont {Riotto},\ and\ \citenamefont {Vitale}}]{Franciolini:2021tla}%
  \BibitemOpen
  \bibfield  {author} {\bibinfo {author} {\bibfnamefont {G.}~\bibnamefont {Franciolini}}, \bibinfo {author} {\bibfnamefont {V.}~\bibnamefont {Baibhav}}, \bibinfo {author} {\bibfnamefont {V.}~\bibnamefont {De~Luca}}, \bibinfo {author} {\bibfnamefont {K.~K.~Y.}\ \bibnamefont {Ng}}, \bibinfo {author} {\bibfnamefont {K.~W.~K.}\ \bibnamefont {Wong}}, \bibinfo {author} {\bibfnamefont {E.}~\bibnamefont {Berti}}, \bibinfo {author} {\bibfnamefont {P.}~\bibnamefont {Pani}}, \bibinfo {author} {\bibfnamefont {A.}~\bibnamefont {Riotto}},\ and\ \bibinfo {author} {\bibfnamefont {S.}~\bibnamefont {Vitale}},\ }\bibfield  {title} {\bibinfo {title} {{Searching for a subpopulation of primordial black holes in LIGO-Virgo gravitational-wave data}},\ }\href {https://doi.org/10.1103/PhysRevD.105.083526} {\bibfield  {journal} {\bibinfo  {journal} {Phys. Rev. D}\ }\textbf {\bibinfo {volume} {105}},\ \bibinfo {pages} {083526} (\bibinfo {year} {2022}{\natexlab{a}})},\ \Eprint {https://arxiv.org/abs/2105.03349} {arXiv:2105.03349 [gr-qc]}
  \BibitemShut {NoStop}%
\bibitem [{\citenamefont {De~Luca}\ \emph {et~al.}(2021)\citenamefont {De~Luca}, \citenamefont {Franciolini}, \citenamefont {Pani},\ and\ \citenamefont {Riotto}}]{DeLuca:2021wjr}%
  \BibitemOpen
  \bibfield  {author} {\bibinfo {author} {\bibfnamefont {V.}~\bibnamefont {De~Luca}}, \bibinfo {author} {\bibfnamefont {G.}~\bibnamefont {Franciolini}}, \bibinfo {author} {\bibfnamefont {P.}~\bibnamefont {Pani}},\ and\ \bibinfo {author} {\bibfnamefont {A.}~\bibnamefont {Riotto}},\ }\bibfield  {title} {\bibinfo {title} {{Bayesian Evidence for Both Astrophysical and Primordial Black Holes: Mapping the GWTC-2 Catalog to Third-Generation Detectors}},\ }\href {https://doi.org/10.1088/1475-7516/2021/05/003} {\bibfield  {journal} {\bibinfo  {journal} {JCAP}\ }\textbf {\bibinfo {volume} {05}},\ \bibinfo {pages} {003}},\ \Eprint {https://arxiv.org/abs/2102.03809} {arXiv:2102.03809 [astro-ph.CO]} \BibitemShut {NoStop}%
\bibitem [{\citenamefont {Franciolini}\ \emph {et~al.}(2022{\natexlab{b}})\citenamefont {Franciolini}, \citenamefont {Musco}, \citenamefont {Pani},\ and\ \citenamefont {Urbano}}]{Franciolini:2022tfm}%
  \BibitemOpen
  \bibfield  {author} {\bibinfo {author} {\bibfnamefont {G.}~\bibnamefont {Franciolini}}, \bibinfo {author} {\bibfnamefont {I.}~\bibnamefont {Musco}}, \bibinfo {author} {\bibfnamefont {P.}~\bibnamefont {Pani}},\ and\ \bibinfo {author} {\bibfnamefont {A.}~\bibnamefont {Urbano}},\ }\bibfield  {title} {\bibinfo {title} {{From inflation to black hole mergers and back again: Gravitational-wave data-driven constraints on inflationary scenarios with a first-principle model of primordial black holes across the QCD epoch}},\ }\href {https://doi.org/10.1103/PhysRevD.106.123526} {\bibfield  {journal} {\bibinfo  {journal} {Phys. Rev. D}\ }\textbf {\bibinfo {volume} {106}},\ \bibinfo {pages} {123526} (\bibinfo {year} {2022}{\natexlab{b}})},\ \Eprint {https://arxiv.org/abs/2209.05959} {arXiv:2209.05959 [astro-ph.CO]} \BibitemShut {NoStop}%
\bibitem [{\citenamefont {Andr\'es-Carcasona}\ \emph {et~al.}(2024{\natexlab{a}})\citenamefont {Andr\'es-Carcasona}, \citenamefont {Iovino}, \citenamefont {Vaskonen}, \citenamefont {Veerm\"ae}, \citenamefont {Mart\'\i{}nez}, \citenamefont {Pujol\`as},\ and\ \citenamefont {Mir}}]{Andres-Carcasona:2024wqk}%
  \BibitemOpen
  \bibfield  {author} {\bibinfo {author} {\bibfnamefont {M.}~\bibnamefont {Andr\'es-Carcasona}}, \bibinfo {author} {\bibfnamefont {A.~J.}\ \bibnamefont {Iovino}}, \bibinfo {author} {\bibfnamefont {V.}~\bibnamefont {Vaskonen}}, \bibinfo {author} {\bibfnamefont {H.}~\bibnamefont {Veerm\"ae}}, \bibinfo {author} {\bibfnamefont {M.}~\bibnamefont {Mart\'\i{}nez}}, \bibinfo {author} {\bibfnamefont {O.}~\bibnamefont {Pujol\`as}},\ and\ \bibinfo {author} {\bibfnamefont {L.~M.}\ \bibnamefont {Mir}},\ }\bibfield  {title} {\bibinfo {title} {{Constraints on primordial black holes from LIGO-Virgo-KAGRA O3 events}},\ }\href {https://doi.org/10.1103/PhysRevD.110.023040} {\bibfield  {journal} {\bibinfo  {journal} {Phys. Rev. D}\ }\textbf {\bibinfo {volume} {110}},\ \bibinfo {pages} {023040} (\bibinfo {year} {2024}{\natexlab{a}})},\ \Eprint {https://arxiv.org/abs/2405.05732} {arXiv:2405.05732 [astro-ph.CO]} \BibitemShut {NoStop}%
\bibitem [{\citenamefont {Abbott}\ \emph {et~al.}(2018)\citenamefont {Abbott} \emph {et~al.}}]{LIGOScientific:2018glc}%
  \BibitemOpen
  \bibfield  {author} {\bibinfo {author} {\bibfnamefont {B.~P.}\ \bibnamefont {Abbott}} \emph {et~al.} (\bibinfo {collaboration} {LIGO Scientific Collaboration, Virgo Collaboration}),\ }\bibfield  {title} {\bibinfo {title} {{Search for Subsolar-Mass Ultracompact Binaries in Advanced LIGO's First Observing Run}},\ }\href {https://doi.org/10.1103/PhysRevLett.121.231103} {\bibfield  {journal} {\bibinfo  {journal} {Phys. Rev. Lett.}\ }\textbf {\bibinfo {volume} {121}},\ \bibinfo {pages} {231103} (\bibinfo {year} {2018})},\ \Eprint {https://arxiv.org/abs/1808.04771} {arXiv:1808.04771 [astro-ph.CO]} \BibitemShut {NoStop}%
\bibitem [{\citenamefont {Abbott}\ \emph {et~al.}(2019{\natexlab{b}})\citenamefont {Abbott} \emph {et~al.}}]{LIGOScientific:2019kan}%
  \BibitemOpen
  \bibfield  {author} {\bibinfo {author} {\bibfnamefont {B.~P.}\ \bibnamefont {Abbott}} \emph {et~al.} (\bibinfo {collaboration} {LIGO Scientific Collaboration, Virgo Collaboration}),\ }\bibfield  {title} {\bibinfo {title} {{Search for Subsolar Mass Ultracompact Binaries in Advanced LIGO's Second Observing Run}},\ }\href {https://doi.org/10.1103/PhysRevLett.123.161102} {\bibfield  {journal} {\bibinfo  {journal} {Phys. Rev. Lett.}\ }\textbf {\bibinfo {volume} {123}},\ \bibinfo {pages} {161102} (\bibinfo {year} {2019}{\natexlab{b}})},\ \Eprint {https://arxiv.org/abs/1904.08976} {arXiv:1904.08976 [astro-ph.CO]} \BibitemShut {NoStop}%
\bibitem [{\citenamefont {Nitz}\ and\ \citenamefont {Wang}(2021{\natexlab{a}})}]{Nitz:2020bdb}%
  \BibitemOpen
  \bibfield  {author} {\bibinfo {author} {\bibfnamefont {A.~H.}\ \bibnamefont {Nitz}}\ and\ \bibinfo {author} {\bibfnamefont {Y.-F.}\ \bibnamefont {Wang}},\ }\bibfield  {title} {\bibinfo {title} {{Search for Gravitational Waves from High-Mass-Ratio Compact-Binary Mergers of Stellar Mass and Subsolar Mass Black Holes}},\ }\href {https://doi.org/10.1103/PhysRevLett.126.021103} {\bibfield  {journal} {\bibinfo  {journal} {Phys. Rev. Lett.}\ }\textbf {\bibinfo {volume} {126}},\ \bibinfo {pages} {021103} (\bibinfo {year} {2021}{\natexlab{a}})},\ \Eprint {https://arxiv.org/abs/2007.03583} {arXiv:2007.03583 [astro-ph.HE]} \BibitemShut {NoStop}%
\bibitem [{\citenamefont {Phukon}\ \emph {et~al.}(2021)\citenamefont {Phukon}, \citenamefont {Baltus}, \citenamefont {Caudill}, \citenamefont {Clesse}, \citenamefont {Depasse}, \citenamefont {Fays}, \citenamefont {Fong}, \citenamefont {Kapadia}, \citenamefont {Magee},\ and\ \citenamefont {Tanasijczuk}}]{Phukon:2021cus}%
  \BibitemOpen
  \bibfield  {author} {\bibinfo {author} {\bibfnamefont {K.~S.}\ \bibnamefont {Phukon}}, \bibinfo {author} {\bibfnamefont {G.}~\bibnamefont {Baltus}}, \bibinfo {author} {\bibfnamefont {S.}~\bibnamefont {Caudill}}, \bibinfo {author} {\bibfnamefont {S.}~\bibnamefont {Clesse}}, \bibinfo {author} {\bibfnamefont {A.}~\bibnamefont {Depasse}}, \bibinfo {author} {\bibfnamefont {M.}~\bibnamefont {Fays}}, \bibinfo {author} {\bibfnamefont {H.}~\bibnamefont {Fong}}, \bibinfo {author} {\bibfnamefont {S.~J.}\ \bibnamefont {Kapadia}}, \bibinfo {author} {\bibfnamefont {R.}~\bibnamefont {Magee}},\ and\ \bibinfo {author} {\bibfnamefont {A.~J.}\ \bibnamefont {Tanasijczuk}},\ }\bibfield  {title} {\bibinfo {title} {{The hunt for sub-solar primordial black holes in low mass ratio binaries is open}},\ }\href@noop {} {\  (\bibinfo {year} {2021})},\ \Eprint {https://arxiv.org/abs/2105.11449} {arXiv:2105.11449 [astro-ph.CO]} \BibitemShut {NoStop}%
\bibitem [{\citenamefont {Nitz}\ and\ \citenamefont {Wang}(2021{\natexlab{b}})}]{Nitz:2021mzz}%
  \BibitemOpen
  \bibfield  {author} {\bibinfo {author} {\bibfnamefont {A.~H.}\ \bibnamefont {Nitz}}\ and\ \bibinfo {author} {\bibfnamefont {Y.-F.}\ \bibnamefont {Wang}},\ }\bibfield  {title} {\bibinfo {title} {{Search for gravitational waves from the coalescence of sub-solar mass and eccentric compact binaries}},\ }\href {https://doi.org/10.3847/1538-4357/ac01d9} {\bibfield  {journal} {\bibinfo  {journal} {\apj}\ }\textbf {\bibinfo {volume} {915}},\ \bibinfo {eid} {54} (\bibinfo {year} {2021}{\natexlab{b}})},\ \Eprint {https://arxiv.org/abs/2102.00868} {arXiv:2102.00868 [astro-ph.HE]} \BibitemShut {NoStop}%
\bibitem [{\citenamefont {Nitz}\ and\ \citenamefont {Wang}(2021{\natexlab{c}})}]{Nitz:2021vqh}%
  \BibitemOpen
  \bibfield  {author} {\bibinfo {author} {\bibfnamefont {A.~H.}\ \bibnamefont {Nitz}}\ and\ \bibinfo {author} {\bibfnamefont {Y.-F.}\ \bibnamefont {Wang}},\ }\bibfield  {title} {\bibinfo {title} {{Search for Gravitational Waves from the Coalescence of Subsolar-Mass Binaries in the First Half of Advanced LIGO and Virgo\textquoteright{}s Third Observing Run}},\ }\href {https://doi.org/10.1103/PhysRevLett.127.151101} {\bibfield  {journal} {\bibinfo  {journal} {Phys. Rev. Lett.}\ }\textbf {\bibinfo {volume} {127}},\ \bibinfo {pages} {151101} (\bibinfo {year} {2021}{\natexlab{c}})},\ \Eprint {https://arxiv.org/abs/2106.08979} {arXiv:2106.08979 [astro-ph.HE]} \BibitemShut {NoStop}%
\bibitem [{\citenamefont {Nitz}\ and\ \citenamefont {Wang}(2022)}]{Nitz:2022ltl}%
  \BibitemOpen
  \bibfield  {author} {\bibinfo {author} {\bibfnamefont {A.~H.}\ \bibnamefont {Nitz}}\ and\ \bibinfo {author} {\bibfnamefont {Y.-F.}\ \bibnamefont {Wang}},\ }\bibfield  {title} {\bibinfo {title} {{Broad search for gravitational waves from subsolar-mass binary black hole mergers}},\ }\href {https://doi.org/10.1103/PhysRevD.105.062008} {\bibfield  {journal} {\bibinfo  {journal} {Phys. Rev. D}\ }\textbf {\bibinfo {volume} {105}},\ \bibinfo {pages} {062008} (\bibinfo {year} {2022})},\ \Eprint {https://arxiv.org/abs/2202.11024} {arXiv:2202.11024 [gr-qc]} \BibitemShut {NoStop}%
\bibitem [{\citenamefont {Miller}\ \emph {et~al.}(2021{\natexlab{a}})\citenamefont {Miller}, \citenamefont {Clesse}, \citenamefont {De~Lillo}, \citenamefont {Bruno}, \citenamefont {Depasse},\ and\ \citenamefont {Tanasijczuk}}]{Miller:2020kmv}%
  \BibitemOpen
  \bibfield  {author} {\bibinfo {author} {\bibfnamefont {A.~L.}\ \bibnamefont {Miller}}, \bibinfo {author} {\bibfnamefont {S.}~\bibnamefont {Clesse}}, \bibinfo {author} {\bibfnamefont {F.}~\bibnamefont {De~Lillo}}, \bibinfo {author} {\bibfnamefont {G.}~\bibnamefont {Bruno}}, \bibinfo {author} {\bibfnamefont {A.}~\bibnamefont {Depasse}},\ and\ \bibinfo {author} {\bibfnamefont {A.}~\bibnamefont {Tanasijczuk}},\ }\bibfield  {title} {\bibinfo {title} {{Probing planetary-mass primordial black holes with continuous gravitational waves}},\ }\href {https://doi.org/10.1016/j.dark.2021.100836} {\bibfield  {journal} {\bibinfo  {journal} {Phys. Dark Univ.}\ }\textbf {\bibinfo {volume} {32}},\ \bibinfo {pages} {100836} (\bibinfo {year} {2021}{\natexlab{a}})},\ \Eprint {https://arxiv.org/abs/2012.12983} {arXiv:2012.12983 [astro-ph.HE]} \BibitemShut {NoStop}%
\bibitem [{\citenamefont {Miller}\ \emph {et~al.}(2022)\citenamefont {Miller}, \citenamefont {Aggarwal}, \citenamefont {Clesse},\ and\ \citenamefont {De~Lillo}}]{Miller:2021knj}%
  \BibitemOpen
  \bibfield  {author} {\bibinfo {author} {\bibfnamefont {A.~L.}\ \bibnamefont {Miller}}, \bibinfo {author} {\bibfnamefont {N.}~\bibnamefont {Aggarwal}}, \bibinfo {author} {\bibfnamefont {S.}~\bibnamefont {Clesse}},\ and\ \bibinfo {author} {\bibfnamefont {F.}~\bibnamefont {De~Lillo}},\ }\bibfield  {title} {\bibinfo {title} {{Constraints on planetary and asteroid-mass primordial black holes from continuous gravitational-wave searches}},\ }\href {https://doi.org/10.1103/PhysRevD.105.062008} {\bibfield  {journal} {\bibinfo  {journal} {Phys. Rev. D}\ }\textbf {\bibinfo {volume} {105}},\ \bibinfo {pages} {062008} (\bibinfo {year} {2022})},\ \Eprint {https://arxiv.org/abs/2110.06188} {arXiv:2110.06188 [gr-qc]} \BibitemShut {NoStop}%
\bibitem [{\citenamefont {Morras}\ \emph {et~al.}(2023)\citenamefont {Morras} \emph {et~al.}}]{Morras:2023jvb}%
  \BibitemOpen
  \bibfield  {author} {\bibinfo {author} {\bibfnamefont {G.}~\bibnamefont {Morras}} \emph {et~al.},\ }\bibfield  {title} {\bibinfo {title} {{Analysis of a subsolar-mass compact binary candidate from the second observing run of Advanced LIGO}},\ }\href {https://doi.org/10.1016/j.dark.2023.101285} {\bibfield  {journal} {\bibinfo  {journal} {Phys. Dark Univ.}\ }\textbf {\bibinfo {volume} {42}},\ \bibinfo {pages} {101285} (\bibinfo {year} {2023})},\ \Eprint {https://arxiv.org/abs/2301.11619} {arXiv:2301.11619 [gr-qc]} \BibitemShut {NoStop}%
\bibitem [{\citenamefont {Mukherjee}\ and\ \citenamefont {Silk}(2021)}]{Mukherjee:2021ags}%
  \BibitemOpen
  \bibfield  {author} {\bibinfo {author} {\bibfnamefont {S.}~\bibnamefont {Mukherjee}}\ and\ \bibinfo {author} {\bibfnamefont {J.}~\bibnamefont {Silk}},\ }\bibfield  {title} {\bibinfo {title} {{Can we distinguish astrophysical from primordial black holes via the stochastic gravitational wave background?}},\ }\href {https://doi.org/10.1093/mnras/stab1932} {\bibfield  {journal} {\bibinfo  {journal} {Mon. Not. Roy. Astron. Soc.}\ }\textbf {\bibinfo {volume} {506}},\ \bibinfo {pages} {3977} (\bibinfo {year} {2021})},\ \Eprint {https://arxiv.org/abs/2105.11139} {arXiv:2105.11139 [gr-qc]} \BibitemShut {NoStop}%
\bibitem [{\citenamefont {Mukherjee}\ \emph {et~al.}(2022)\citenamefont {Mukherjee}, \citenamefont {Meinema},\ and\ \citenamefont {Silk}}]{Mukherjee:2021itf}%
  \BibitemOpen
  \bibfield  {author} {\bibinfo {author} {\bibfnamefont {S.}~\bibnamefont {Mukherjee}}, \bibinfo {author} {\bibfnamefont {M.~S.~P.}\ \bibnamefont {Meinema}},\ and\ \bibinfo {author} {\bibfnamefont {J.}~\bibnamefont {Silk}},\ }\bibfield  {title} {\bibinfo {title} {{Prospects of discovering subsolar primordial black holes using the stochastic gravitational wave background from third-generation detectors}},\ }\href {https://doi.org/10.1093/mnras/stab3756} {\bibfield  {journal} {\bibinfo  {journal} {Mon. Not. Roy. Astron. Soc.}\ }\textbf {\bibinfo {volume} {510}},\ \bibinfo {pages} {6218} (\bibinfo {year} {2022})},\ \Eprint {https://arxiv.org/abs/2107.02181} {arXiv:2107.02181 [astro-ph.CO]} \BibitemShut {NoStop}%
\bibitem [{\citenamefont {Andres-Carcasona}\ \emph {et~al.}(2023)\citenamefont {Andres-Carcasona}, \citenamefont {Menendez-Vazquez}, \citenamefont {Martinez},\ and\ \citenamefont {Mir}}]{Andres-Carcasona:2022prl}%
  \BibitemOpen
  \bibfield  {author} {\bibinfo {author} {\bibfnamefont {M.}~\bibnamefont {Andres-Carcasona}}, \bibinfo {author} {\bibfnamefont {A.}~\bibnamefont {Menendez-Vazquez}}, \bibinfo {author} {\bibfnamefont {M.}~\bibnamefont {Martinez}},\ and\ \bibinfo {author} {\bibfnamefont {L.~M.}\ \bibnamefont {Mir}},\ }\bibfield  {title} {\bibinfo {title} {{Searches for mass-asymmetric compact binary coalescence events using neural networks in the LIGO/Virgo third observation period}},\ }\href {https://doi.org/10.1103/PhysRevD.107.082003} {\bibfield  {journal} {\bibinfo  {journal} {Phys. Rev. D}\ }\textbf {\bibinfo {volume} {107}},\ \bibinfo {pages} {082003} (\bibinfo {year} {2023})},\ \Eprint {https://arxiv.org/abs/2212.02829} {arXiv:2212.02829 [gr-qc]} \BibitemShut {NoStop}%
\bibitem [{\citenamefont {Miller}\ \emph {et~al.}(2024{\natexlab{a}})\citenamefont {Miller}, \citenamefont {Aggarwal}, \citenamefont {Clesse}, \citenamefont {De~Lillo}, \citenamefont {Sachdev}, \citenamefont {Astone}, \citenamefont {Palomba}, \citenamefont {Piccinni},\ and\ \citenamefont {Pierini}}]{Miller:2024fpo}%
  \BibitemOpen
  \bibfield  {author} {\bibinfo {author} {\bibfnamefont {A.~L.}\ \bibnamefont {Miller}}, \bibinfo {author} {\bibfnamefont {N.}~\bibnamefont {Aggarwal}}, \bibinfo {author} {\bibfnamefont {S.}~\bibnamefont {Clesse}}, \bibinfo {author} {\bibfnamefont {F.}~\bibnamefont {De~Lillo}}, \bibinfo {author} {\bibfnamefont {S.}~\bibnamefont {Sachdev}}, \bibinfo {author} {\bibfnamefont {P.}~\bibnamefont {Astone}}, \bibinfo {author} {\bibfnamefont {C.}~\bibnamefont {Palomba}}, \bibinfo {author} {\bibfnamefont {O.~J.}\ \bibnamefont {Piccinni}},\ and\ \bibinfo {author} {\bibfnamefont {L.}~\bibnamefont {Pierini}},\ }\bibfield  {title} {\bibinfo {title} {{Gravitational Wave Constraints on Planetary-Mass Primordial Black Holes Using LIGO O3a Data}},\ }\href {https://doi.org/10.1103/PhysRevLett.133.111401} {\bibfield  {journal} {\bibinfo  {journal} {Phys. Rev. Lett.}\ }\textbf {\bibinfo {volume} {133}},\ \bibinfo {pages} {111401} (\bibinfo {year} {2024}{\natexlab{a}})},\ \Eprint {https://arxiv.org/abs/2402.19468}
  {arXiv:2402.19468 [gr-qc]} \BibitemShut {NoStop}%
\bibitem [{\citenamefont {Miller}\ \emph {et~al.}(2024{\natexlab{b}})\citenamefont {Miller}, \citenamefont {Aggarwal}, \citenamefont {Clesse}, \citenamefont {De~Lillo}, \citenamefont {Sachdev}, \citenamefont {Astone}, \citenamefont {Palomba}, \citenamefont {Piccinni},\ and\ \citenamefont {Pierini}}]{Miller:2024jpo_2}%
  \BibitemOpen
  \bibfield  {author} {\bibinfo {author} {\bibfnamefont {A.~L.}\ \bibnamefont {Miller}}, \bibinfo {author} {\bibfnamefont {N.}~\bibnamefont {Aggarwal}}, \bibinfo {author} {\bibfnamefont {S.}~\bibnamefont {Clesse}}, \bibinfo {author} {\bibfnamefont {F.}~\bibnamefont {De~Lillo}}, \bibinfo {author} {\bibfnamefont {S.}~\bibnamefont {Sachdev}}, \bibinfo {author} {\bibfnamefont {P.}~\bibnamefont {Astone}}, \bibinfo {author} {\bibfnamefont {C.}~\bibnamefont {Palomba}}, \bibinfo {author} {\bibfnamefont {O.~J.}\ \bibnamefont {Piccinni}},\ and\ \bibinfo {author} {\bibfnamefont {L.}~\bibnamefont {Pierini}},\ }\bibfield  {title} {\bibinfo {title} {{Method to search for inspiraling planetary-mass ultracompact binaries using the generalized frequency-Hough transform in LIGO O3a data}},\ }\href {https://doi.org/10.1103/PhysRevD.110.082004} {\bibfield  {journal} {\bibinfo  {journal} {Phys. Rev. D}\ }\textbf {\bibinfo {volume} {110}},\ \bibinfo {pages} {082004} (\bibinfo {year} {2024}{\natexlab{b}})},\ \Eprint
  {https://arxiv.org/abs/2407.17052} {arXiv:2407.17052 [astro-ph.IM]} \BibitemShut {NoStop}%
\bibitem [{\citenamefont {Niikura}\ \emph {et~al.}(2019)\citenamefont {Niikura} \emph {et~al.}}]{Niikura:2019kqi}%
  \BibitemOpen
  \bibfield  {author} {\bibinfo {author} {\bibfnamefont {H.}~\bibnamefont {Niikura}} \emph {et~al.},\ }\bibfield  {title} {\bibinfo {title} {{Microlensing constraints on primordial black holes with Subaru/HSC Andromeda observations}},\ }\href {https://doi.org/10.1038/s41550-019-0723-1} {\bibfield  {journal} {\bibinfo  {journal} {Nature Astron.}\ }\textbf {\bibinfo {volume} {3}},\ \bibinfo {pages} {524} (\bibinfo {year} {2019})},\ \Eprint {https://arxiv.org/abs/1701.02151} {arXiv:1701.02151 [astro-ph.CO]} \BibitemShut {NoStop}%
\bibitem [{\citenamefont {Tisserand}\ \emph {et~al.}(2007)\citenamefont {Tisserand} \emph {et~al.}}]{EROS-2:2006ryy}%
  \BibitemOpen
  \bibfield  {author} {\bibinfo {author} {\bibfnamefont {P.}~\bibnamefont {Tisserand}} \emph {et~al.} (\bibinfo {collaboration} {EROS-2}),\ }\bibfield  {title} {\bibinfo {title} {{Limits on the Macho Content of the Galactic Halo from the EROS-2 Survey of the Magellanic Clouds}},\ }\href {https://doi.org/10.1051/0004-6361:20066017} {\bibfield  {journal} {\bibinfo  {journal} {Astron. Astrophys.}\ }\textbf {\bibinfo {volume} {469}},\ \bibinfo {pages} {387} (\bibinfo {year} {2007})},\ \Eprint {https://arxiv.org/abs/astro-ph/0607207} {arXiv:astro-ph/0607207} \BibitemShut {NoStop}%
\bibitem [{\citenamefont {Alcock}\ \emph {et~al.}(2000)\citenamefont {Alcock} \emph {et~al.}}]{MACHO:2000qbb}%
  \BibitemOpen
  \bibfield  {author} {\bibinfo {author} {\bibfnamefont {C.}~\bibnamefont {Alcock}} \emph {et~al.} (\bibinfo {collaboration} {MACHO}),\ }\bibfield  {title} {\bibinfo {title} {{The MACHO project: Microlensing results from 5.7 years of LMC observations}},\ }\href {https://doi.org/10.1086/309512} {\bibfield  {journal} {\bibinfo  {journal} {Astrophys. J.}\ }\textbf {\bibinfo {volume} {542}},\ \bibinfo {pages} {281} (\bibinfo {year} {2000})},\ \Eprint {https://arxiv.org/abs/astro-ph/0001272} {arXiv:astro-ph/0001272} \BibitemShut {NoStop}%
\bibitem [{\citenamefont {Balick}\ and\ \citenamefont {Brown}(1974)}]{balick1974intense}%
  \BibitemOpen
  \bibfield  {author} {\bibinfo {author} {\bibfnamefont {B.}~\bibnamefont {Balick}}\ and\ \bibinfo {author} {\bibfnamefont {R.~L.}\ \bibnamefont {Brown}},\ }\bibfield  {title} {\bibinfo {title} {Intense sub-arcsecond structure in the galactic center},\ }\href {https://doi.org/10.1086/153242} {\bibfield  {journal} {\bibinfo  {journal} {Astrophysical Journal}\ }\textbf {\bibinfo {volume} {194}},\ \bibinfo {pages} {265} (\bibinfo {year} {1974})}\BibitemShut {NoStop}%
\bibitem [{\citenamefont {Ghez}\ \emph {et~al.}(2003)\citenamefont {Ghez} \emph {et~al.}}]{ghez2003first}%
  \BibitemOpen
  \bibfield  {author} {\bibinfo {author} {\bibfnamefont {A.}~\bibnamefont {Ghez}} \emph {et~al.},\ }\bibfield  {title} {\bibinfo {title} {The first measurement of spectral lines in a short-period star bound to the galaxy’s central black hole: a paradox of youth},\ }\href {https://doi.org/10.1086/374804} {\bibfield  {journal} {\bibinfo  {journal} {\apj}\ }\textbf {\bibinfo {volume} {586}},\ \bibinfo {pages} {L127} (\bibinfo {year} {2003})}\BibitemShut {NoStop}%
\bibitem [{\citenamefont {Genzel}\ \emph {et~al.}(2010)\citenamefont {Genzel}, \citenamefont {Eisenhauer},\ and\ \citenamefont {Gillessen}}]{Genzel:2010:SgrA}%
  \BibitemOpen
  \bibfield  {author} {\bibinfo {author} {\bibfnamefont {R.}~\bibnamefont {Genzel}}, \bibinfo {author} {\bibfnamefont {F.}~\bibnamefont {Eisenhauer}},\ and\ \bibinfo {author} {\bibfnamefont {S.}~\bibnamefont {Gillessen}},\ }\bibfield  {title} {\bibinfo {title} {The galactic center massive black hole and nuclear star cluster},\ }\href {https://doi.org/10.1103/RevModPhys.82.3121} {\bibfield  {journal} {\bibinfo  {journal} {Rev. Mod. Phys.}\ }\textbf {\bibinfo {volume} {82}},\ \bibinfo {pages} {3121} (\bibinfo {year} {2010})}\BibitemShut {NoStop}%
\bibitem [{\citenamefont {Akiyama}\ \emph {et~al.}(2019)\citenamefont {Akiyama} \emph {et~al.}}]{EventHorizonTelescope:2019dse}%
  \BibitemOpen
  \bibfield  {author} {\bibinfo {author} {\bibfnamefont {K.}~\bibnamefont {Akiyama}} \emph {et~al.} (\bibinfo {collaboration} {Event Horizon Telescope}),\ }\bibfield  {title} {\bibinfo {title} {{First M87 Event Horizon Telescope Results. I. The Shadow of the Supermassive Black Hole}},\ }\bibfield  {journal} {\bibinfo  {journal} {Astrophys. J. Lett.}\ }\textbf {\bibinfo {volume} {875}},\ \href {https://doi.org/10.3847/2041-8213/ab0ec7} {10.3847/2041-8213/ab0ec7} (\bibinfo {year} {2019}),\ \Eprint {https://arxiv.org/abs/1906.11238} {arXiv:1906.11238 [astro-ph.GA]} \BibitemShut {NoStop}%
\bibitem [{\citenamefont {Pujolas}\ \emph {et~al.}(2021)\citenamefont {Pujolas}, \citenamefont {Vaskonen},\ and\ \citenamefont {Veerm\"ae}}]{Pujolas:2021yaw}%
  \BibitemOpen
  \bibfield  {author} {\bibinfo {author} {\bibfnamefont {O.}~\bibnamefont {Pujolas}}, \bibinfo {author} {\bibfnamefont {V.}~\bibnamefont {Vaskonen}},\ and\ \bibinfo {author} {\bibfnamefont {H.}~\bibnamefont {Veerm\"ae}},\ }\bibfield  {title} {\bibinfo {title} {{Prospects for probing gravitational waves from primordial black hole binaries}},\ }\href {https://doi.org/10.1103/PhysRevD.104.083521} {\bibfield  {journal} {\bibinfo  {journal} {Phys. Rev. D}\ }\textbf {\bibinfo {volume} {104}},\ \bibinfo {pages} {083521} (\bibinfo {year} {2021})},\ \Eprint {https://arxiv.org/abs/2107.03379} {arXiv:2107.03379 [astro-ph.CO]} \BibitemShut {NoStop}%
\bibitem [{\citenamefont {Maggiore}(2007)}]{MaggioreBible}%
  \BibitemOpen
  \bibfield  {author} {\bibinfo {author} {\bibfnamefont {M.}~\bibnamefont {Maggiore}},\ }\href {https://doi.org/10.1093/acprof:oso/9780198570745.001.0001} {\emph {\bibinfo {title} {{Gravitational Waves: Volume 1: Theory and Experiments}}}}\ (\bibinfo  {publisher} {Oxford University Press},\ \bibinfo {year} {2007})\BibitemShut {NoStop}%
\bibitem [{\citenamefont {Riles}(2023)}]{Riles_Rev}%
  \BibitemOpen
  \bibfield  {author} {\bibinfo {author} {\bibfnamefont {K.}~\bibnamefont {Riles}},\ }\bibfield  {title} {\bibinfo {title} {{Searches for continuous-wave gravitational radiation}},\ }\href {https://doi.org/10.1007/s41114-023-00044-3} {\bibfield  {journal} {\bibinfo  {journal} {Living Rev. Rel.}\ }\textbf {\bibinfo {volume} {26}},\ \bibinfo {pages} {3} (\bibinfo {year} {2023})}\BibitemShut {NoStop}%
\bibitem [{\citenamefont {Wette}(2023)}]{Wette20yrReview}%
  \BibitemOpen
  \bibfield  {author} {\bibinfo {author} {\bibfnamefont {K.}~\bibnamefont {Wette}},\ }\bibfield  {title} {\bibinfo {title} {{Searches for continuous gravitational waves from neutron stars: A twenty-year retrospective}},\ }\href {https://doi.org/https://doi.org/10.1016/j.astropartphys.2023.102880} {\bibfield  {journal} {\bibinfo  {journal} {Astropart. Phys.}\ }\textbf {\bibinfo {volume} {153}},\ \bibinfo {pages} {102880} (\bibinfo {year} {2023})}\BibitemShut {NoStop}%
\bibitem [{\citenamefont {Tenorio}\ \emph {et~al.}(2021)\citenamefont {Tenorio}, \citenamefont {Keitel},\ and\ \citenamefont {Sintes}}]{Tenorio_Rev}%
  \BibitemOpen
  \bibfield  {author} {\bibinfo {author} {\bibfnamefont {R.}~\bibnamefont {Tenorio}}, \bibinfo {author} {\bibfnamefont {D.}~\bibnamefont {Keitel}},\ and\ \bibinfo {author} {\bibfnamefont {A.~M.}\ \bibnamefont {Sintes}},\ }\bibfield  {title} {\bibinfo {title} {{Search Methods for Continuous Gravitational-Wave Signals from Unknown Sources in the Advanced-Detector Era}},\ }\href {https://doi.org/10.3390/universe7120474} {\bibfield  {journal} {\bibinfo  {journal} {Universe}\ }\textbf {\bibinfo {volume} {7}},\ \bibinfo {pages} {474} (\bibinfo {year} {2021})}\BibitemShut {NoStop}%
\bibitem [{\citenamefont {Piccinni}(2022)}]{Ornella_Review}%
  \BibitemOpen
  \bibfield  {author} {\bibinfo {author} {\bibfnamefont {O.~J.}\ \bibnamefont {Piccinni}},\ }\bibfield  {title} {\bibinfo {title} {{Status and Perspectives of Continuous Gravitational Wave Searches}},\ }\href {https://doi.org/10.3390/galaxies10030072} {\bibfield  {journal} {\bibinfo  {journal} {Galaxies}\ }\textbf {\bibinfo {volume} {10}},\ \bibinfo {pages} {72} (\bibinfo {year} {2022})}\BibitemShut {NoStop}%
\bibitem [{\citenamefont {Piccinni}\ \emph {et~al.}(2020)\citenamefont {Piccinni} \emph {et~al.}}]{CW_GC_O2}%
  \BibitemOpen
  \bibfield  {author} {\bibinfo {author} {\bibfnamefont {O.}~\bibnamefont {Piccinni}} \emph {et~al.},\ }\bibfield  {title} {\bibinfo {title} {{Directed search for continuous gravitational-wave signals from the Galactic Center in the Advanced LIGO second observing run}},\ }\href {https://doi.org/10.1103/PhysRevD.101.082004} {\bibfield  {journal} {\bibinfo  {journal} {Phys. Rev. D}\ }\textbf {\bibinfo {volume} {101}},\ \bibinfo {pages} {082004} (\bibinfo {year} {2020})}\BibitemShut {NoStop}%
\bibitem [{\citenamefont {Abbott}\ \emph {et~al.}(2021{\natexlab{b}})\citenamefont {Abbott} \emph {et~al.}}]{LIGOScientific:2021mwx}%
  \BibitemOpen
  \bibfield  {author} {\bibinfo {author} {\bibfnamefont {R.}~\bibnamefont {Abbott}} \emph {et~al.} (\bibinfo {collaboration} {LIGO Scientific, Virgo, KAGRA, Virgo}),\ }\bibfield  {title} {\bibinfo {title} {{Searches for Continuous Gravitational Waves from Young Supernova Remnants in the Early Third Observing Run of Advanced LIGO and Virgo}},\ }\href {https://doi.org/10.3847/1538-4357/ac17ea} {\bibfield  {journal} {\bibinfo  {journal} {Astrophys. J.}\ }\textbf {\bibinfo {volume} {921}},\ \bibinfo {pages} {80} (\bibinfo {year} {2021}{\natexlab{b}})},\ \Eprint {https://arxiv.org/abs/2105.11641} {arXiv:2105.11641 [astro-ph.HE]} \BibitemShut {NoStop}%
\bibitem [{\citenamefont {Abbott}\ \emph {et~al.}(2021{\natexlab{c}})\citenamefont {Abbott} \emph {et~al.}}]{LIGOScientific:2021yby}%
  \BibitemOpen
  \bibfield  {author} {\bibinfo {author} {\bibfnamefont {R.}~\bibnamefont {Abbott}} \emph {et~al.} (\bibinfo {collaboration} {LIGO Scientific, Virgo, KAGRA}),\ }\bibfield  {title} {\bibinfo {title} {{Constraints from LIGO O3 Data on Gravitational-wave Emission Due to R-modes in the Glitching Pulsar PSR J0537\textendash{}6910}},\ }\href {https://doi.org/10.3847/1538-4357/ac0d52} {\bibfield  {journal} {\bibinfo  {journal} {Astrophys. J.}\ }\textbf {\bibinfo {volume} {922}},\ \bibinfo {pages} {71} (\bibinfo {year} {2021}{\natexlab{c}})},\ \Eprint {https://arxiv.org/abs/2104.14417} {arXiv:2104.14417 [astro-ph.HE]} \BibitemShut {NoStop}%
\bibitem [{\citenamefont {Abbott}\ \emph {et~al.}(2021{\natexlab{d}})\citenamefont {Abbott} \emph {et~al.}}]{LIGOScientific:2020lkw}%
  \BibitemOpen
  \bibfield  {author} {\bibinfo {author} {\bibfnamefont {R.}~\bibnamefont {Abbott}} \emph {et~al.} (\bibinfo {collaboration} {LIGO Scientific, Virgo, KAGRA}),\ }\bibfield  {title} {\bibinfo {title} {{Diving below the Spin-down Limit: Constraints on Gravitational Waves from the Energetic Young Pulsar PSR J0537-6910}},\ }\href {https://doi.org/10.3847/2041-8213/abffcd} {\bibfield  {journal} {\bibinfo  {journal} {Astrophys. J.}\ }\textbf {\bibinfo {volume} {913}},\ \bibinfo {pages} {L27} (\bibinfo {year} {2021}{\natexlab{d}})},\ \Eprint {https://arxiv.org/abs/2012.12926} {arXiv:2012.12926 [astro-ph.HE]} \BibitemShut {NoStop}%
\bibitem [{\citenamefont {Abbott}\ \emph {et~al.}(2021{\natexlab{e}})\citenamefont {Abbott} \emph {et~al.}}]{LIGOScientific:2020qhb}%
  \BibitemOpen
  \bibfield  {author} {\bibinfo {author} {\bibfnamefont {R.}~\bibnamefont {Abbott}} \emph {et~al.} (\bibinfo {collaboration} {LIGO Scientific, Virgo}),\ }\bibfield  {title} {\bibinfo {title} {{All-sky search in early O3 LIGO data for continuous gravitational-wave signals from unknown neutron stars in binary systems}},\ }\href {https://doi.org/10.1103/PhysRevD.103.064017} {\bibfield  {journal} {\bibinfo  {journal} {Phys. Rev. D}\ }\textbf {\bibinfo {volume} {103}},\ \bibinfo {pages} {064017} (\bibinfo {year} {2021}{\natexlab{e}})},\ \bibinfo {note} {[Erratum: Phys.Rev.D 108, 069901 (2023)]},\ \Eprint {https://arxiv.org/abs/2012.12128} {arXiv:2012.12128 [gr-qc]} \BibitemShut {NoStop}%
\bibitem [{\citenamefont {Abbott}\ \emph {et~al.}(2021{\natexlab{f}})\citenamefont {Abbott} \emph {et~al.}}]{KAGRA:2021una}%
  \BibitemOpen
  \bibfield  {author} {\bibinfo {author} {\bibfnamefont {R.}~\bibnamefont {Abbott}} \emph {et~al.} (\bibinfo {collaboration} {LIGO Scientific, Virgo, KAGRA}),\ }\bibfield  {title} {\bibinfo {title} {{All-sky search for continuous gravitational waves from isolated neutron stars in the early O3 LIGO data}},\ }\href {https://doi.org/10.1103/PhysRevD.104.082004} {\bibfield  {journal} {\bibinfo  {journal} {Phys. Rev. D}\ }\textbf {\bibinfo {volume} {104}},\ \bibinfo {pages} {082004} (\bibinfo {year} {2021}{\natexlab{f}})},\ \Eprint {https://arxiv.org/abs/2107.00600} {arXiv:2107.00600 [gr-qc]} \BibitemShut {NoStop}%
\bibitem [{\citenamefont {Abbott}\ \emph {et~al.}(2022{\natexlab{a}})\citenamefont {Abbott} \emph {et~al.}}]{LIGOScientific:2021ozr}%
  \BibitemOpen
  \bibfield  {author} {\bibinfo {author} {\bibfnamefont {R.}~\bibnamefont {Abbott}} \emph {et~al.} (\bibinfo {collaboration} {LIGO Scientific, Virgo, KAGRA}),\ }\bibfield  {title} {\bibinfo {title} {{Search for continuous gravitational waves from 20 accreting millisecond x-ray pulsars in O3 LIGO data}},\ }\href {https://doi.org/10.1103/PhysRevD.105.022002} {\bibfield  {journal} {\bibinfo  {journal} {Phys. Rev. D}\ }\textbf {\bibinfo {volume} {105}},\ \bibinfo {pages} {022002} (\bibinfo {year} {2022}{\natexlab{a}})},\ \Eprint {https://arxiv.org/abs/2109.09255} {arXiv:2109.09255 [astro-ph.HE]} \BibitemShut {NoStop}%
\bibitem [{\citenamefont {Abbott}\ \emph {et~al.}(2022{\natexlab{b}})\citenamefont {Abbott} \emph {et~al.}}]{LIGOScientific:2021inr}%
  \BibitemOpen
  \bibfield  {author} {\bibinfo {author} {\bibfnamefont {R.}~\bibnamefont {Abbott}} \emph {et~al.} (\bibinfo {collaboration} {LIGO Scientific, Virgo}),\ }\bibfield  {title} {\bibinfo {title} {{Search of the early O3 LIGO data for continuous gravitational waves from the Cassiopeia A and Vela Jr. supernova remnants}},\ }\href {https://doi.org/10.1103/PhysRevD.105.082005} {\bibfield  {journal} {\bibinfo  {journal} {Phys. Rev. D}\ }\textbf {\bibinfo {volume} {105}},\ \bibinfo {pages} {082005} (\bibinfo {year} {2022}{\natexlab{b}})},\ \Eprint {https://arxiv.org/abs/2111.15116} {arXiv:2111.15116 [gr-qc]} \BibitemShut {NoStop}%
\bibitem [{\citenamefont {Abbott}\ \emph {et~al.}(2022{\natexlab{c}})\citenamefont {Abbott} \emph {et~al.}}]{KAGRA:2022dqk}%
  \BibitemOpen
  \bibfield  {author} {\bibinfo {author} {\bibfnamefont {R.}~\bibnamefont {Abbott}} \emph {et~al.} (\bibinfo {collaboration} {LIGO Scientific, Virgo, KAGRA}),\ }\bibfield  {title} {\bibinfo {title} {{Search for gravitational waves from Scorpius X-1 with a hidden Markov model in O3 LIGO data}},\ }\href {https://doi.org/10.1103/PhysRevD.106.062002} {\bibfield  {journal} {\bibinfo  {journal} {Phys. Rev. D}\ }\textbf {\bibinfo {volume} {106}},\ \bibinfo {pages} {062002} (\bibinfo {year} {2022}{\natexlab{c}})},\ \Eprint {https://arxiv.org/abs/2201.10104} {arXiv:2201.10104 [gr-qc]} \BibitemShut {NoStop}%
\bibitem [{\citenamefont {Abbott}\ \emph {et~al.}(2022{\natexlab{d}})\citenamefont {Abbott} \emph {et~al.}}]{LIGOScientific:2022enz}%
  \BibitemOpen
  \bibfield  {author} {\bibinfo {author} {\bibfnamefont {R.}~\bibnamefont {Abbott}} \emph {et~al.} (\bibinfo {collaboration} {LIGO Scientific, Virgo, KAGRA}),\ }\bibfield  {title} {\bibinfo {title} {{Model-based Cross-correlation Search for Gravitational Waves from the Low-mass X-Ray Binary Scorpius X-1 in LIGO O3 Data}},\ }\href {https://doi.org/10.3847/2041-8213/aca1b0} {\bibfield  {journal} {\bibinfo  {journal} {Astrophys. J. Lett.}\ }\textbf {\bibinfo {volume} {941}},\ \bibinfo {pages} {L30} (\bibinfo {year} {2022}{\natexlab{d}})},\ \Eprint {https://arxiv.org/abs/2209.02863} {arXiv:2209.02863 [astro-ph.HE]} \BibitemShut {NoStop}%
\bibitem [{\citenamefont {Abbott}\ \emph {et~al.}(2022{\natexlab{e}})\citenamefont {Abbott} \emph {et~al.}}]{CW_allSky_O3}%
  \BibitemOpen
  \bibfield  {author} {\bibinfo {author} {\bibfnamefont {R.}~\bibnamefont {Abbott}} \emph {et~al.} (\bibinfo {collaboration} {LIGO Scientific Collaboration, Virgo Collaboration, KAGRA Collaboration}),\ }\bibfield  {title} {\bibinfo {title} {{All-sky search for continuous gravitational waves from isolated neutron stars using Advanced LIGO and Advanced Virgo O3 data}},\ }\href {https://doi.org/10.1103/PhysRevD.106.102008} {\bibfield  {journal} {\bibinfo  {journal} {Phys. Rev. D}\ }\textbf {\bibinfo {volume} {106}},\ \bibinfo {pages} {102008} (\bibinfo {year} {2022}{\natexlab{e}})}\BibitemShut {NoStop}%
\bibitem [{\citenamefont {Abbott}\ \emph {et~al.}(2022{\natexlab{f}})\citenamefont {Abbott} \emph {et~al.}}]{CW_GC_O3}%
  \BibitemOpen
  \bibfield  {author} {\bibinfo {author} {\bibfnamefont {R.}~\bibnamefont {Abbott}} \emph {et~al.} (\bibinfo {collaboration} {LIGO Scientific Collaboration, Virgo Collaboration, KAGRA Collaboration}),\ }\bibfield  {title} {\bibinfo {title} {{Search for continuous gravitational wave emission from the Milky~Way center in O3 LIGO-Virgo data}},\ }\href {https://doi.org/10.1103/PhysRevD.106.042003} {\bibfield  {journal} {\bibinfo  {journal} {Phys. Rev. D}\ }\textbf {\bibinfo {volume} {106}},\ \bibinfo {pages} {042003} (\bibinfo {year} {2022}{\natexlab{f}})}\BibitemShut {NoStop}%
\bibitem [{\citenamefont {Abbott}\ \emph {et~al.}(2022{\natexlab{g}})\citenamefont {Abbott} \emph {et~al.}}]{CW_Pulsars_O3}%
  \BibitemOpen
  \bibfield  {author} {\bibinfo {author} {\bibfnamefont {R.}~\bibnamefont {Abbott}} \emph {et~al.} (\bibinfo {collaboration} {LIGO Scientific Collaboration, Virgo Collaboration, KAGRA Collaboration}),\ }\bibfield  {title} {\bibinfo {title} {{Searches for Gravitational Waves from Known Pulsars at Two Harmonics in the Second and Third LIGO-Virgo Observing Runs}},\ }\href {https://doi.org/10.3847/1538-4357/ac6acf} {\bibfield  {journal} {\bibinfo  {journal} {Astrophys. J.}\ }\textbf {\bibinfo {volume} {935}},\ \bibinfo {pages} {1} (\bibinfo {year} {2022}{\natexlab{g}})}\BibitemShut {NoStop}%
\bibitem [{\citenamefont {Pitkin}\ \emph {et~al.}(2015)\citenamefont {Pitkin} \emph {et~al.}}]{Pitkin_DualHarmonic}%
  \BibitemOpen
  \bibfield  {author} {\bibinfo {author} {\bibfnamefont {M.}~\bibnamefont {Pitkin}} \emph {et~al.},\ }\bibfield  {title} {\bibinfo {title} {{First results and future prospects for dual-harmonic searches for gravitational waves from spinning neutron stars}},\ }\href {https://doi.org/10.1093/mnras/stv1931} {\bibfield  {journal} {\bibinfo  {journal} {Mon. Not. Roy. Astron. Soc.}\ }\textbf {\bibinfo {volume} {453}},\ \bibinfo {pages} {4399} (\bibinfo {year} {2015})}\BibitemShut {NoStop}%
\bibitem [{\citenamefont {Abbott}\ \emph {et~al.}(2019{\natexlab{c}})\citenamefont {Abbott} \emph {et~al.}}]{CW_TwoHarmonics}%
  \BibitemOpen
  \bibfield  {author} {\bibinfo {author} {\bibfnamefont {B.~P.}\ \bibnamefont {Abbott}} \emph {et~al.} (\bibinfo {collaboration} {LIGO Scientific Collaboration, Virgo Collaboration}),\ }\bibfield  {title} {\bibinfo {title} {{Searches for Gravitational Waves from Known Pulsars at Two Harmonics in 2015-2017 LIGO Data}},\ }\href {https://doi.org/10.3847/1538-4357/ab20cb} {\bibfield  {journal} {\bibinfo  {journal} {Astrophys. J.}\ }\textbf {\bibinfo {volume} {879}},\ \bibinfo {pages} {10} (\bibinfo {year} {2019}{\natexlab{c}})}\BibitemShut {NoStop}%
\bibitem [{\citenamefont {Abbott}\ \emph {et~al.}(2017{\natexlab{a}})\citenamefont {Abbott} \emph {et~al.}}]{CW_TwoHarmonics2}%
  \BibitemOpen
  \bibfield  {author} {\bibinfo {author} {\bibfnamefont {B.~P.}\ \bibnamefont {Abbott}} \emph {et~al.} (\bibinfo {collaboration} {LIGO Scientific Collaboration, Virgo Collaboration}),\ }\bibfield  {title} {\bibinfo {title} {{First search for gravitational waves from known pulsars with Advanced LIGO}},\ }\href {https://doi.org/10.3847/1538-4357/aa677f} {\bibfield  {journal} {\bibinfo  {journal} {Astrophys. J.}\ }\textbf {\bibinfo {volume} {839}},\ \bibinfo {pages} {12} (\bibinfo {year} {2017}{\natexlab{a}})}\BibitemShut {NoStop}%
\bibitem [{\citenamefont {Abbott}\ \emph {et~al.}(2022{\natexlab{h}})\citenamefont {Abbott} \emph {et~al.}}]{CW_LongTransients_O3}%
  \BibitemOpen
  \bibfield  {author} {\bibinfo {author} {\bibfnamefont {R.}~\bibnamefont {Abbott}} \emph {et~al.} (\bibinfo {collaboration} {LIGO Scientific Collaboration, Virgo Collaboration, KAGRA Collaboration}),\ }\bibfield  {title} {\bibinfo {title} {{Narrowband Searches for Continuous and Long-duration Transient Gravitational Waves from Known Pulsars in the LIGO-Virgo Third Observing Run}},\ }\href {https://doi.org/10.3847/1538-4357/ac6ad0} {\bibfield  {journal} {\bibinfo  {journal} {Astrophys. J.}\ }\textbf {\bibinfo {volume} {932}},\ \bibinfo {pages} {133} (\bibinfo {year} {2022}{\natexlab{h}})}\BibitemShut {NoStop}%
\bibitem [{\citenamefont {Horowitz}\ \emph {et~al.}(2020)\citenamefont {Horowitz}, \citenamefont {Papa},\ and\ \citenamefont {Reddy}}]{Horowitz:2019pru}%
  \BibitemOpen
  \bibfield  {author} {\bibinfo {author} {\bibfnamefont {C.~J.}\ \bibnamefont {Horowitz}}, \bibinfo {author} {\bibfnamefont {M.~A.}\ \bibnamefont {Papa}},\ and\ \bibinfo {author} {\bibfnamefont {S.}~\bibnamefont {Reddy}},\ }\bibfield  {title} {\bibinfo {title} {{Gravitational waves from compact dark matter objects in the solar system}},\ }\href {https://doi.org/10.1016/j.physletb.2019.135072} {\bibfield  {journal} {\bibinfo  {journal} {Phys. Lett. B}\ }\textbf {\bibinfo {volume} {800}},\ \bibinfo {pages} {135072} (\bibinfo {year} {2020})},\ \Eprint {https://arxiv.org/abs/1902.08273} {arXiv:1902.08273 [gr-qc]} \BibitemShut {NoStop}%
\bibitem [{\citenamefont {Miller}\ \emph {et~al.}(2021{\natexlab{b}})\citenamefont {Miller} \emph {et~al.}}]{Miller_2021}%
  \BibitemOpen
  \bibfield  {author} {\bibinfo {author} {\bibfnamefont {A.~L.}\ \bibnamefont {Miller}} \emph {et~al.},\ }\bibfield  {title} {\bibinfo {title} {{Probing planetary-mass primordial black holes with continuous gravitational waves}},\ }\href {https://doi.org/10.1016/j.dark.2021.100836} {\bibfield  {journal} {\bibinfo  {journal} {Phys. Dark Univ.}\ }\textbf {\bibinfo {volume} {32}},\ \bibinfo {pages} {100836} (\bibinfo {year} {2021}{\natexlab{b}})},\ \Eprint {https://arxiv.org/abs/2012.12983} {arXiv:2012.12983 [astro-ph.HE]} \BibitemShut {NoStop}%
\bibitem [{\citenamefont {Alestas}\ \emph {et~al.}(2024)\citenamefont {Alestas}, \citenamefont {Morras}, \citenamefont {Yamamoto}, \citenamefont {Garcia-Bellido}, \citenamefont {Kuroyanagi},\ and\ \citenamefont {Nesseris}}]{Alestas:2024ubs}%
  \BibitemOpen
  \bibfield  {author} {\bibinfo {author} {\bibfnamefont {G.}~\bibnamefont {Alestas}}, \bibinfo {author} {\bibfnamefont {G.}~\bibnamefont {Morras}}, \bibinfo {author} {\bibfnamefont {T.~S.}\ \bibnamefont {Yamamoto}}, \bibinfo {author} {\bibfnamefont {J.}~\bibnamefont {Garcia-Bellido}}, \bibinfo {author} {\bibfnamefont {S.}~\bibnamefont {Kuroyanagi}},\ and\ \bibinfo {author} {\bibfnamefont {S.}~\bibnamefont {Nesseris}},\ }\bibfield  {title} {\bibinfo {title} {{Applying the Viterbi algorithm to planetary-mass black hole searches}},\ }\href {https://doi.org/10.1103/PhysRevD.109.123516} {\bibfield  {journal} {\bibinfo  {journal} {Phys. Rev. D}\ }\textbf {\bibinfo {volume} {109}},\ \bibinfo {pages} {123516} (\bibinfo {year} {2024})},\ \Eprint {https://arxiv.org/abs/2401.02314} {arXiv:2401.02314 [astro-ph.CO]} \BibitemShut {NoStop}%
\bibitem [{\citenamefont {Velcani}(2024)}]{velcani2024thesis}%
  \BibitemOpen
  \bibfield  {author} {\bibinfo {author} {\bibfnamefont {E.}~\bibnamefont {Velcani}},\ }\emph {\bibinfo {title} {Study of data analysis methods for the search of gravitational waves from primordial black hole binaries}},\ \href@noop {} {Master's thesis},\ \bibinfo  {school} {Sapienza Università di Roma} (\bibinfo {year} {2024})\BibitemShut {NoStop}%
\bibitem [{\citenamefont {Miller}(2024)}]{Miller:2024khl}%
  \BibitemOpen
  \bibfield  {author} {\bibinfo {author} {\bibfnamefont {A.~L.}\ \bibnamefont {Miller}},\ }\bibfield  {title} {\bibinfo {title} {{Prospects for detecting asteroid-mass primordial black holes in extreme-mass-ratio inspirals with continuous gravitational waves}},\ }\href@noop {} {\  (\bibinfo {year} {2024})},\ \Eprint {https://arxiv.org/abs/2410.01348} {arXiv:2410.01348 [gr-qc]} \BibitemShut {NoStop}%
\bibitem [{\citenamefont {Cutler}\ and\ \citenamefont {Flanagan}(1994)}]{Cutler:1994ys}%
  \BibitemOpen
  \bibfield  {author} {\bibinfo {author} {\bibfnamefont {C.}~\bibnamefont {Cutler}}\ and\ \bibinfo {author} {\bibfnamefont {E.~E.}\ \bibnamefont {Flanagan}},\ }\bibfield  {title} {\bibinfo {title} {{Gravitational waves from merging compact binaries: How accurately can one extract the binary's parameters from the inspiral wave form?}},\ }\href {https://doi.org/10.1103/PhysRevD.49.2658} {\bibfield  {journal} {\bibinfo  {journal} {Phys. Rev. D}\ }\textbf {\bibinfo {volume} {49}},\ \bibinfo {pages} {2658} (\bibinfo {year} {1994})},\ \Eprint {https://arxiv.org/abs/gr-qc/9402014} {arXiv:gr-qc/9402014} \BibitemShut {NoStop}%
\bibitem [{\citenamefont {Poisson}\ and\ \citenamefont {Will}(1995)}]{Poisson:1995ef}%
  \BibitemOpen
  \bibfield  {author} {\bibinfo {author} {\bibfnamefont {E.}~\bibnamefont {Poisson}}\ and\ \bibinfo {author} {\bibfnamefont {C.~M.}\ \bibnamefont {Will}},\ }\bibfield  {title} {\bibinfo {title} {{Gravitational waves from inspiraling compact binaries: Parameter estimation using second postNewtonian wave forms}},\ }\href {https://doi.org/10.1103/PhysRevD.52.848} {\bibfield  {journal} {\bibinfo  {journal} {Phys. Rev. D}\ }\textbf {\bibinfo {volume} {52}},\ \bibinfo {pages} {848} (\bibinfo {year} {1995})},\ \Eprint {https://arxiv.org/abs/gr-qc/9502040} {arXiv:gr-qc/9502040} \BibitemShut {NoStop}%
\bibitem [{\citenamefont {Piccinni}\ \emph {et~al.}(2019)\citenamefont {Piccinni} \emph {et~al.}}]{BSD}%
  \BibitemOpen
  \bibfield  {author} {\bibinfo {author} {\bibfnamefont {O.~J.}\ \bibnamefont {Piccinni}} \emph {et~al.},\ }\bibfield  {title} {\bibinfo {title} {{A new data analysis framework for the search of continuous gravitational wave signals}},\ }\href {https://doi.org/10.1088/1361-6382/aaefb5} {\bibfield  {journal} {\bibinfo  {journal} {Class. Quantum Grav.}\ }\textbf {\bibinfo {volume} {36}},\ \bibinfo {pages} {015008} (\bibinfo {year} {2019})}\BibitemShut {NoStop}%
\bibitem [{\citenamefont {Dupuis}\ and\ \citenamefont {Woan}(2005)}]{Bayesian_CW}%
  \BibitemOpen
  \bibfield  {author} {\bibinfo {author} {\bibfnamefont {R.~J.}\ \bibnamefont {Dupuis}}\ and\ \bibinfo {author} {\bibfnamefont {G.}~\bibnamefont {Woan}},\ }\bibfield  {title} {\bibinfo {title} {{Bayesian estimation of pulsar parameters from gravitational wave data}},\ }\href {https://doi.org/10.1103/PhysRevD.72.102002} {\bibfield  {journal} {\bibinfo  {journal} {Phys. Rev. D}\ }\textbf {\bibinfo {volume} {72}},\ \bibinfo {pages} {102002} (\bibinfo {year} {2005})}\BibitemShut {NoStop}%
\bibitem [{\citenamefont {Sun}\ \emph {et~al.}(2020)\citenamefont {Sun} \emph {et~al.}}]{Sun:2020wke}%
  \BibitemOpen
  \bibfield  {author} {\bibinfo {author} {\bibfnamefont {L.}~\bibnamefont {Sun}} \emph {et~al.},\ }\bibfield  {title} {\bibinfo {title} {{Characterization of systematic error in Advanced LIGO calibration}},\ }\href {https://doi.org/10.1088/1361-6382/abb14e} {\bibfield  {journal} {\bibinfo  {journal} {Class. Quant. Grav.}\ }\textbf {\bibinfo {volume} {37}},\ \bibinfo {pages} {225008} (\bibinfo {year} {2020})},\ \Eprint {https://arxiv.org/abs/2005.02531} {arXiv:2005.02531 [astro-ph.IM]} \BibitemShut {NoStop}%
\bibitem [{\citenamefont {Abbott}\ \emph {et~al.}(2023)\citenamefont {Abbott} \emph {et~al.}}]{GWOSCO3}%
  \BibitemOpen
  \bibfield  {author} {\bibinfo {author} {\bibfnamefont {R.}~\bibnamefont {Abbott}} \emph {et~al.} (\bibinfo {collaboration} {KAGRA, VIRGO, LIGO Scientific}),\ }\bibfield  {title} {\bibinfo {title} {{Open Data from the Third Observing Run of LIGO, Virgo, KAGRA, and GEO}},\ }\href {https://doi.org/10.3847/1538-4365/acdc9f} {\bibfield  {journal} {\bibinfo  {journal} {Astrophys. J. Suppl.}\ }\textbf {\bibinfo {volume} {267}},\ \bibinfo {pages} {29} (\bibinfo {year} {2023})},\ \Eprint {https://arxiv.org/abs/2302.03676} {arXiv:2302.03676 [gr-qc]} \BibitemShut {NoStop}%
\bibitem [{\citenamefont {{Zweizig}}\ and\ \citenamefont {{Riles}}(2021)}]{T2000384}%
  \BibitemOpen
  \bibfield  {author} {\bibinfo {author} {\bibfnamefont {J.}~\bibnamefont {{Zweizig}}}\ and\ \bibinfo {author} {\bibfnamefont {K.}~\bibnamefont {{Riles}}},\ }\href {https://dcc.ligo.org/T2000384/public} {\emph {\bibinfo {title} {Information on self-gating of h(t) used in {O3} continuous-wave and stochastic searches}}},\ \bibinfo {type} {Tech. Rep.}\ \bibinfo {number} {LIGO-T2000384}\ (\bibinfo  {institution} {LIGO Laboratory},\ \bibinfo {year} {2021})\BibitemShut {NoStop}%
\bibitem [{\citenamefont {Vajente}\ \emph {et~al.}(2020)\citenamefont {Vajente}, \citenamefont {Huang}, \citenamefont {Isi}, \citenamefont {Driggers}, \citenamefont {Kissel}, \citenamefont {Szczepa\ifmmode~\acute{n}\else \'{n}\fi{}czyk},\ and\ \citenamefont {Vitale}}]{PhysRevD.101.042003}%
  \BibitemOpen
  \bibfield  {author} {\bibinfo {author} {\bibfnamefont {G.}~\bibnamefont {Vajente}}, \bibinfo {author} {\bibfnamefont {Y.}~\bibnamefont {Huang}}, \bibinfo {author} {\bibfnamefont {M.}~\bibnamefont {Isi}}, \bibinfo {author} {\bibfnamefont {J.~C.}\ \bibnamefont {Driggers}}, \bibinfo {author} {\bibfnamefont {J.~S.}\ \bibnamefont {Kissel}}, \bibinfo {author} {\bibfnamefont {M.~J.}\ \bibnamefont {Szczepa\ifmmode~\acute{n}\else \'{n}\fi{}czyk}},\ and\ \bibinfo {author} {\bibfnamefont {S.}~\bibnamefont {Vitale}},\ }\bibfield  {title} {\bibinfo {title} {Machine-learning nonstationary noise out of gravitational-wave detectors},\ }\href {https://doi.org/10.1103/PhysRevD.101.042003} {\bibfield  {journal} {\bibinfo  {journal} {Phys. Rev. D}\ }\textbf {\bibinfo {volume} {101}},\ \bibinfo {pages} {042003} (\bibinfo {year} {2020})}\BibitemShut {NoStop}%
\bibitem [{\citenamefont {{Goetz}}\ and\ \citenamefont {{Riles}}(2021)}]{T2300068}%
  \BibitemOpen
  \bibfield  {author} {\bibinfo {author} {\bibfnamefont {E.}~\bibnamefont {{Goetz}}}\ and\ \bibinfo {author} {\bibfnamefont {K.}~\bibnamefont {{Riles}}},\ }\href {https://dcc.ligo.org/T2300068/public} {\emph {\bibinfo {title} {Segments used for creating standard {SFTs} in {O3} data}}},\ \bibinfo {type} {Tech. Rep.}\ \bibinfo {number} {LIGO-T2300068}\ (\bibinfo  {institution} {LIGO Laboratory},\ \bibinfo {year} {2021})\BibitemShut {NoStop}%
\bibitem [{\citenamefont {Astone}\ \emph {et~al.}(2005)\citenamefont {Astone}, \citenamefont {Frasca},\ and\ \citenamefont {Palomba}}]{Astone:2005fj}%
  \BibitemOpen
  \bibfield  {author} {\bibinfo {author} {\bibfnamefont {P.}~\bibnamefont {Astone}}, \bibinfo {author} {\bibfnamefont {S.}~\bibnamefont {Frasca}},\ and\ \bibinfo {author} {\bibfnamefont {C.}~\bibnamefont {Palomba}},\ }\bibfield  {title} {\bibinfo {title} {{The short FFT database and the peak map for the hierarchical search of periodic sources}},\ }\href {https://doi.org/10.1088/0264-9381/22/18/S34} {\bibfield  {journal} {\bibinfo  {journal} {Class. Quant. Grav.}\ }\textbf {\bibinfo {volume} {22}},\ \bibinfo {pages} {S1197} (\bibinfo {year} {2005})}\BibitemShut {NoStop}%
\bibitem [{\citenamefont {Abbott}\ \emph {et~al.}(2021{\natexlab{g}})\citenamefont {Abbott} \emph {et~al.}}]{CW_SNR_O3a}%
  \BibitemOpen
  \bibfield  {author} {\bibinfo {author} {\bibfnamefont {R.}~\bibnamefont {Abbott}} \emph {et~al.} (\bibinfo {collaboration} {LIGO Scientific Collaboration, Virgo Collaboration, KAGRA Collaboration}),\ }\bibfield  {title} {\bibinfo {title} {{Searches for Continuous Gravitational Waves from Young Supernova Remnants in the Early Third Observing Run of Advanced LIGO and Virgo}},\ }\href {https://doi.org/10.3847/1538-4357/ac17ea} {\bibfield  {journal} {\bibinfo  {journal} {Astrophys. J.}\ }\textbf {\bibinfo {volume} {921}},\ \bibinfo {pages} {80} (\bibinfo {year} {2021}{\natexlab{g}})}\BibitemShut {NoStop}%
\bibitem [{\citenamefont {Abbott}\ \emph {et~al.}(2020{\natexlab{b}})\citenamefont {Abbott} \emph {et~al.}}]{CW_known_O1_O2_O3}%
  \BibitemOpen
  \bibfield  {author} {\bibinfo {author} {\bibfnamefont {R.}~\bibnamefont {Abbott}} \emph {et~al.} (\bibinfo {collaboration} {LIGO Scientific, Virgo}),\ }\bibfield  {title} {\bibinfo {title} {{Gravitational-wave Constraints on the Equatorial Ellipticity of Millisecond Pulsars}},\ }\href {https://doi.org/10.3847/2041-8213/abb655} {\bibfield  {journal} {\bibinfo  {journal} {Astrophys. J. Lett.}\ }\textbf {\bibinfo {volume} {902}},\ \bibinfo {pages} {L21} (\bibinfo {year} {2020}{\natexlab{b}})},\ \Eprint {https://arxiv.org/abs/2007.14251} {arXiv:2007.14251 [astro-ph.HE]} \BibitemShut {NoStop}%
\bibitem [{\citenamefont {Abbott}\ \emph {et~al.}(2022{\natexlab{i}})\citenamefont {Abbott} \emph {et~al.}}]{CW_known_O2_O3}%
  \BibitemOpen
  \bibfield  {author} {\bibinfo {author} {\bibfnamefont {R.}~\bibnamefont {Abbott}} \emph {et~al.} (\bibinfo {collaboration} {LIGO Scientific, VIRGO, KAGRA}),\ }\bibfield  {title} {\bibinfo {title} {{Searches for Gravitational Waves from Known Pulsars at Two Harmonics in the Second and Third LIGO-Virgo Observing Runs}},\ }\href {https://doi.org/10.3847/1538-4357/ac6acf} {\bibfield  {journal} {\bibinfo  {journal} {Astrophys. J.}\ }\textbf {\bibinfo {volume} {935}},\ \bibinfo {pages} {1} (\bibinfo {year} {2022}{\natexlab{i}})},\ \Eprint {https://arxiv.org/abs/2111.13106} {arXiv:2111.13106 [astro-ph.HE]} \BibitemShut {NoStop}%
\bibitem [{\citenamefont {D'Onofrio}\ \emph {et~al.}(2023)\citenamefont {D'Onofrio} \emph {et~al.}}]{CW_ensemble_O3}%
  \BibitemOpen
  \bibfield  {author} {\bibinfo {author} {\bibfnamefont {L.}~\bibnamefont {D'Onofrio}} \emph {et~al.},\ }\bibfield  {title} {\bibinfo {title} {{Search for gravitational wave signals from known pulsars in LIGO-Virgo O3 data using the 5n-vector ensemble method}},\ }\href {https://doi.org/10.1103/PhysRevD.108.122002} {\bibfield  {journal} {\bibinfo  {journal} {Phys. Rev. D}\ }\textbf {\bibinfo {volume} {108}},\ \bibinfo {pages} {122002} (\bibinfo {year} {2023})},\ \Eprint {https://arxiv.org/abs/2311.08229} {arXiv:2311.08229 [gr-qc]} \BibitemShut {NoStop}%
\bibitem [{\citenamefont {Abbott}\ \emph {et~al.}(2022{\natexlab{j}})\citenamefont {Abbott} \emph {et~al.}}]{CW_BC_O3}%
  \BibitemOpen
  \bibfield  {author} {\bibinfo {author} {\bibfnamefont {R.}~\bibnamefont {Abbott}} \emph {et~al.} (\bibinfo {collaboration} {LIGO Scientific Collaboration, Virgo Collaboration, KAGRA Collaboration}),\ }\bibfield  {title} {\bibinfo {title} {{All-sky search for gravitational wave emission from scalar boson clouds around spinning black holes in LIGO O3 data}},\ }\href {https://doi.org/10.1103/PhysRevD.105.102001} {\bibfield  {journal} {\bibinfo  {journal} {Phys. Rev. D}\ }\textbf {\bibinfo {volume} {105}},\ \bibinfo {pages} {102001} (\bibinfo {year} {2022}{\natexlab{j}})}\BibitemShut {NoStop}%
\bibitem [{\citenamefont {Abbott}\ \emph {et~al.}(2022{\natexlab{k}})\citenamefont {Abbott} \emph {et~al.}}]{CW_DPDM_O3}%
  \BibitemOpen
  \bibfield  {author} {\bibinfo {author} {\bibfnamefont {R.}~\bibnamefont {Abbott}} \emph {et~al.} (\bibinfo {collaboration} {LIGO Scientific Collaboration, Virgo Collaboration, KAGRA Collaboration}),\ }\bibfield  {title} {\bibinfo {title} {{Constraints on dark photon dark matter using data from LIGO\textquoteright{}s and Virgo\textquoteright{}s third observing run}},\ }\href {https://doi.org/10.1103/PhysRevD.105.063030} {\bibfield  {journal} {\bibinfo  {journal} {Phys. Rev. D}\ }\textbf {\bibinfo {volume} {105}},\ \bibinfo {pages} {063030} (\bibinfo {year} {2022}{\natexlab{k}})}\BibitemShut {NoStop}%
\bibitem [{\citenamefont {Andr\'es-Carcasona}\ \emph {et~al.}(2024{\natexlab{b}})\citenamefont {Andr\'es-Carcasona}, \citenamefont {Piccinni}, \citenamefont {Mart\'\i{}nez},\ and\ \citenamefont {Mir}}]{Andres-Carcasona:2023PoS}%
  \BibitemOpen
  \bibfield  {author} {\bibinfo {author} {\bibfnamefont {M.}~\bibnamefont {Andr\'es-Carcasona}}, \bibinfo {author} {\bibfnamefont {O.~J.}\ \bibnamefont {Piccinni}}, \bibinfo {author} {\bibfnamefont {M.}~\bibnamefont {Mart\'\i{}nez}},\ and\ \bibinfo {author} {\bibfnamefont {L.-M.}\ \bibnamefont {Mir}},\ }\bibfield  {title} {\bibinfo {title} {{BSD-COBI: New search pipeline to target inspiraling light dark compact objects.}},\ }\href {https://doi.org/10.22323/1.449.0067} {\bibfield  {journal} {\bibinfo  {journal} {PoS}\ }\textbf {\bibinfo {volume} {EPS-HEP2023}},\ \bibinfo {pages} {067} (\bibinfo {year} {2024}{\natexlab{b}})}\BibitemShut {NoStop}%
\bibitem [{\citenamefont {Astone}\ \emph {et~al.}(2014)\citenamefont {Astone} \emph {et~al.}}]{FreqHough}%
  \BibitemOpen
  \bibfield  {author} {\bibinfo {author} {\bibfnamefont {P.}~\bibnamefont {Astone}} \emph {et~al.},\ }\bibfield  {title} {\bibinfo {title} {{Method for all-sky searches of continuous gravitational wave signals using the frequency-Hough transform}},\ }\href {https://doi.org/10.1103/PhysRevD.90.042002} {\bibfield  {journal} {\bibinfo  {journal} {Phys. Rev. D}\ }\textbf {\bibinfo {volume} {90}},\ \bibinfo {pages} {042002} (\bibinfo {year} {2014})}\BibitemShut {NoStop}%
\bibitem [{\citenamefont {Acernese}\ \emph {et~al.}(2023{\natexlab{a}})\citenamefont {Acernese} \emph {et~al.}}]{Virgo:2022kwz}%
  \BibitemOpen
  \bibfield  {author} {\bibinfo {author} {\bibfnamefont {F.}~\bibnamefont {Acernese}} \emph {et~al.} (\bibinfo {collaboration} {Virgo}),\ }\bibfield  {title} {\bibinfo {title} {{Virgo detector characterization and data quality: tools}},\ }\href {https://doi.org/10.1088/1361-6382/acdf36} {\bibfield  {journal} {\bibinfo  {journal} {Class. Quant. Grav.}\ }\textbf {\bibinfo {volume} {40}},\ \bibinfo {pages} {185005} (\bibinfo {year} {2023}{\natexlab{a}})},\ \Eprint {https://arxiv.org/abs/2210.15634} {arXiv:2210.15634 [gr-qc]} \BibitemShut {NoStop}%
\bibitem [{\citenamefont {Acernese}\ \emph {et~al.}(2023{\natexlab{b}})\citenamefont {Acernese} \emph {et~al.}}]{Virgo:2022ysc}%
  \BibitemOpen
  \bibfield  {author} {\bibinfo {author} {\bibfnamefont {F.}~\bibnamefont {Acernese}} \emph {et~al.} (\bibinfo {collaboration} {Virgo}),\ }\bibfield  {title} {\bibinfo {title} {{Virgo detector characterization and data quality: results from the O3 run}},\ }\href {https://doi.org/10.1088/1361-6382/acd92d} {\bibfield  {journal} {\bibinfo  {journal} {Class. Quant. Grav.}\ }\textbf {\bibinfo {volume} {40}},\ \bibinfo {pages} {185006} (\bibinfo {year} {2023}{\natexlab{b}})},\ \Eprint {https://arxiv.org/abs/2210.15633} {arXiv:2210.15633 [gr-qc]} \BibitemShut {NoStop}%
\bibitem [{\citenamefont {{Davis}}\ and\ \citenamefont {et~al.}(2021)}]{DavisdetcharO3}%
  \BibitemOpen
  \bibfield  {author} {\bibinfo {author} {\bibfnamefont {D.}~\bibnamefont {{Davis}}}\ and\ \bibinfo {author} {\bibnamefont {et~al.}},\ }\bibfield  {title} {\bibinfo {title} {{LIGO} detector characterization in the second and third observing runs},\ }\href {https://doi.org/10.1088/1361-6382/abfd85} {\bibfield  {journal} {\bibinfo  {journal} {Classical and Quantum Gravity}\ }\textbf {\bibinfo {volume} {38}},\ \bibinfo {eid} {135014} (\bibinfo {year} {2021})},\ \Eprint {https://arxiv.org/abs/2101.11673} {arXiv:2101.11673 [astro-ph.IM]} \BibitemShut {NoStop}%
\bibitem [{\citenamefont {{Covas}}\ and\ \citenamefont {et~al.}(2018)}]{CovaslinesO1O2}%
  \BibitemOpen
  \bibfield  {author} {\bibinfo {author} {\bibfnamefont {P.~B.}\ \bibnamefont {{Covas}}}\ and\ \bibinfo {author} {\bibnamefont {et~al.}},\ }\bibfield  {title} {\bibinfo {title} {Identification and mitigation of narrow spectral artifacts that degrade searches for persistent gravitational waves in the first two observing runs of advanced {LIGO}},\ }\href {https://doi.org/10.1103/PhysRevD.97.082002} {\bibfield  {journal} {\bibinfo  {journal} {Physical Review D.}\ }\textbf {\bibinfo {volume} {97}},\ \bibinfo {eid} {082002} (\bibinfo {year} {2018})},\ \Eprint {https://arxiv.org/abs/1801.07204} {arXiv:1801.07204 [astro-ph.IM]} \BibitemShut {NoStop}%
\bibitem [{\citenamefont {Sun}\ \emph {et~al.}(2021)\citenamefont {Sun} \emph {et~al.}}]{Sun:2021qcg}%
  \BibitemOpen
  \bibfield  {author} {\bibinfo {author} {\bibfnamefont {L.}~\bibnamefont {Sun}} \emph {et~al.},\ }\bibfield  {title} {\bibinfo {title} {{Characterization of systematic error in Advanced LIGO calibration in the second half of O3}},\ }\href@noop {} {\  (\bibinfo {year} {2021})},\ \Eprint {https://arxiv.org/abs/2107.00129} {arXiv:2107.00129 [astro-ph.IM]} \BibitemShut {NoStop}%
\bibitem [{\citenamefont {Soni}\ \emph {et~al.}(2024)\citenamefont {Soni} \emph {et~al.}}]{LIGO:2024kkz}%
  \BibitemOpen
  \bibfield  {author} {\bibinfo {author} {\bibfnamefont {S.}~\bibnamefont {Soni}} \emph {et~al.} (\bibinfo {collaboration} {LIGO}),\ }\bibfield  {title} {\bibinfo {title} {{LIGO Detector Characterization in the first half of the fourth Observing run}},\ }\href@noop {} {\  (\bibinfo {year} {2024})},\ \Eprint {https://arxiv.org/abs/2409.02831} {arXiv:2409.02831 [astro-ph.IM]} \BibitemShut {NoStop}%
\bibitem [{\citenamefont {Davis}\ \emph {et~al.}(2021)\citenamefont {Davis} \emph {et~al.}}]{LIGO:2021ppb}%
  \BibitemOpen
  \bibfield  {author} {\bibinfo {author} {\bibfnamefont {D.}~\bibnamefont {Davis}} \emph {et~al.} (\bibinfo {collaboration} {LIGO}),\ }\bibfield  {title} {\bibinfo {title} {{LIGO detector characterization in the second and third observing runs}},\ }\href {https://doi.org/10.1088/1361-6382/abfd85} {\bibfield  {journal} {\bibinfo  {journal} {Class. Quant. Grav.}\ }\textbf {\bibinfo {volume} {38}},\ \bibinfo {pages} {135014} (\bibinfo {year} {2021})},\ \Eprint {https://arxiv.org/abs/2101.11673} {arXiv:2101.11673 [astro-ph.IM]} \BibitemShut {NoStop}%
\bibitem [{\citenamefont {{ET science team}}(2011)}]{ETcds}%
  \BibitemOpen
  \bibfield  {author} {\bibinfo {author} {\bibnamefont {{ET science team}}},\ }\href@noop {} {\emph {\bibinfo {title} {Einstein gravitational wave telescope conceptual design study}}},\ \bibinfo {type} {Tech. Rep.}\ \bibinfo {number} {ET-0106C-10}\ (\bibinfo  {institution} {ET},\ \bibinfo {year} {2011})\BibitemShut {NoStop}%
\bibitem [{\citenamefont {{ET steering committee}}(2020)}]{ETdesign}%
  \BibitemOpen
  \bibfield  {author} {\bibinfo {author} {\bibnamefont {{ET steering committee}}},\ }\href@noop {} {\emph {\bibinfo {title} {{ET design report update 2020}}}},\ \bibinfo {type} {Tech. Rep.}\ \bibinfo {number} {ET-0007B-20}\ (\bibinfo  {institution} {ET},\ \bibinfo {year} {2020})\BibitemShut {NoStop}%
\bibitem [{\citenamefont {Abbott}\ \emph {et~al.}(2017{\natexlab{b}})\citenamefont {Abbott} \emph {et~al.}}]{CE1}%
  \BibitemOpen
  \bibfield  {author} {\bibinfo {author} {\bibfnamefont {B.~P.}\ \bibnamefont {Abbott}} \emph {et~al.} (\bibinfo {collaboration} {LIGO Scientific Collaboration}),\ }\bibfield  {title} {\bibinfo {title} {{Exploring the sensitivity of next generation gravitational wave detectors}},\ }\href {https://doi.org/10.1088/1361-6382/aa51f4} {\bibfield  {journal} {\bibinfo  {journal} {Class. Quantum Grav.}\ }\textbf {\bibinfo {volume} {34}},\ \bibinfo {pages} {044001} (\bibinfo {year} {2017}{\natexlab{b}})}\BibitemShut {NoStop}%
\bibitem [{\citenamefont {Reitze}\ \emph {et~al.}(2019{\natexlab{a}})\citenamefont {Reitze} \emph {et~al.}}]{CE2}%
  \BibitemOpen
  \bibfield  {author} {\bibinfo {author} {\bibfnamefont {D.}~\bibnamefont {Reitze}} \emph {et~al.},\ }\bibfield  {title} {\bibinfo {title} {{The US Program in Ground-Based Gravitational Wave Science: Contribution from the LIGO Laboratory}},\ }\href {https://doi.org/10.48550/arXiv.1903.04615} {\bibfield  {journal} {\bibinfo  {journal} {Bull. Am. Astron. Soc.}\ }\textbf {\bibinfo {volume} {51}},\ \bibinfo {pages} {141} (\bibinfo {year} {2019}{\natexlab{a}})}\BibitemShut {NoStop}%
\bibitem [{\citenamefont {Reitze}\ \emph {et~al.}(2019{\natexlab{b}})\citenamefont {Reitze} \emph {et~al.}}]{CE3}%
  \BibitemOpen
  \bibfield  {author} {\bibinfo {author} {\bibfnamefont {D.}~\bibnamefont {Reitze}} \emph {et~al.},\ }\bibfield  {title} {\bibinfo {title} {{Cosmic Explorer: The U.S. Contribution to Gravitational-Wave Astronomy beyond LIGO}},\ }\href {https://doi.org/10.48550/arXiv.1903.04615} {\bibfield  {journal} {\bibinfo  {journal} {Bull. Am. Astron. Soc.}\ }\textbf {\bibinfo {volume} {51}},\ \bibinfo {pages} {035} (\bibinfo {year} {2019}{\natexlab{b}})}\BibitemShut {NoStop}%
\bibitem [{\citenamefont {Barsotti}\ \emph {et~al.}(2018)\citenamefont {Barsotti}, \citenamefont {McCuller}, \citenamefont {Evans},\ and\ \citenamefont {Fritschel}}]{A+sensitivity}%
  \BibitemOpen
  \bibfield  {author} {\bibinfo {author} {\bibfnamefont {L.}~\bibnamefont {Barsotti}}, \bibinfo {author} {\bibfnamefont {L.}~\bibnamefont {McCuller}}, \bibinfo {author} {\bibfnamefont {M.}~\bibnamefont {Evans}},\ and\ \bibinfo {author} {\bibfnamefont {P.}~\bibnamefont {Fritschel}},\ }\href@noop {} {\emph {\bibinfo {title} {{The A+ design curve}}}},\ \bibinfo {type} {Tech. Rep.}\ \bibinfo {number} {LIGO-T1800042-v4}\ (\bibinfo  {institution} {LIGO},\ \bibinfo {year} {2018})\BibitemShut {NoStop}%
\bibitem [{\citenamefont {{Abuter}}\ \emph {et~al.}(2019)\citenamefont {{Abuter}} \emph {et~al.}}]{distGC1}%
  \BibitemOpen
  \bibfield  {author} {\bibinfo {author} {\bibfnamefont {R.}~\bibnamefont {{Abuter}}} \emph {et~al.} (\bibinfo {collaboration} {GRAVITY}),\ }\bibfield  {title} {\bibinfo {title} {{A geometric distance measurement to the Galactic center black hole with 0.3\% uncertainty}},\ }\href {https://doi.org/10.1051/0004-6361/201935656} {\bibfield  {journal} {\bibinfo  {journal} {Astron. Astrophys.}\ }\textbf {\bibinfo {volume} {625}},\ \bibinfo {eid} {L10} (\bibinfo {year} {2019})},\ \Eprint {https://arxiv.org/abs/1904.05721} {arXiv:1904.05721 [astro-ph.GA]} \BibitemShut {NoStop}%
\bibitem [{\citenamefont {Francis}\ and\ \citenamefont {Anderson}(2014)}]{distGC3}%
  \BibitemOpen
  \bibfield  {author} {\bibinfo {author} {\bibfnamefont {C.}~\bibnamefont {Francis}}\ and\ \bibinfo {author} {\bibfnamefont {E.}~\bibnamefont {Anderson}},\ }\bibfield  {title} {\bibinfo {title} {{Two estimates of the distance to the Galactic centre}},\ }\href {https://doi.org/10.1093/mnras/stu631} {\bibfield  {journal} {\bibinfo  {journal} {Mon. Not. Roy. Astron. Soc.}\ }\textbf {\bibinfo {volume} {441}},\ \bibinfo {pages} {1105} (\bibinfo {year} {2014})},\ \Eprint {https://arxiv.org/abs/1309.2629} {arXiv:1309.2629 [astro-ph.GA]} \BibitemShut {NoStop}%
\bibitem [{\citenamefont {Hirota}\ \emph {et~al.}(2020)\citenamefont {Hirota} \emph {et~al.}}]{distGC4}%
  \BibitemOpen
  \bibfield  {author} {\bibinfo {author} {\bibfnamefont {T.}~\bibnamefont {Hirota}} \emph {et~al.} (\bibinfo {collaboration} {VERA}),\ }\bibfield  {title} {\bibinfo {title} {{The First VERA Astrometry Catalog}},\ }\href {https://doi.org/10.1093/pasj/psaa018} {\bibfield  {journal} {\bibinfo  {journal} {Publ. Astron. Soc. Jpn}\ }\textbf {\bibinfo {volume} {72}},\ \bibinfo {pages} {50} (\bibinfo {year} {2020})}\BibitemShut {NoStop}%
\bibitem [{\citenamefont {Astone}\ \emph {et~al.}(2010)\citenamefont {Astone}, \citenamefont {D'Antonio}, \citenamefont {Frasca},\ and\ \citenamefont {Palomba}}]{Astone:2010zz}%
  \BibitemOpen
  \bibfield  {author} {\bibinfo {author} {\bibfnamefont {P.}~\bibnamefont {Astone}}, \bibinfo {author} {\bibfnamefont {S.}~\bibnamefont {D'Antonio}}, \bibinfo {author} {\bibfnamefont {S.}~\bibnamefont {Frasca}},\ and\ \bibinfo {author} {\bibfnamefont {C.}~\bibnamefont {Palomba}},\ }\bibfield  {title} {\bibinfo {title} {{A method for detection of known sources of continuous gravitational wave signals in non-stationary data}},\ }\href {https://doi.org/10.1088/0264-9381/27/19/194016} {\bibfield  {journal} {\bibinfo  {journal} {Class. Quant. Grav.}\ }\textbf {\bibinfo {volume} {27}},\ \bibinfo {pages} {194016} (\bibinfo {year} {2010})}\BibitemShut {NoStop}%
\bibitem [{\citenamefont {D'Onofrio}\ \emph {et~al.}(2022)\citenamefont {D'Onofrio}, \citenamefont {De~Rosa}, \citenamefont {Errico}, \citenamefont {Palomba}, \citenamefont {Sequino},\ and\ \citenamefont {Trozzo}}]{DOnofrio:2022zvw}%
  \BibitemOpen
  \bibfield  {author} {\bibinfo {author} {\bibfnamefont {L.}~\bibnamefont {D'Onofrio}}, \bibinfo {author} {\bibfnamefont {R.}~\bibnamefont {De~Rosa}}, \bibinfo {author} {\bibfnamefont {L.}~\bibnamefont {Errico}}, \bibinfo {author} {\bibfnamefont {C.}~\bibnamefont {Palomba}}, \bibinfo {author} {\bibfnamefont {V.}~\bibnamefont {Sequino}},\ and\ \bibinfo {author} {\bibfnamefont {L.}~\bibnamefont {Trozzo}},\ }\bibfield  {title} {\bibinfo {title} {{5n-vector ensemble method for detecting gravitational waves from known pulsars}},\ }\href {https://doi.org/10.1103/PhysRevD.105.063012} {\bibfield  {journal} {\bibinfo  {journal} {Phys. Rev. D}\ }\textbf {\bibinfo {volume} {105}},\ \bibinfo {pages} {063012} (\bibinfo {year} {2022})}\BibitemShut {NoStop}%
\bibitem [{\citenamefont {D'Antonio}\ \emph {et~al.}(2023)\citenamefont {D'Antonio} \emph {et~al.}}]{DAntonio:2023jxm}%
  \BibitemOpen
  \bibfield  {author} {\bibinfo {author} {\bibfnamefont {S.}~\bibnamefont {D'Antonio}} \emph {et~al.},\ }\bibfield  {title} {\bibinfo {title} {{Semicoherent method to search for continuous gravitational waves}},\ }\href {https://doi.org/10.1103/PhysRevD.108.122001} {\bibfield  {journal} {\bibinfo  {journal} {Phys. Rev. D}\ }\textbf {\bibinfo {volume} {108}},\ \bibinfo {pages} {122001} (\bibinfo {year} {2023})},\ \Eprint {https://arxiv.org/abs/2311.06021} {arXiv:2311.06021 [gr-qc]} \BibitemShut {NoStop}%
\bibitem [{\citenamefont {{Frasca}}\ \emph {et~al.}(2005)\citenamefont {{Frasca}} \emph {et~al.}}]{Frasca2005}%
  \BibitemOpen
  \bibfield  {author} {\bibinfo {author} {\bibfnamefont {S.}~\bibnamefont {{Frasca}}} \emph {et~al.},\ }\bibfield  {title} {\bibinfo {title} {{Evaluation of sensitivity and computing power for the Virgo hierarchical search for periodic sources}},\ }\href {https://doi.org/10.1088/0264-9381/22/18/S15} {\bibfield  {journal} {\bibinfo  {journal} {Class. Quantum Grav.}\ }\textbf {\bibinfo {volume} {22}},\ \bibinfo {pages} {S1013} (\bibinfo {year} {2005})}\BibitemShut {NoStop}%
\end{thebibliography}%
\end{document}